\documentclass[final,5p,times,twocolumn,authoryear]{elsarticle}

\usepackage{lineno}

\usepackage{cleveref}
\usepackage{hyperref} 

\usepackage[fleqn]{amsmath}
\usepackage{graphicx}
\usepackage{color}
\usepackage{caption}
\usepackage{subcaption}
\usepackage{wasysym}
\usepackage{morefloats}
\usepackage{balance}
\usepackage{longtable}
\usepackage{booktabs}
\usepackage{tabu}
\usepackage{xcolor}
\usepackage{placeins}
\usepackage[]{units}
\usepackage{supertabular,booktabs}
\usepackage[export]{adjustbox} 

\definecolor{hlinecolor}{RGB}{0,71,171}
\definecolor{headercolor}{gray}{0.85}

\usepackage{siunitx}
\usepackage{xcolor}
\usepackage{booktabs,colortbl, array}
\usepackage{pgfplotstable}
\pgfplotsset{compat=1.5}
\definecolor{rulecolor}{RGB}{0,71,171}
\definecolor{tableheadcolor}{gray}{0.92}
\newcommand{\topline}{ %
        \arrayrulecolor{rulecolor}\specialrule{0.1em}{\abovetopsep}{0pt}%
        \arrayrulecolor{tableheadcolor}\specialrule{\belowrulesep}{0pt}{0pt}%
        \arrayrulecolor{rulecolor}}
\newcommand{\midtopline}{ %
        \arrayrulecolor{tableheadcolor}\specialrule{\aboverulesep}{0pt}{0pt}%
        \arrayrulecolor{rulecolor}\specialrule{\lightrulewidth}{0pt}{0pt}%
        \arrayrulecolor{white}\specialrule{\belowrulesep}{0pt}{0pt}%
        \arrayrulecolor{rulecolor}}
\newcommand{\bottomline}{ %
        \arrayrulecolor{white}\specialrule{\aboverulesep}{0pt}{0pt}%
        \arrayrulecolor{rulecolor} %
        \specialrule{\heavyrulewidth}{0pt}{\belowbottomsep}}%

\pgfplotstableset{
        normal/.style={%
        header=true,
        string type,
        font=\tiny,
        column type=l,
        every odd row/.style={
            before row=
        },
        every head row/.style={
            before row={\topline\rowcolor{tableheadcolor}},
            after row={\midtopline}
        },
        every last row/.style={
            after row=\bottomline
        },
        col sep=&,
        row sep=\\
        }
}

\journal{Icarus}

\begin{document}

\begin{frontmatter}



\title{The effects of short-lived radionuclides and porosity on the early thermo-mechanical evolution of planetesimals}


\author[inst1,inst2]{Tim Lichtenberg\corref{cor1}}
\author[inst2,inst3]{Gregor J. Golabek}
\author[inst2]{Taras V. Gerya}
\author[inst1]{Michael R. Meyer}

\cortext[cor1]{Corresponding author. E-mail: \href{mailto:tim.lichtenberg@phys.ethz.ch}{tim.lichtenberg@phys.ethz.ch}.}

\address[inst1]{Institute for Astronomy, ETH Z{\"u}rich, Wolfgang-Pauli-Strasse 27, 8093 Z{\"u}rich, Switzerland}
\address[inst2]{Institute of Geophysics, ETH Z{\"u}rich, Sonneggstrasse 5, 8092 Z{\"u}rich, Switzerland}
\address[inst3]{Bayerisches Geoinstitut, University of Bayreuth, Universit{\"a}tsstrasse 30, 95440 Bayreuth, Germany}


\begin{abstract}

The thermal history and internal structure of chondritic planetesimals, assembled before the giant impact phase of chaotic growth, potentially yield important implications for the final composition and evolution of terrestrial planets. These parameters critically depend on the internal balance of heating versus cooling, which is mostly determined by the presence of short-lived radionuclides (SLRs), such as $^{26}$Al and $^{60}$Fe, as well as the heat conductivity of the material. The heating by SLRs depends on their initial abundances, the formation time of the planetesimal and its size. It has been argued that the cooling history is determined by the porosity of the granular material, which undergoes dramatic changes via compaction processes and tends to decrease with time. In this study we assess the influence of these parameters on the thermo-mechanical evolution of young planetesimals with both 2D and 3D simulations. Using the code family \textsc{i2elvis/i3elvis} we have run numerous 2D and 3D numerical finite-difference fluid dynamic models with varying planetesimal radius, formation time and initial porosity. Our results indicate that powdery materials lowered the threshold for melting and convection in planetesimals, depending on the amount of SLRs present. A subset of planetesimals retained a powdery surface layer which lowered the thermal conductivity and hindered cooling. The effect of initial porosity was small, however, compared to those of planetesimal size and formation time, which dominated the thermo-mechanical evolution and were the primary factors for the onset of melting and differentiation. We comment on the implications of this work concerning the structure and evolution of these planetesimals, as well as their behavior as possible building blocks of terrestrial planets.

\end{abstract}

\begin{keyword}
Planetary formation
\sep Terrestrial planets
\sep Planetesimals
\sep Interiors
\sep Thermal histories




\end{keyword}

\end{frontmatter}



\section{Introduction}
\label{sec:introduction}

During the early stages of planet formation the building material of terrestrial planets like Earth or Mars is distributed within planetesimals with sizes of $\sim$ $10^1$--$10^2$ km \citep{2006mess.book..473W}. It remains unclear how these bodies assembled from sub-micron grains in a circumstellar disk in detail. First order constraints from the standard collisional model for growth relate the doubling time $t_{\mathrm{s}} \sim \rho_{\mathrm{p}} R_{\mathrm{p}} / (\Sigma_{\mathrm{disk}} \Omega_{\mathrm{K}})$ of a growing planetesimal to its size $R_{\mathrm{p}}$ and density $\rho_{\mathrm{p}}$ as well as to the properties of the disk, namely mass surface density $\Sigma_{\mathrm{disk}}$ and Keplerian frequency $\Omega_{\mathrm{K}}$ \citep{2004ARAA..42..549G}. This formula, however, essentially a cross-section calculation, ignores gravitational focusing and limits to growth, such as the bouncing barrier \citep[e.g.,][]{2010AA...513A..57Z} and the radial migration of solids due to gas drag \citep{1977MNRAS.180...57W}. Nonetheless, there are also complex local processes that can enhance the formation of planetesimals with up to several hundred kilometers radii due to particle collection in vortices, pressure bumps, and other effects \citep[e.g.,][]{2007Natur.448.1022J,2008ApJ...687.1432C,2009Icar..204..558M,2010Icar..208..505C,2015SciA....115109J}. These point to rapid formation on the time scale of $\sim 10^5$ yr after the formation of Ca-Al-rich inclusions (CAIs), consistent with findings from geochemical data \citep{2009GeCoA..73.5150K}.

Theoretical models to investigate this epoch after the initial assembly of the planetesimals rely on numerical models of internal dynamics. So far, such models were mostly based on 1D studies, focusing on conductive cooling as the main heat transfer mechanism \citep[e.g., ][]{1998Icar..134..187G,2006MPS...41...95H,sahijpal2007numerical}. Recent work, however, has shown that more mechanisms need to be taken into account. Firstly, these bodies are supposed to be sufficiently big to become heated by decay of short-lived radionuclides (SLRs), most importantly $^{26}$Al and $^{60}$Fe, which would alter their inner structure and evolution dramatically up to the point of silicate melting. For example, bodies greater than $\sim$ 10 km in radius, formed at the time of CAI formation, are supposed to melt completely \citep{2006MPS...41...95H}. Secondly, some meteorite parent bodies seem to have experienced solid-state deformation \citep{2013NatGe...6...93T,2014MPS...49.1202Ta}. These points underline the importance of 2D or 3D thermo-mechanical modeling approaches for the evolution of planetesimals to detect effects such as the differences of the surface-to-volume ratio in 1D, 2D and 3D models or non-axisymmetric advection processes. As a further complicating issue, recent work highlights the potentially important role of porous bulk material on the thermal history of planetesimals, by lowering the thermal conductivity of the silicate material and thus to prevent effective heat transport via conduction \citep{2008ApJ...687.1432C,2014AA...567A.120N}. 

The initial powdery state of the uncompacted material is however reduced in the inner parts of the planetesimals by cold isostatic compaction due to self-gravity \citep{2012AA...537A..45H}, effectively decreasing its influence with increasing size of the body. Another important aspect is the formation time of the body. As outlined above, the accretion time scale of planetesimals is on the order of $10^5$ yr, which is roughly an order of magnitude shorter than the evolutionary time scale of the protoplanetary disk and the thermo-mechanical evolution of planetesimals on the order of $10^6$ yr. Hence, the quasi-instantaneous formation time sets the limit on the amount of SLRs incorporated into the body. 

Additional heat sources for planetesimals can be energy injection during the accretion of the body and later impacts. First, the temperature increase due to the conversion of gravitational energy to heat is low for bodies $<$ 1000 km \citep{schubert1986thermal,2008EPSL.273...94Q,2011E&PSL.305....1E}. Second, during runaway growth, the velocity dispersion of planetesimals is set by the equilibrium between self-stirring and gas drag. Impact velocities are therefore comparable or smaller to the escape velocity \citep{1978Icar...35....1G,2015Icar..258..418M}, which drastically limits the amount of injected energy. The formation time thus dominates the energy budget for heating and sets the pace of internal dynamic processes, such as core formation, to the order of several $^{26}$Al half-lives.

Clearly, the thermo-mechanical evolution of planetesimals needs to be treated adequatly to achieve a consistent theoretical understanding of this stage of planetary assembly. In this study we assessed the role of the initial size, formation time and porosity of planetesimals on their thermo-mechanical history via 2D and 3D numerical models. In Sect. \ref{sec:methods} we describe constraints from earlier work and outline the most important concepts of our numerical model; in Sect. \ref{sec:results} we present the results obtained from the simulation runs, for which we outline the technically inherent limitations in Sect. \ref{sec:limitations}. In Sect. \ref{sec:discussion} we discuss the physical implications and draw conclusions in Sect. \ref{sec:conclusions}. Supplementary material can be found in \ref{sec:supplementary_material} and a list of all simulations is given in \ref{sec:sim_tables}.


\section{Physical and numerical methodology}
\label{sec:methods}

The physical and numerical methods in this work follow earlier work by \citet{2014MPS...49.1083G}, in which an in-depth analysis of observational constraints on the thermal history for the acapulcoite-lodranite parent body is compiled. In contrast to this study, we focused on the general role of planetesimal evolution and seeked to explore the thermo-mechnical regimes before the onset of the giant impact phase in terrestrial planet formation. 
The most important physical constants used in the model are explained in the following sections, all others are listed with their respective references in Table \ref{tab:constants}.

\begin{table*}[t]
\small{
\centering
\begin{tabu}{lllll} 
    \taburulecolor{hlinecolor}
    \toprule
    \rowcolor{headercolor}
    \textsc{Parameter} & \textsc{Symbol} &  \textsc{Value} & \textsc{Unit} & \textsc{Reference} \\
    \midrule
    Density of uncompressed solid silicates  & $\rho_{\mathrm{Si-sol}}$ & 3500 & kg m$^{-3}$ & \cite{1981JGR....86.6261S}; \\
    &&&& \cite{1998PEPI..107...53S} \\
    Density of uncompressed molten silicates & $\rho_{\mathrm{Si-liq}}$ & 2900 & kg m$^{-3}$ &   \cite{1981JGR....86.6261S}  \\
    Temperature of space (sticky air) & $T_{\mathrm{space}}$ & 290 & K &  \cite{1998Icar..134..187G}; \\
    &&&& \cite{1976ARAA..14...81B}  \\
    Activation energy & $E_{\mathrm{a}}$ & 470 & kJ mol$^{-1}$  &  \citet{ranalli1995rheology}  \\
    Dislocation creep onset stress & $\sigma_{\mathrm{0}}$ & $3 \cdot 10^7$ & Pa &  \cite{turcotte2014geodynamics}   \\
    Power law exponent & $n$ & 4 &  & \citet{ranalli1995rheology}  \\
    Latent heat of silicate melting & $L_{\mathrm{Si}}$ & 400 & kJ kg$^{-1}$ &  \cite{1998Icar..134..187G}; \\
    &&&& \cite{turcotte2014geodynamics}   \\
    Silicate melt fraction  & $\varphi_{\mathrm{crit}}$ & 0.4 & non-dim. & \cite{solomatov2015magma};\\
    \hspace{1cm} at rheological transition &&&& \cite{2009GGG....10.3010C}  \\
    Heat capacity of of silicates & $c_{\mathrm{P}}$ & 1000 & J kg$^{-1}$ K$^{-1}$ &  \cite{turcotte2014geodynamics}   \\
    Thermal expansivity of solid silicates & $\alpha_{\mathrm{Si-sol}}$ & $3 \cdot 10^{-5}$ & K$^{-1}$ & \cite{1998PEPI..107...53S}   \\
    Thermal expansivity of molten silicates & $\alpha_{\mathrm{Si-liq}}$ & $6 \cdot 10^{-5}$ & K$^{-1}$ & \cite{1998PEPI..107...53S}   \\
    Thermal conductivity of solid silicates & $k_{\mathrm{}}$ & 3 & W m$^{-1}$ K$^{-1}$ &  \cite{2012Sci...338..939T}   \\
    Thermal expansivity of molten silicates & $k_{\mathrm{eff}}$ & $\le 10^6$ & W m$^{-1}$ K$^{-1}$ &  \cite{2014MPS...49.1083G}   \\
    Minimum thermal conductivity  & $k_{\mathrm{low}}$ & $10^{-3}$ & W m$^{-1}$ K$^{-1}$ &  \cite{1984EPSL..68...34Y}; \\
    \hspace{1cm} of unsintered solid silicates &&&& \cite{2012AA...537A..45H}   \\
    Temperature at onset of hot sintering & $T_{\mathrm{sint}}$ & 700 & K & \cite{1984EPSL..68...34Y}   \\
    \bottomrule
\end{tabu}
    \caption{List of physical parameters in the numerical model.}
     \label{tab:constants}
    }
\end{table*}

\subsection{Fluid flow}

As outlined in Sect. \ref{sec:introduction} we studied the thermo-mechanical evolution of instantaneously and recently formed planetesimals using the \textsc{i2elvis/i3elvis} code family \citep{2007PEPI..163...83G}. The code solves the fluid dynamic conservation equations using the extended Boussinesq approximation, to account for thermal and chemical buoyancy forces, with a conservative finite-differences (FD) approach on a fully staggered-grid \citep{2003PEPI..140..293G}, namely the continuity equation
\begin{align}
\frac{\partial \rho}{\partial t} + \nabla \rho \mathbf{v} = 0,
\end{align}
with density $\rho$, time $t$ and flow velocity $\mathbf{v}$; the Stokes equation
\begin{align}
\nabla \mathbf{\sigma'} - \nabla P + \rho \mathbf{g} = 0,
\end{align}
with deviatoric stress tensor $\mathbf{\sigma}'$, pressure $P$ and directional gravity $\mathbf{g}$ obtained via the location-dependent Poisson equation
\begin{align}
\nabla^2 \Phi = 4 \pi G \rho,
\end{align}
with the gravitational potential $\Phi$ and Newton's constant $G$;
and finally the energy equation
\begin{align} 
\label{eq:energy-conservation}
\rho c_{\mathrm{P}} \left( \frac{\partial T}{\partial t} + v_{\mathrm{i}} \cdot \nabla T \right) = - \frac{\partial q_{\mathrm{i}}}{\partial x_{\mathrm{i}}} + H_{\mathrm{r}} + H_{\mathrm{s}} + H_{\mathrm{L}},
\end{align}
with heat capacity $c_{\mathrm{P}}$, temperature $T$, heat flux $q_{\mathrm{i}} = -k \frac{\partial T}{\partial x_{\mathrm{i}}}$, thermal conductivity $k$, and radioactive ($H_{\mathrm{r}}$), shear ($H_{\mathrm{s}}$) and latent ($H_{\mathrm{L}}$) heat production terms.
The energy equation is advanced using a Lagrangian marker-in-cell technique to minimise numerical diffusion and enable an accurate advection of non-diffusive flow properties during material deformation.
The staggered-grid FD method permits to capture sharp variations of stresses and thermal gradients with strongly variable viscosity and thermal conductivity. For further details on the code's features we refer to \cite{2003PEPI..140..293G,2007PEPI..163...83G}.

\subsection{Heating by short-lived radionuclides}

As discussed earlier, the radiogenic heat source term $H_{\mathrm{r}}$ in Equation 4 is dominant for early formed planetesimals. It is driven by the decay of short-lived isotopes $^{26}$Al and $^{60}$Fe and the long-lived $^{40}$K, $^{235}$U, $^{238}$U and $^{232}$Th. Among these $^{26}$Al is by far the most important one and therefore drives the internal heating of the young bodies, as the abundance of $^{60}$Fe is lower by orders of magnitude \citep{2008Icar..198..163B,2012E&PSL.359..248Ta,2016E&PSL.436...71M}. In this work, we considered time-dependent radiogenic heating by $^{26}$Al and the long-lived radioactive isotopes as input for $H_{\mathrm{r}}$ in Equation 4. For the initial $^{26}$Al/$^{27}$Al ratio we adopted an upper-limit value \citep{jacobsen200826} of $5.85 \cdot 10^{-5}$ \citep{2006ApJ...646L.159T} at CAI formation.

\subsection{Silicate melting model}
\label{sec:melt-model}

For the silicates we assumed a peridotite composition and used the parameterizations by \citet{2000GGG.....1.1051H} and \citet{2005EPSL.236...78W} \citep[based on data of][]{2002EPSL.197..117T} for the solidus  and liquidus temperatures $T_{\mathrm{sol}}$ and $T_{\mathrm{liq}}$, which determine the silicate melt fraction 
\begin{align}
\varphi = \left\{
  \begin{array}{ll}
    0 & : T \le T_{\mathrm{sol}},\\
    \frac{T - T_{\mathrm{sol}}}{T_{\mathrm{liq}} - T_{\mathrm{sol}}} & : T_{\mathrm{sol}} < T < T_{\mathrm{liq}},\\
    1 & : T \ge T_{\mathrm{liq}}.
  \end{array}
\right.
\end{align}

We took into account both consumption and release of latent heat due to melting and freezing of silicates. Silicate density depends on the melt fraction $\varphi$ as
\begin{align}
\rho_{\mathrm{eff}}(P,T,\varphi) = & \; \rho_{\mathrm{Si-sol}}(P,T) & \\
    & - \varphi[\rho_{\mathrm{Si-sol}}(P,T)-\rho_{\mathrm{Si-liq}}(P,T)] &
\end{align}
with solid and liquid silicate densities $\rho_{\mathrm{Si-sol}}$ and $\rho_{\mathrm{Si-liq}}$.
For silicate melt fractions $0.1 < \varphi \lesssim 0.4$ the effective viscosity \citep{1992JVGR...53...47P} is given as
\begin{align}
\eta_{\mathrm{eff}} = \eta_{\mathrm{Si-liq}} \cdot \mathrm{exp} \left(\left[ 2.5 + \left( \frac{1-\varphi}{\varphi} \right)^{0.48} \right] \cdot ( 1 - \varphi ) \right).
\end{align}
Above $\varphi \gtrsim 0.4$ a transition occurs from solid-like structures to low-viscosity crystal suspensions \citep{solomatov2015magma,2009GGG....10.3010C}, with $\eta_{\mathrm{Si-liq}} = 10^{-4} - 10^2$ Pa s \citep{bottinga1972viscosity,2003EPSL.205..239R,2005EPSL.240..589L}. This effectively increases the Rayleigh number
\begin{align}
Ra = \frac{\alpha g (T - T_{\mathrm{surf}})\rho_{\mathrm{eff}}^2 c_{\mathrm{P}}D^3}{k \eta_{\mathrm{Si-liq}}},
\end{align}
with thermal expansivity $\alpha$, surface temperature $T_{\mathrm{surf}}$, depth of the magma ocean $D$ and thermal conductivity $k$ and thus enables an efficient cooling process.

Above melt fractions $\varphi \gtrsim 0.4$ our model is restricted by a lower cut-off viscosity $\eta_{\mathrm{num}} = 10^{17}$ Pa s, which preserves numerical stability, but lies orders of magnitude above realistic values of molten state silicate viscosities. To bypass restrictions of the physical interpretation in this melt regime we employed the soft turbulence model by \citet{1962PhFl....5.1374K} and \citet{1994AnRFM..26..137S}, and estimated the convective heat flux as
\[
q = 0.089 \frac{k(T-T_{\mathrm{surf}})}{D} Ra^{1/3}. 
\label{eq:soft-turb}
\]
Using Equation 10 we derived an increased effective thermal conductivity 
$$
k_{\mathrm{eff}} = \left( \frac{q}{0.089} \right)^{3/2}  \frac{1}{(T - T_{\mathrm{surf}})^2 \rho_{\mathrm{eff}}} \left( \frac{\alpha g c_{\mathrm{P}}}{\eta_{\mathrm{num}}} \right)^{-1/2},
\label{eq:k_eff}
$$
which approximates correct heat flux for a low viscosity magma ocean \citep{2001Icar..149...79T,2006MPS...41...95H,2011Icar..215..346G}. For a more detailed discussion on model limitations due to this issue see Sect. \ref{sec:limitations}.

\subsection{Porosity}

\begin{table*}[t]
\centering
\begin{tabu}{lllll} 
    \taburulecolor{hlinecolor}
    \toprule
    \rowcolor{headercolor}
    \textsc{Parameter} & \textsc{Symbol} &  \textsc{Value range} &  \textsc{Unit} &  \textsc{List of values} \\
    \midrule
    Planetesimal radius & $R_{\mathrm{p}}$ & 20--200 & km & $20, 50, 80, 110, 140, 170, 200$  \\
    Instantaneous formation time & $t_{\mathrm{form}}$ & 0.1--1.75 & Myr & $0.1, 0.5, 1.0, 1.1, 1.2, 1.3, 1.4, 1.5, 1.6, 1.7, 1.75$   \\
    Initial porosity & $\phi_{\mathrm{init}}$ & 0.0--0.75 &  & $0.0, 0.1, 0.2, 0.25, 0.3, 0.4, 0.5, 0.75$ \\
    \bottomrule
\end{tabu}
    \caption{Distinct values of chosen parameter space.}
    \label{tab:parameters}
\end{table*}

As already indicated in Sect. \ref{sec:introduction}, the initial porous state of recently accreted planetesimals is thought to decrease due to cold isostatic pressing with pressure and thus depth into a configuration of closer packing \citep{2012AA...537A..45H}, via
\begin{align}
\phi (P) = 0.42 + 0.46 \cdot \left[ \left( \frac{P}{P_{\mathrm{0}}} \right)^{1.72} +1 \right]^{-1},
\label{eq:cold_pressing}
\end{align}
with $P_{\mathrm{0}} = 0.13$ bar, which effectively introduces an upper cut-off porosity for depths greater than $\sim 10^2$ m, mostly dependent on the size of the body.
Furthermore, the porosity changes the density of the solid material
\begin{align}
\rho_{\mathrm{Si-por}} (P,T,\phi) = \rho_{\mathrm{Si-sol}} (P,T) \cdot (1-\phi),
\end{align}
and the effective thermal conductivity for porous material
\begin{align}
k_{\mathrm{eff,por}} = \left\{
  \begin{array}{ll}
    k_{\mathrm{1}} = k \cdot e^{-\phi/\phi_{\mathrm{0}}} & : \phi < 0.2,\\
    k_{\mathrm{3}} = (k_{\mathrm{1}}^4 + k_{\mathrm{2}}^4)^{1/4} & : 0.2 \le \phi \le 0.4,\\
    k_{\mathrm{2}} = k \cdot e^{a-\phi/\phi_{\mathrm{1}}} & : \phi > 0.4,
  \end{array}
\right.
\end{align}
with constants $a=-1.2$, $\phi_{\mathrm{0}} = 0.08$ and $\phi_{\mathrm{1}} = 0.167$, fitting lab experiments \citep{2012AA...537A..45H,2015AA...576A..60G}. Finally, the material compaction is sensitive to sintering effects via
\begin{align}
\left| \frac{\partial \phi}{\partial t} \right| = A(1-\phi) \frac{\sigma^{3/2}}{\Re^3} \cdot \mathrm{exp} \left[\frac{-E_\mathrm{a}'}{RT} \right],
\label{eq:hot_pressing}
\end{align}
with the effective stress $\sigma$, the effective grain size $\Re$, the gas constant $R$ and experimentally determined factors $A=4 \cdot 10^{-5}$ and activation energy $E_a' = 85$ kcal mol$^{-1}$ \citep{2012AA...537A..45H}.

\subsection{Initial conditions}
\label{sec:initial_conditions}

The spherical planetesimals in our model box were supposed to be completely composed of silicates. Olivine outrules pyroxene minerals in controlling deformation processes due to its mechanical weakness \citep{1991GeoRL..18.2027M}. Thus, we apply an olivine rheology \citep{ranalli1995rheology} to be able to follow thermo-mechanical processes, i.e., melting and mixing due to internal heating. Each body was built up by several rheologically identical silicate layers, which could be followed by an internal tracking of the corresponding markers. This enabled us to distinguish different silicate layers and reconstruct their mixing history. Illustrative examples are given in Sect. \ref{sec:results}.

As indicated before, the energy release during the accretion phase is only minor for the size of bodies we addressed here \citep{schubert1986thermal}. Therefore, we started from a constant temperature distribution all over the model grid in accordance with values in a typical protoplanetary disk $T_{\mathrm{space}} = 290$ K \citep{1998Icar..134..187G}.

The surrounding of the bodies was made up of a so-called sticky-air layer \citep{2008PEPI..171..198S}, with near zero density, constant temperature $T_{\mathrm{SA}} = T_{\mathrm{space}}$ and constant viscosity $\eta_{\mathrm{SA}} = 10^{19}$ Pa. Such a layer allows for simulation of free surfaces and serves as infinite reservoir to absorb released heat from the planetesimal \citep{2011Icar..215..346G,2012GeoJI.189...38C,2013NatGe...6...93T}.

The numerical model boxes had physical dimensions of 500 km in each direction in 2D and 3D, represented by $501^2$ grid points in 2D, respectively $261^3$ grid points in 3D, which results in physical resolutions of 1 km in 2D and $\sim$ 2 km in 3D.

\subsection{Parameter space}

The goal of this work was to assess the combined effect of radiogenic heating by SLRs and initial porosity on the subsequent evolution of planetesimals. Hence, the parameter space was based on varying the planetesimal radius $R_{\mathrm{p}} =$ 20--200 km, the instantaneous formation time $t_{\mathrm{form}} =$ 0.1--1.75 Myr after CAI formation and the initial porosity $\phi_{\mathrm{init}} =$ 0.0--0.75, in total a set of 616 2D simulations. A full list of all applied values is given in Table \ref{tab:parameters}.

Due to the heavy computational cost of 3D simulations we first analyzed the 2D simulations, categorized them and then performed selected 3D simulations to verify the 2D results.

From our varied parameters, both $R_{\mathrm{p}}$ and $t_{\mathrm{form}}$ directly influenced the amount of SLRs present in the body.
A list of all simulation runs with corresponding parameters and categories can be found for the 2D simulations in Table \ref{tab:2d_runs} and for the 3D simulations in Table \ref{tab:3d_runs}.


\section{Results}
\label{sec:results}

\subsection{Thermo-mechanical evolution}
\label{sec:therm_evo}

In this section we analyze the thermo-mechanical outcome of the simulations. In Sect. \ref{sec:mat_prop} we focus on the temporal evolution of the material properties, i.e., solid or molten, and categorize the 2D results accordingly. Each category is then described and examples are given. In Sect. \ref{sec:T-t} we investigate the time-dependent maximum temperatures of the bodies and assess the influence of each of the varied parameters on it by constructing $R_{\mathrm{p}}$, $t_{\mathrm{form}}$ and $\phi_{\mathrm{init}}$ isolines. Also, we analyze the influence of $\phi_{\mathrm{init}}$ on the temperature profile for fixed formation time and planetesimal size.

\begin{figure*}[tbh]
\centering
\includegraphics[width=0.85\textwidth]{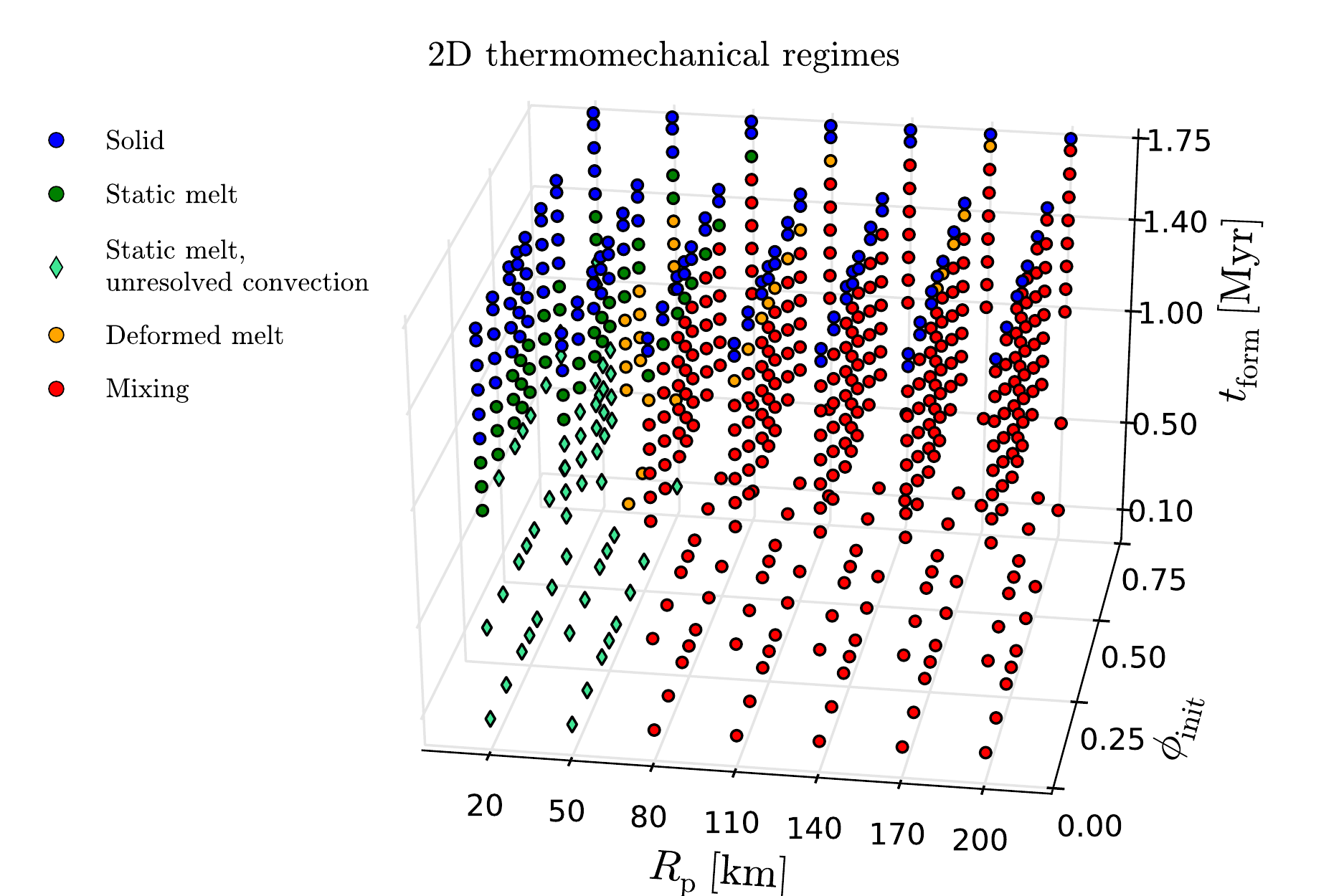}
\caption{3D parameter space covered by the two-dimensional simulation runs, each dot represents one simulation, with $R_{\mathrm{p}}$ in km, $\phi_{\mathrm{init}}$ non-dimensional and $t_{\mathrm{form}}$ in Myr. The colors indicate  which thermo-mechanical state was reached during the time evolution. \emph{Blue}: all silicates were in solid form during all times (Fig. \ref{fig:2d_solid}); \emph{green}: the silicates in the planetesimal were partially or fully molten at some stage during the temporal evolution (Fig. \ref{fig:2d_melt}), green simulations with diamonds indicate that the numerical restrictions in our model did not capture fluid motion due to extremely low viscosities, see Sect. \ref{sec:limitations} for an in-depth discussion of this issue; \emph{orange}: the silicate layers were partially deformed, but the heating was not sufficient for convection (Fig. \ref{fig:2d_deformation}); \emph{red}: convectional mixing occurs during the temporal evolution of the planetesimal (Fig. \ref{fig:2d_mixing}).}
\label{fig:2d_grid_comp}
\end{figure*}

\begin{figure}[bt!]
    \centering
    \begin{subfigure}[b]{0.235\textwidth}
            \includegraphics[width=\textwidth]{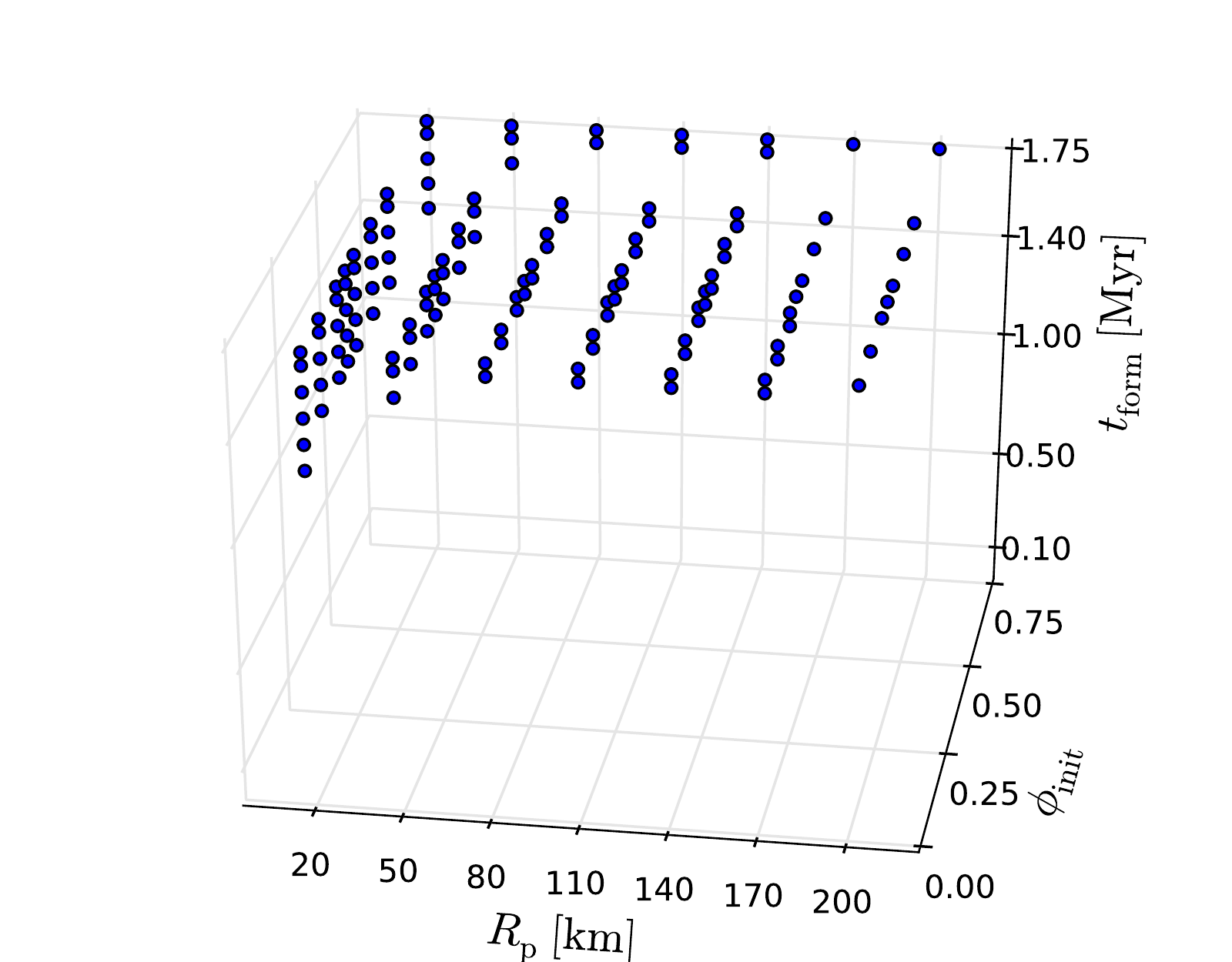}
    \end{subfigure} 
    \begin{subfigure}[b]{0.235\textwidth}
            \includegraphics[width=0.9\textwidth,left]{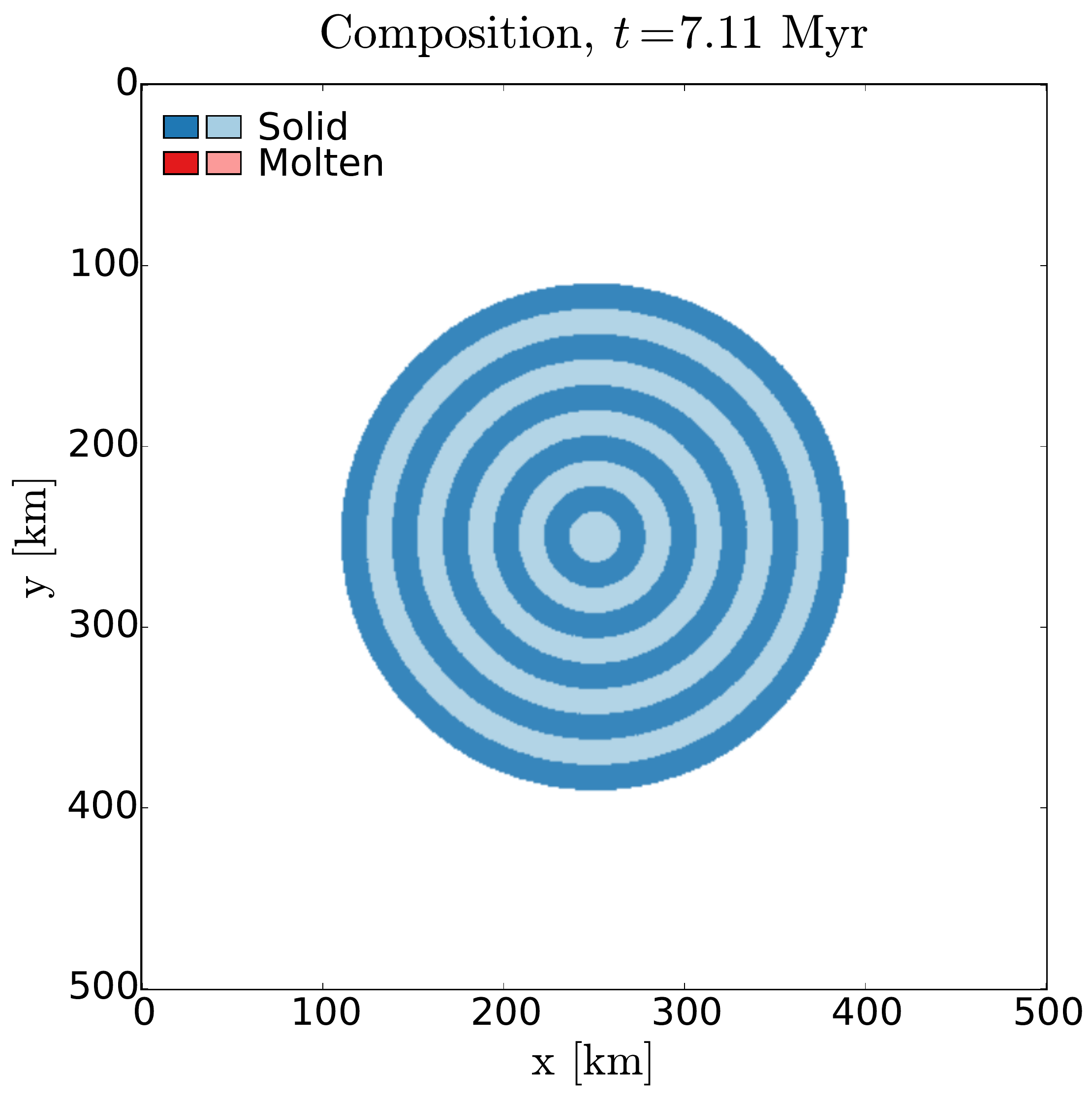}
    \end{subfigure}\\
    \begin{subfigure}[b]{0.235\textwidth}
            \includegraphics[width=\textwidth]{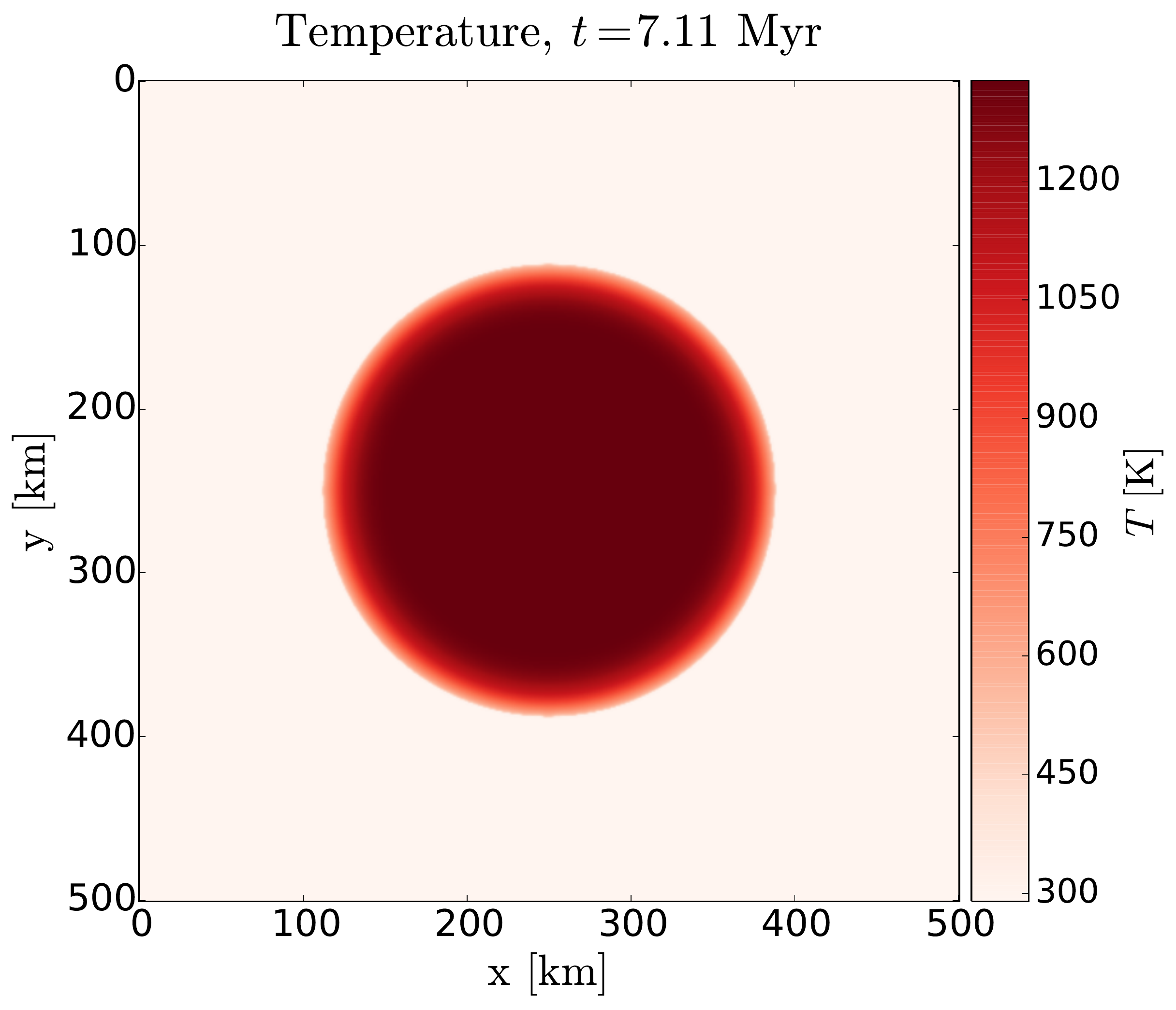}
    \end{subfigure}
    \begin{subfigure}[b]{0.235\textwidth}
            \includegraphics[width=\textwidth]{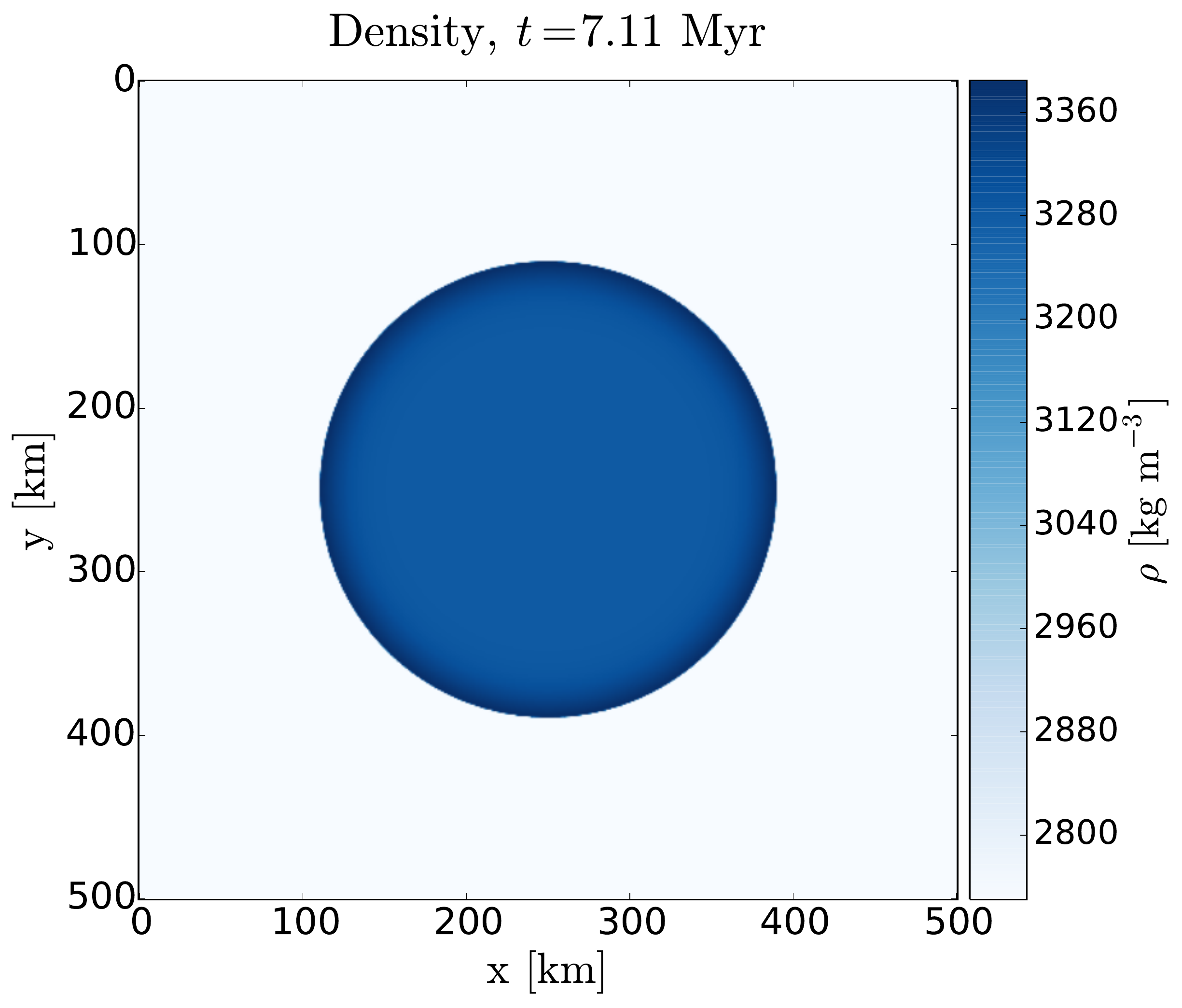}
    \end{subfigure}
    \caption{Example of a \emph{solid} model, i.e., without any melting throughout the temporal evolution, with $R_{\mathrm{p}} = 140$ km, $t_{\mathrm{form}} = 1.7$ Myr, $\phi_{\mathrm{init}} = 0.5$ at $t = 7.11$ Myr. The all-solid (rheologically identical) layers did not deform throughout the simulation.}
    \label{fig:2d_solid}
\end{figure}

\begin{figure}[tb]
    \centering
    \begin{subfigure}[b]{0.235\textwidth}
        \includegraphics[width=\textwidth]{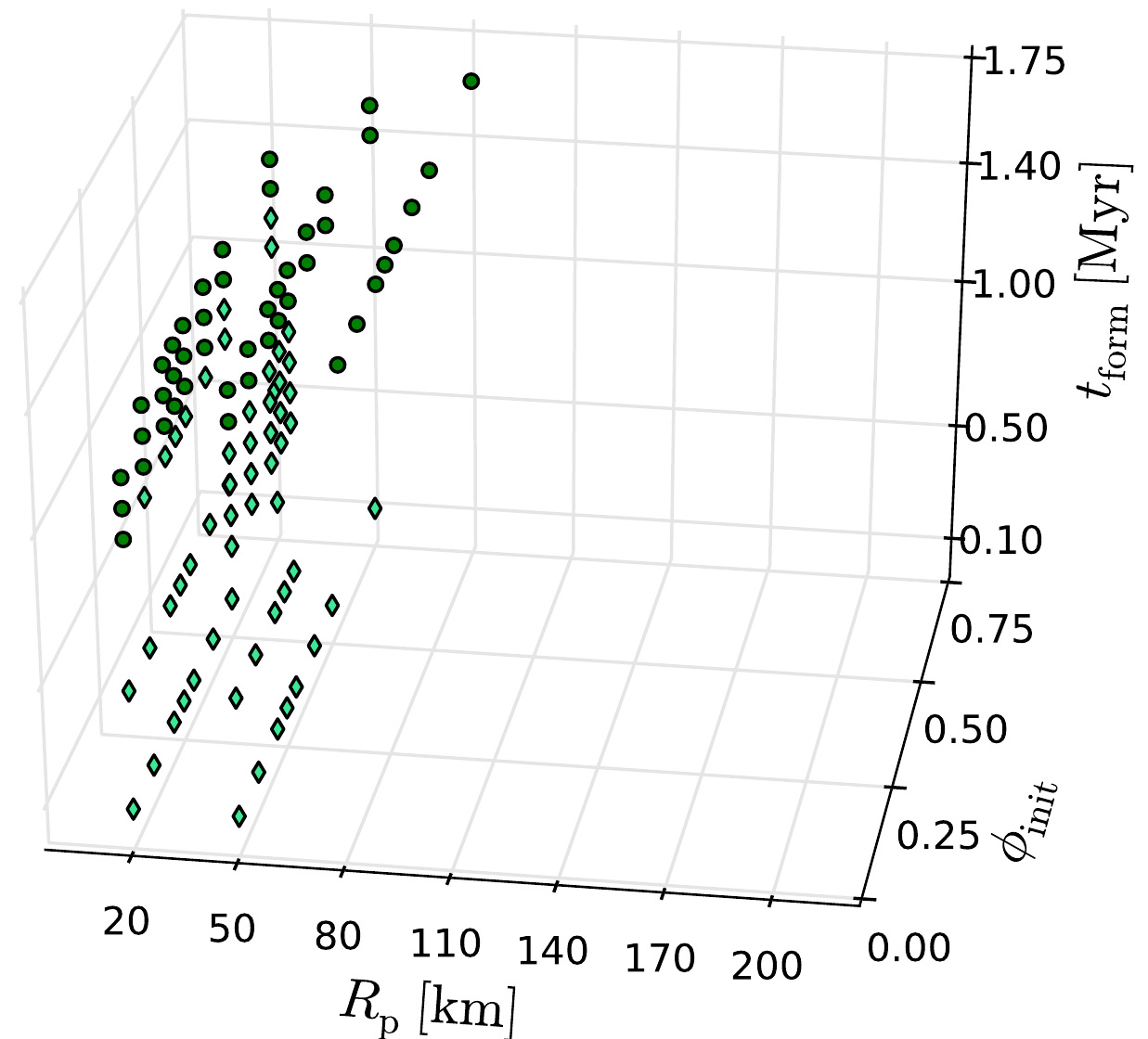}
    \end{subfigure}
    \begin{subfigure}[b]{0.235\textwidth}
        \includegraphics[width=0.9\textwidth,left]{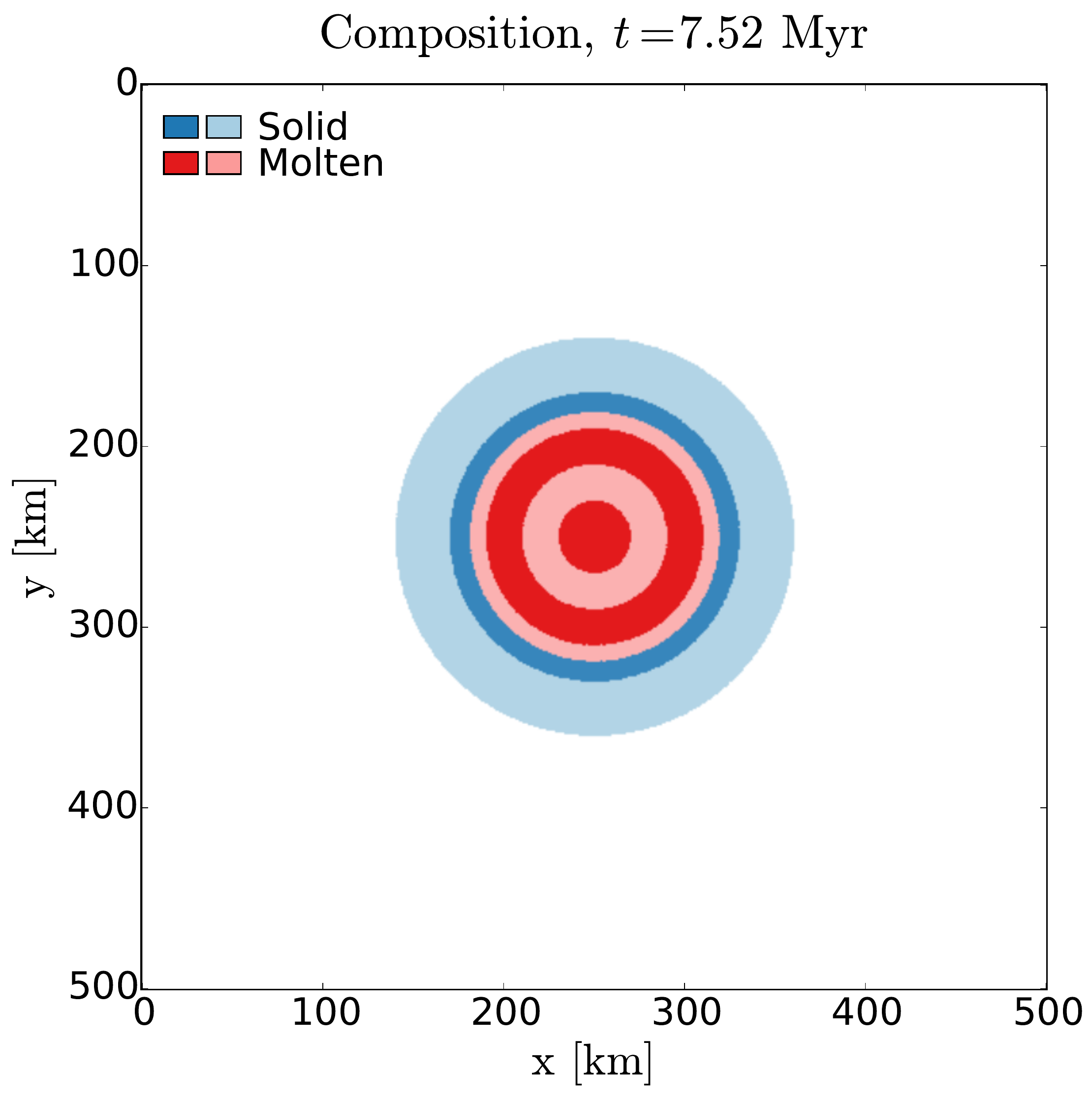}
    \end{subfigure} \\
    \begin{subfigure}[b]{0.235\textwidth}
            \includegraphics[width=\textwidth]{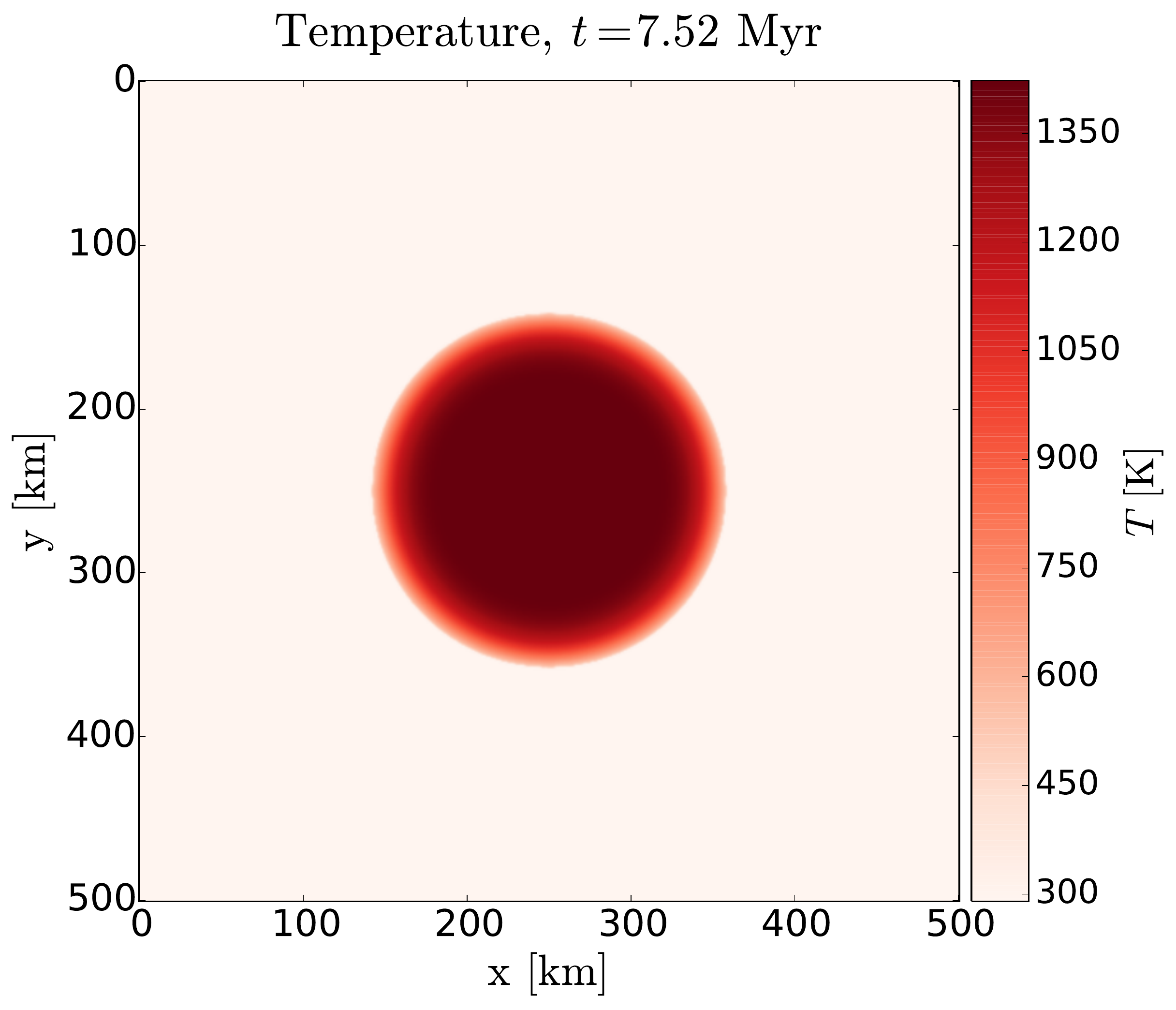}
    \end{subfigure}
    \begin{subfigure}[b]{0.235\textwidth}
            \includegraphics[width=\textwidth]{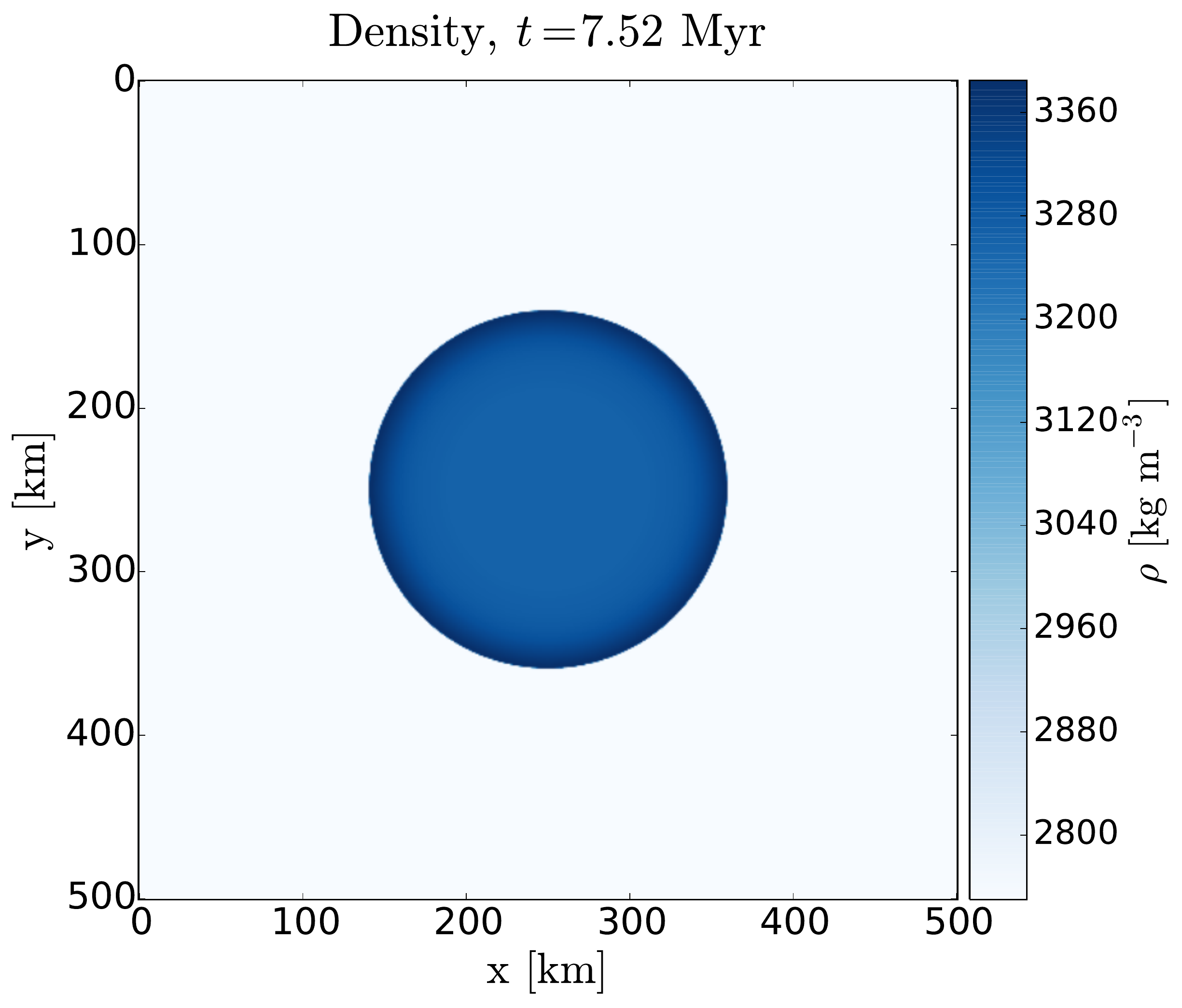}
    \end{subfigure} 
    \caption{Example of a \emph{static melt} model, with $R_{\mathrm{p}} = 110$ km, $t_{\mathrm{form}} = 1.6$ Myr, $\phi_{\mathrm{init}} = 0.25$ at $t = 7.52$ Myr. The molten layers are differently shaded to be able to track the onset of convection (see Fig. \ref{fig:2d_deformation}). The inner parts were hotter and less dense than the upper layers and the temperatures were high enough to partially melt the silicates for a limited time period.}
    \label{fig:2d_melt}
\end{figure}

\begin{figure}[tb]
    \centering
    \begin{subfigure}[b]{0.235\textwidth}
            \includegraphics[width=\textwidth]{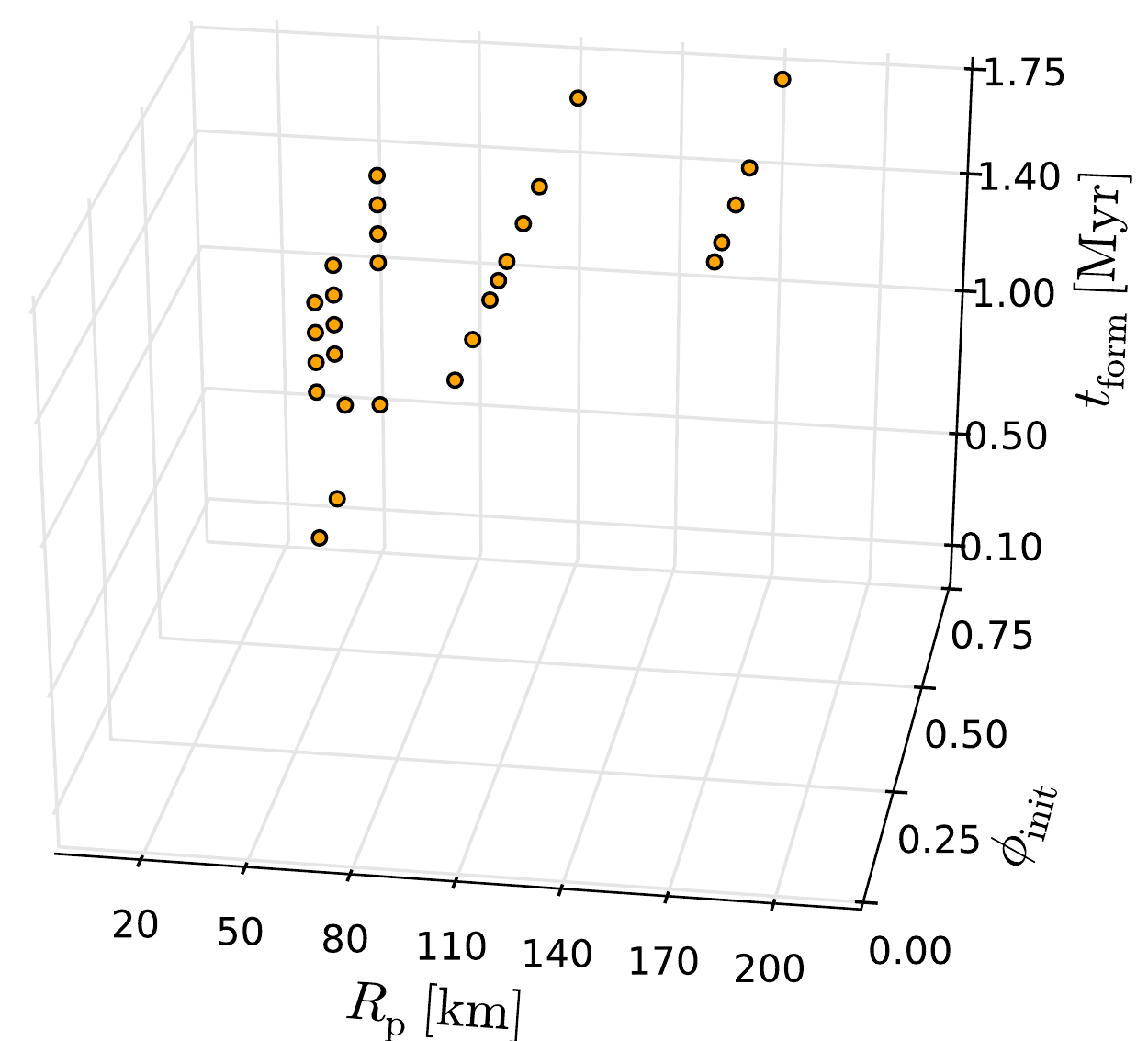}
    \end{subfigure} 
    \begin{subfigure}[b]{0.235\textwidth}
            \includegraphics[width=0.9\textwidth,left]{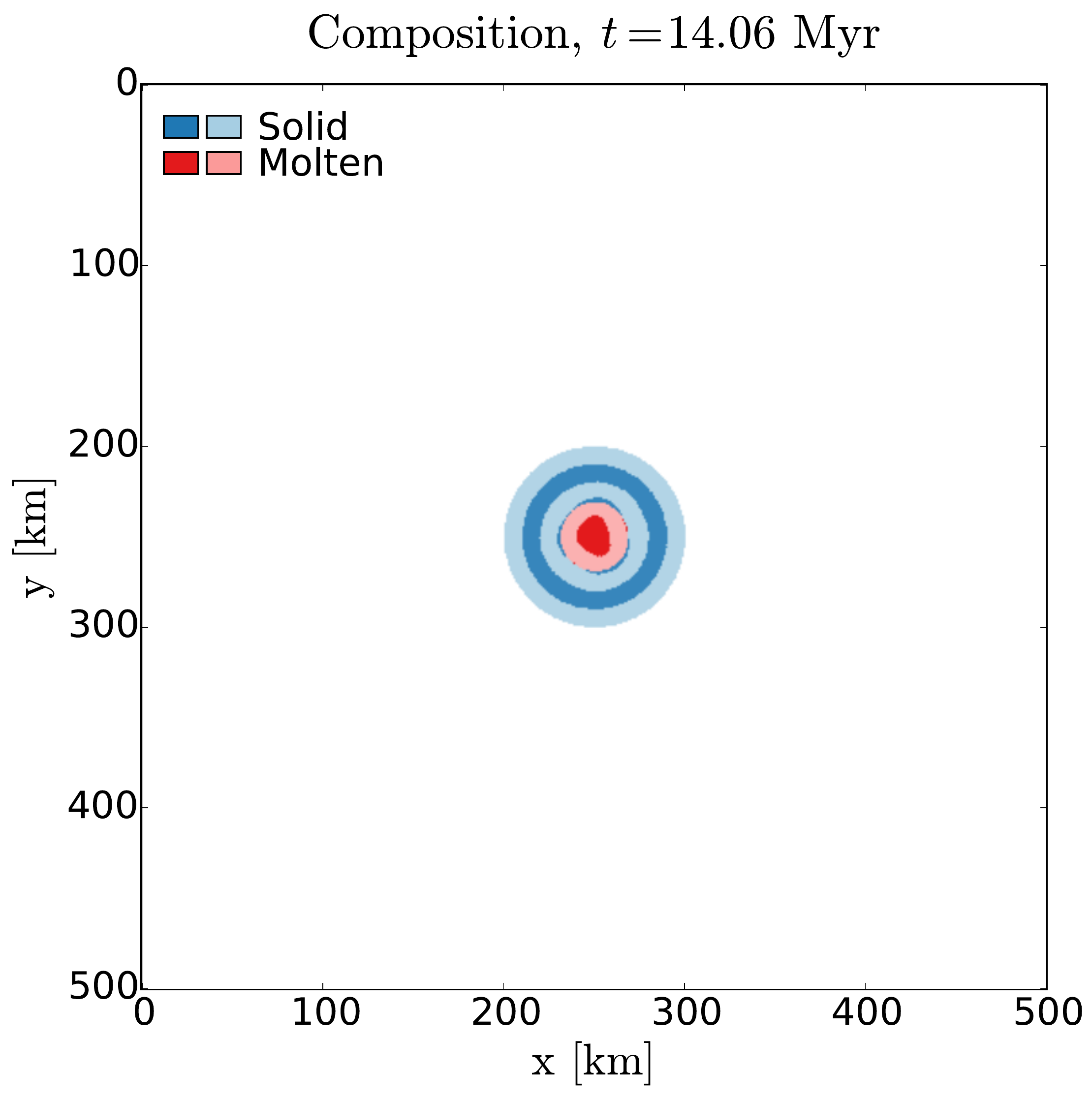}
    \end{subfigure} \\
    \begin{subfigure}[b]{0.235\textwidth}
            \includegraphics[width=\textwidth]{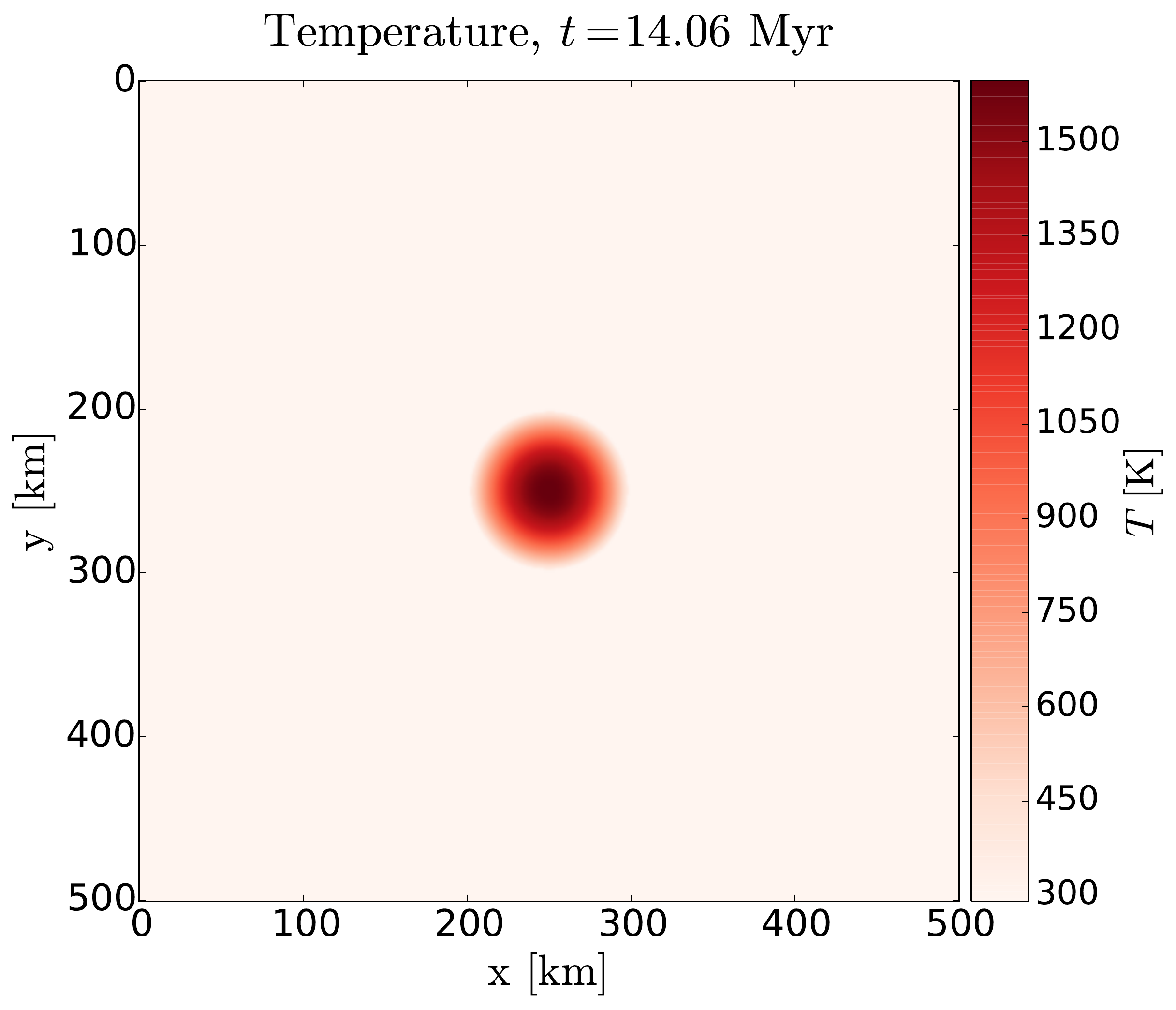}
    \end{subfigure}
    \begin{subfigure}[b]{0.235\textwidth}
            \includegraphics[width=\textwidth]{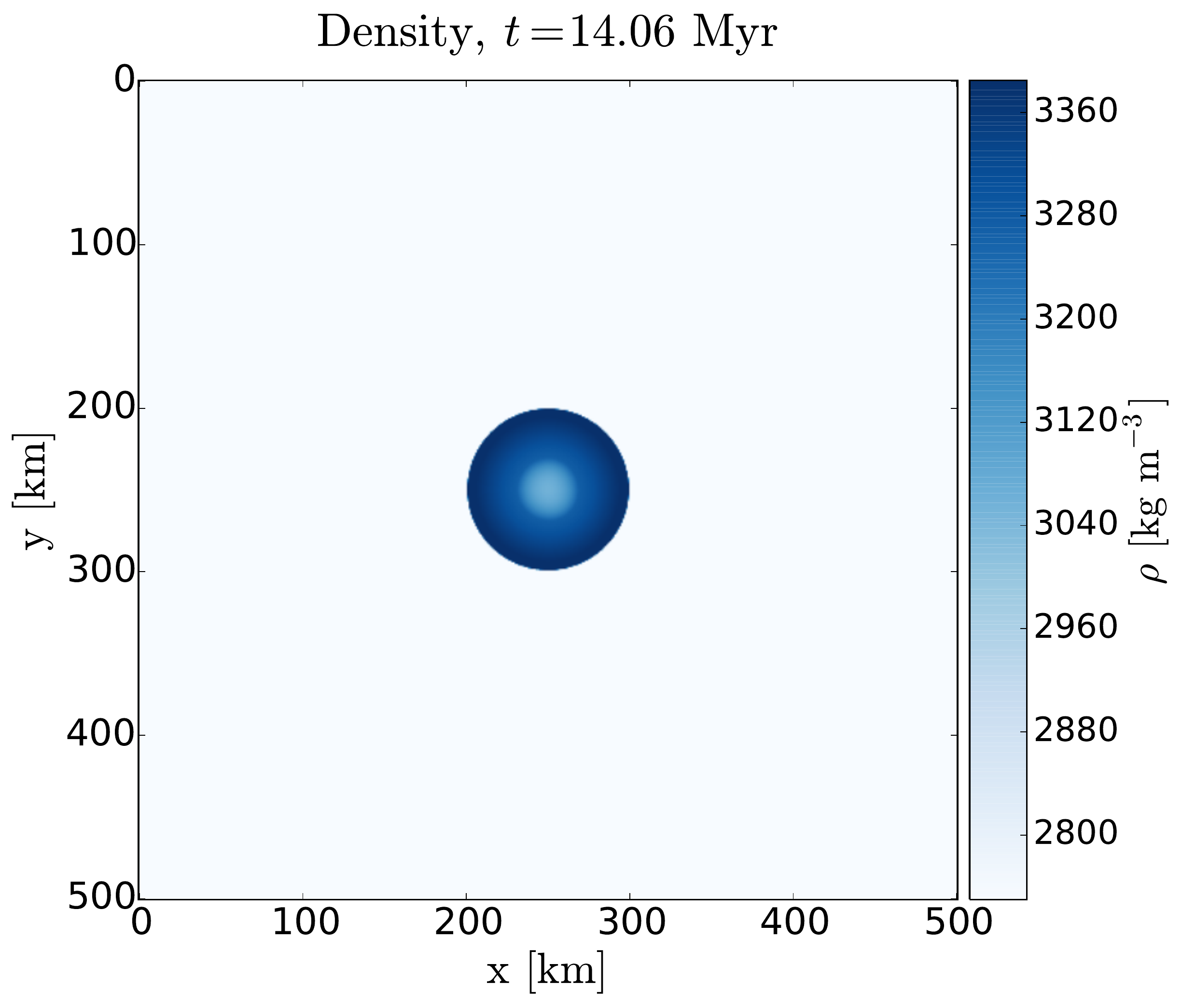}
    \end{subfigure} 
    \caption{Example of a \emph{deformation} model, with $R_{\mathrm{p}} = 50$ km, $t_{\mathrm{form}} = 1.0$ Myr, $\phi_{\mathrm{init}} = 0.75$ at $t = 14.06$ Myr. The temperatures were high enough to initate the onset of convection but could not sustain these temperatures long enough for mixing to occur.}
    \label{fig:2d_deformation}
\end{figure}

\begin{figure}[bt!]
    \centering
    \begin{subfigure}[b]{0.235\textwidth}
            \includegraphics[width=\textwidth]{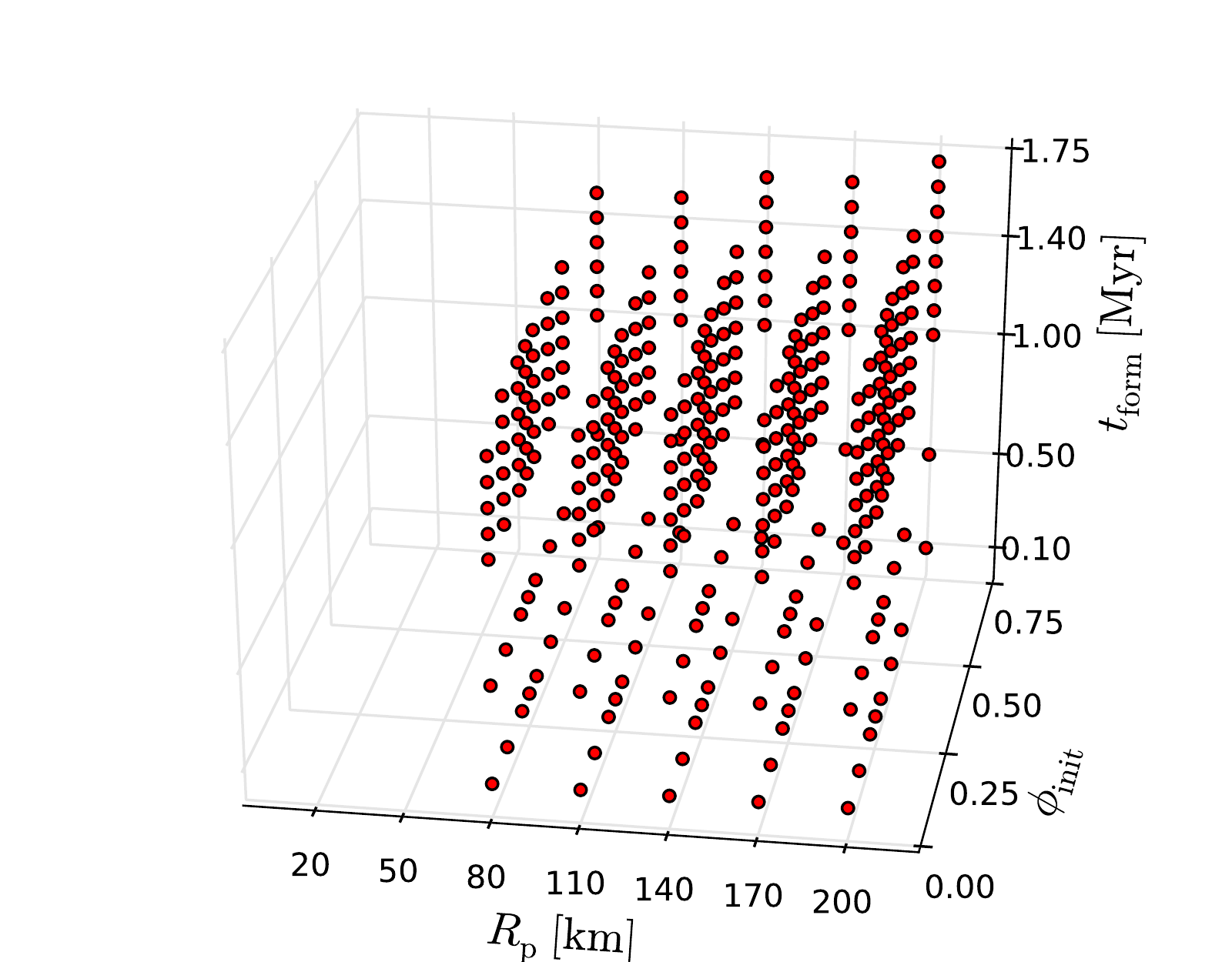}
    \end{subfigure} 
    \begin{subfigure}[b]{0.235\textwidth}
            \includegraphics[width=0.9\textwidth,left]{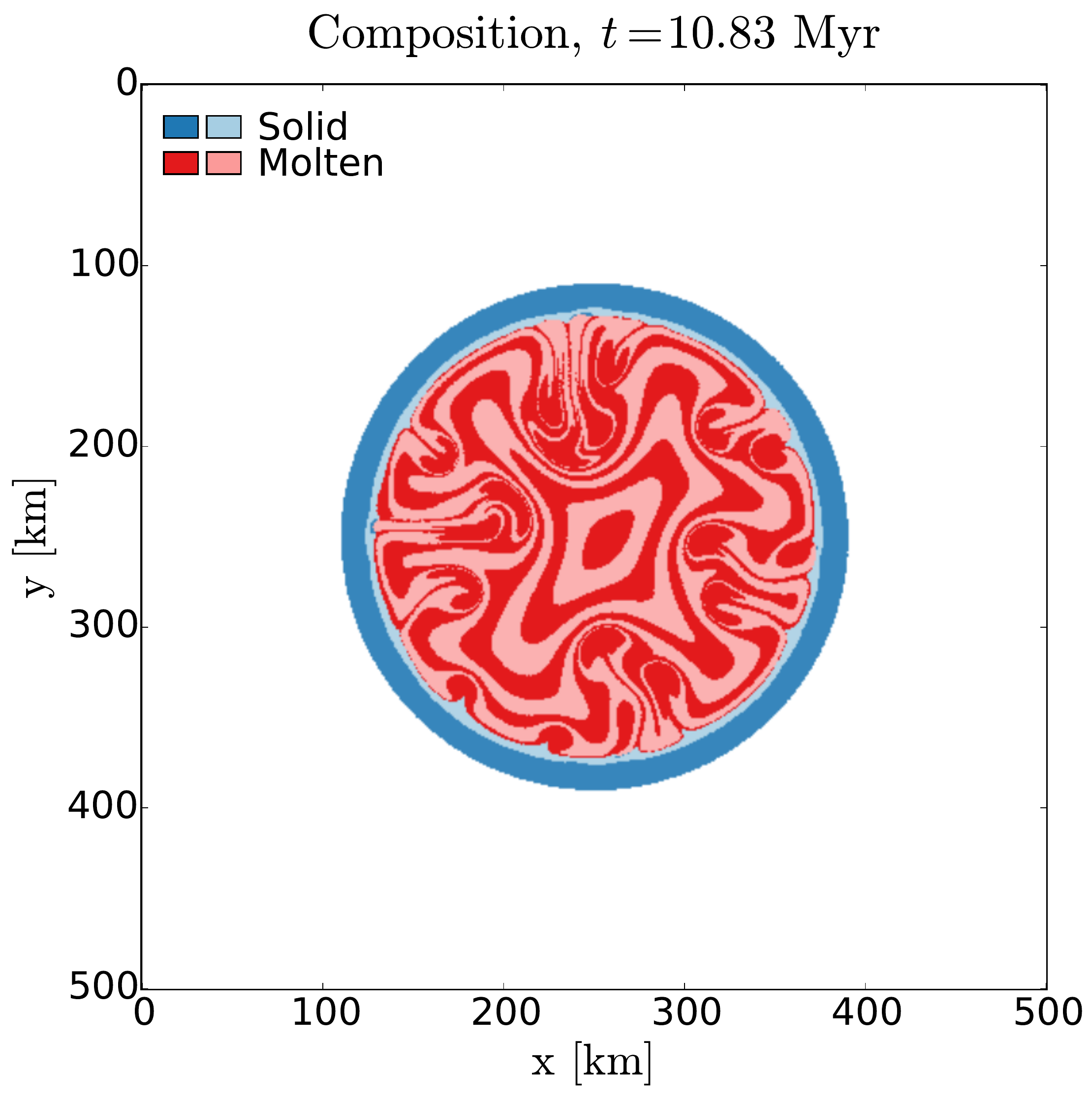}
    \end{subfigure} \\
    \begin{subfigure}[b]{0.235\textwidth}
            \includegraphics[width=\textwidth]{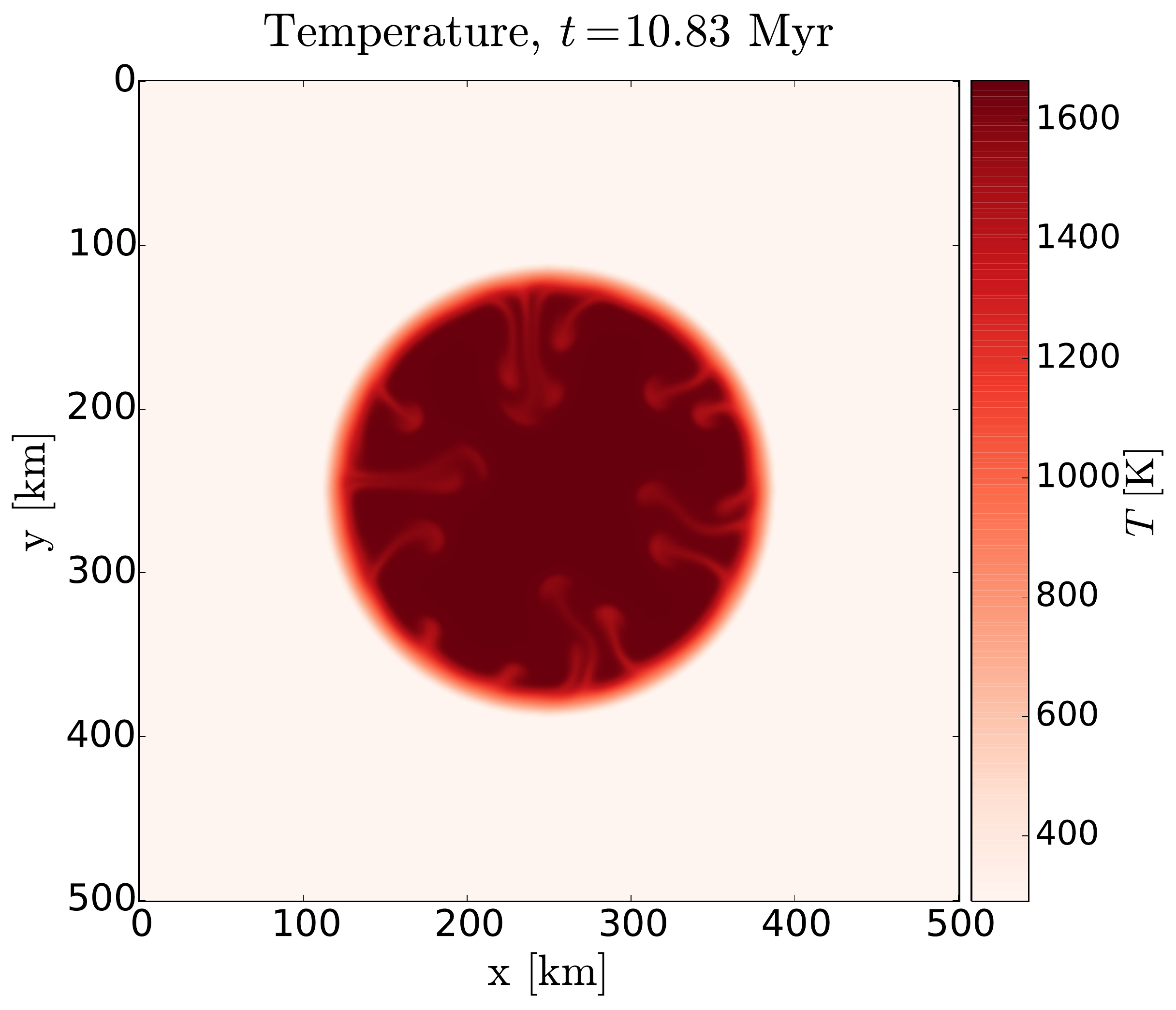}
    \end{subfigure}
    \begin{subfigure}[b]{0.235\textwidth}
            \includegraphics[width=\textwidth]{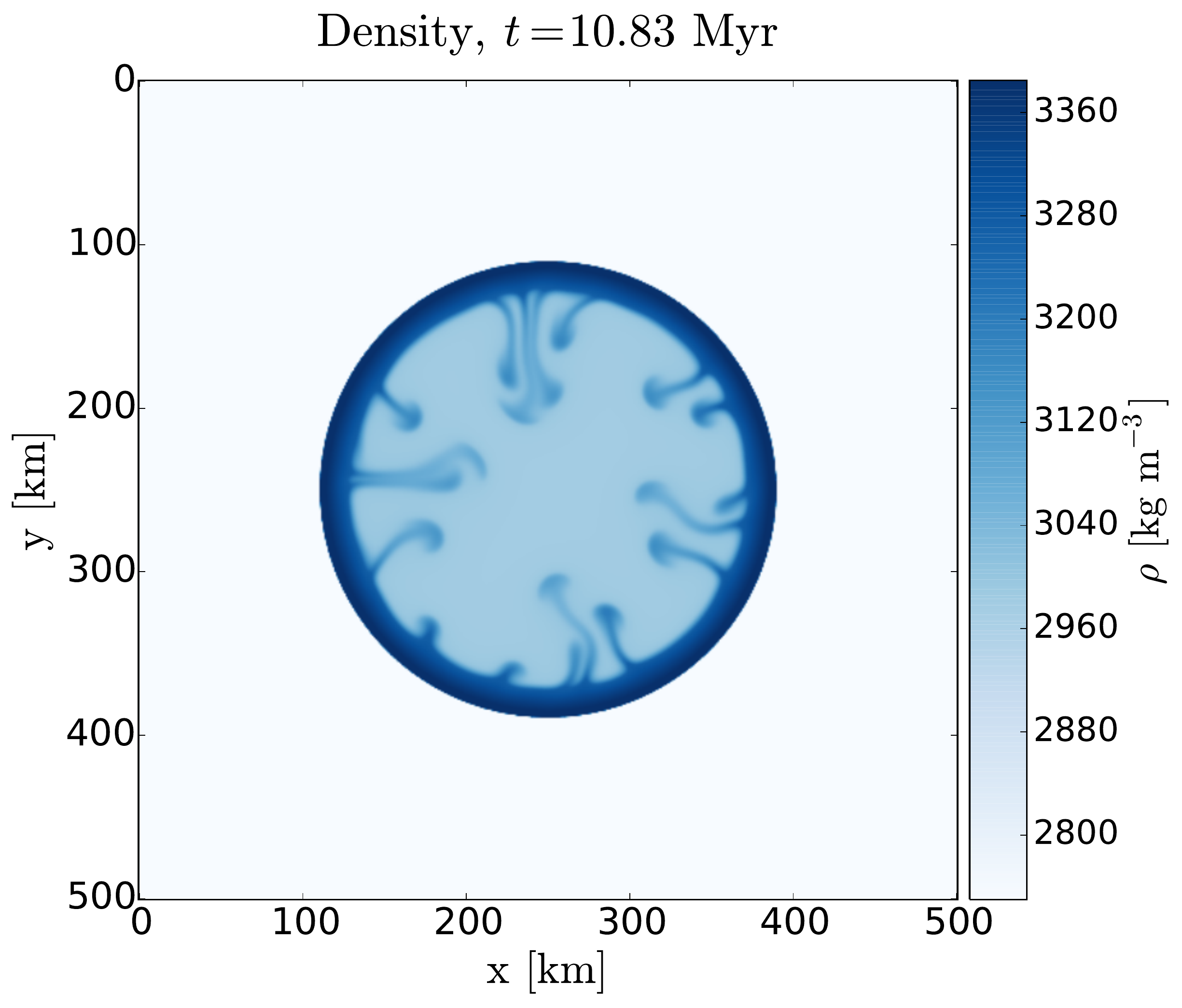}
    \end{subfigure} 
    \caption{Example of a \emph{mixing} model, with $R_{\mathrm{p}} = 140$ km, $t_{\mathrm{form}} = 0.5$ Myr, $\phi_{\mathrm{init}} = 0.4$ at $t = 10.83$ Myr. The density contrast of inner and outer layers drove convection.}
    \label{fig:2d_mixing}
\end{figure}

\subsubsection{Material properties}
\label{sec:mat_prop}

Figure \ref{fig:2d_grid_comp} illustrates the thermo-mechanical results of the material properties within each 2D simulation run. Each dot represents a single simulation and color indicates in which kind of regime we categorize the simulation. Each of these regimes is described below and an example, illustrating the state for $\phi$, $T$ and $\rho$ at a certain time, is given. Illustrating video files for each of the described regimes below can be found in the supplementary material (see \ref{sec:supplementary_material}).

\paragraph{Solid regime}

The blue rendered simulations in Fig. \ref{fig:2d_grid_comp} build the class of \emph{solid} models. These are models which lacked enough heat production by SLRs to experience any sign of transition from the solid silicate to a partially molten silicate state. An example of this kind is given in Fig. \ref{fig:2d_solid}. The upper left image shows all simulation runs of this class. The composition plot illustrates the unperturbed layered structure of the silicates it is composed of. Because the body never experienced enough heat, no transition to a molten state occurred and therefore the layers resided with their original ordering. The temperature and density plots illustrate these parameters at the same time during the evolution. Since the body experienceed some heat from SLRs it heated up and cooled down to the temperature of the surrounding $T_{\mathrm{space}}$ on the order of several tens of Myr.
As shown in Fig. \ref{fig:2d_solid} these kinds of models can be found for all tested radii for $t_{\mathrm{form}} \gtrsim 1.7$ Myr, i.e., when the initial amount of $^{26}$Al has significantly reduced. Additionally, planetesimals with $R_{\mathrm{p}} = 50$ km already belong to this class for earlier formation times $t_{\mathrm{form}} \gtrsim 1.6$ Myr and for $t_{\mathrm{form}} \gtrsim 1.3$ Myr for bodies with $R_{\mathrm{p}} = 20$ km  since they cooled more efficiently. Comparison of figures \ref{fig:2d_grid_comp} and \ref{fig:2d_solid} for bodies with $R_{\mathrm{p}} = 20$ km reveals the influence of $\phi_{\mathrm{init}}$. For $t_{\mathrm{form}} = 1.3$ Myr, the models were \emph{solid} for $\phi_{\mathrm{init}} \le 0.3$ and \emph{molten} for $\phi_{\mathrm{init}} \ge 0.4$. Hence, the effects of initial porosity only affected this transitional stage for the smallest bodies in our parameter space.

\paragraph{Static melt regime}

This class of simulations showed characteristics of phase transitions from solid to molten states, indicated with green circles and diamonds in Fig. \ref{fig:2d_grid_comp}. For the deviations between these we refer to the discussion of our model limitations in Sect. \ref{sec:limitations}. An example of a \emph{static melt} model is given in Fig. \ref{fig:2d_melt}. In the composition Fig. we see molten silicate phases shown in red. As the material in the inner parts could not cool as efficiently as the outer parts higher temperatures occurred and thus silicates in this region tended to melt. Hence, the density in the outer shells was higher than in the inner parts. Simulations of this class were dominant for bodies with $R_{\mathrm{p}} \le 50$ km. For $R_{\mathrm{p}} = 20$ km the boundary for the transition from solid to melt was $t_{\mathrm{form}} \approx 1.3$ Myr, for $R_{\mathrm{p}} = 50$ km it was $t_{\mathrm{form}} \approx 1.6$ Myr. In bodies with $R_{\mathrm{p}} = 80$ km this class could be found solely for $t_{\mathrm{form}} = 1.6$ Myr, marking the boundary to the transition from solid models to more dynamic models displaying convection.

\paragraph{Deformed melt regime}

This class marked the transition from the \emph{static melt} to the \emph{mixing} regime in the three-dimensional parameter space. A \emph{deformation} example is given in Fig. \ref{fig:2d_deformation} for an evolutionary stage with molten silicate phases and deformed layers, which clearly deviated from the initial circular structure. This kind of models reached higher temperatures than their \emph{static melt}-bearing counterparts. Due to the larger density contrast this leaded to the onset of mass segregation within the body. An interesting case is given for $R_{\mathrm{p}} = 50$ km. These bodies were dominated by \emph{deformation} for $\phi_{\mathrm{init}} \ge 0.4$ and $t_{\mathrm{form}} \lesssim 1.3$ Myr. This type is categorized differently as it indicates the restrictions of our model: if the viscosities fell below $\eta_{\mathrm{num}}$, fluid motions could not always be correctly resolved, in spite of accurately modeling the heat flux. Again, for a more detailed discussion on this issue see Sect. \ref{sec:limitations}. 

\paragraph{Mixing regime}

The class of \emph{mixing} models was the most dynamic of all types. An example is given in Fig. \ref{fig:2d_mixing}, showing the onset of convection due to extreme heating conditions within the body due to high SLR abundances. In these cases the density contrast of inner and outer layers initiated and drove convectional motion. The subsequent downwellings from the surface layers (cool and dense) to the inner parts (hot and buoyant) are reflected in the composition, temperature and density plots. We will discuss the time evolution of this in Sect. \ref{sec:T-t}. Models of this kind were only found for bodies with $R_{\mathrm{p}} \ge 80$ km. The formation time is less important than the radius, but showed significant effects by lowering the threshold $t_{\mathrm{form}}$ for smaller bodies, i.e., $R_{\mathrm{p}} \le 140$ km models did not \emph{mix} anymore above $t_{\mathrm{form}} \ge 1.6$ Myr, whereas $R_{\mathrm{p}} \ge 170$ km models did. Even less influential for the qualitative evolution were changes in initial porosity, for which no significant variance was observed.

\subsubsection{Heat balance}
\label{sec:T-t}

\begin{figure*}[bth]
        \centering
                \includegraphics[width=0.49\textwidth]{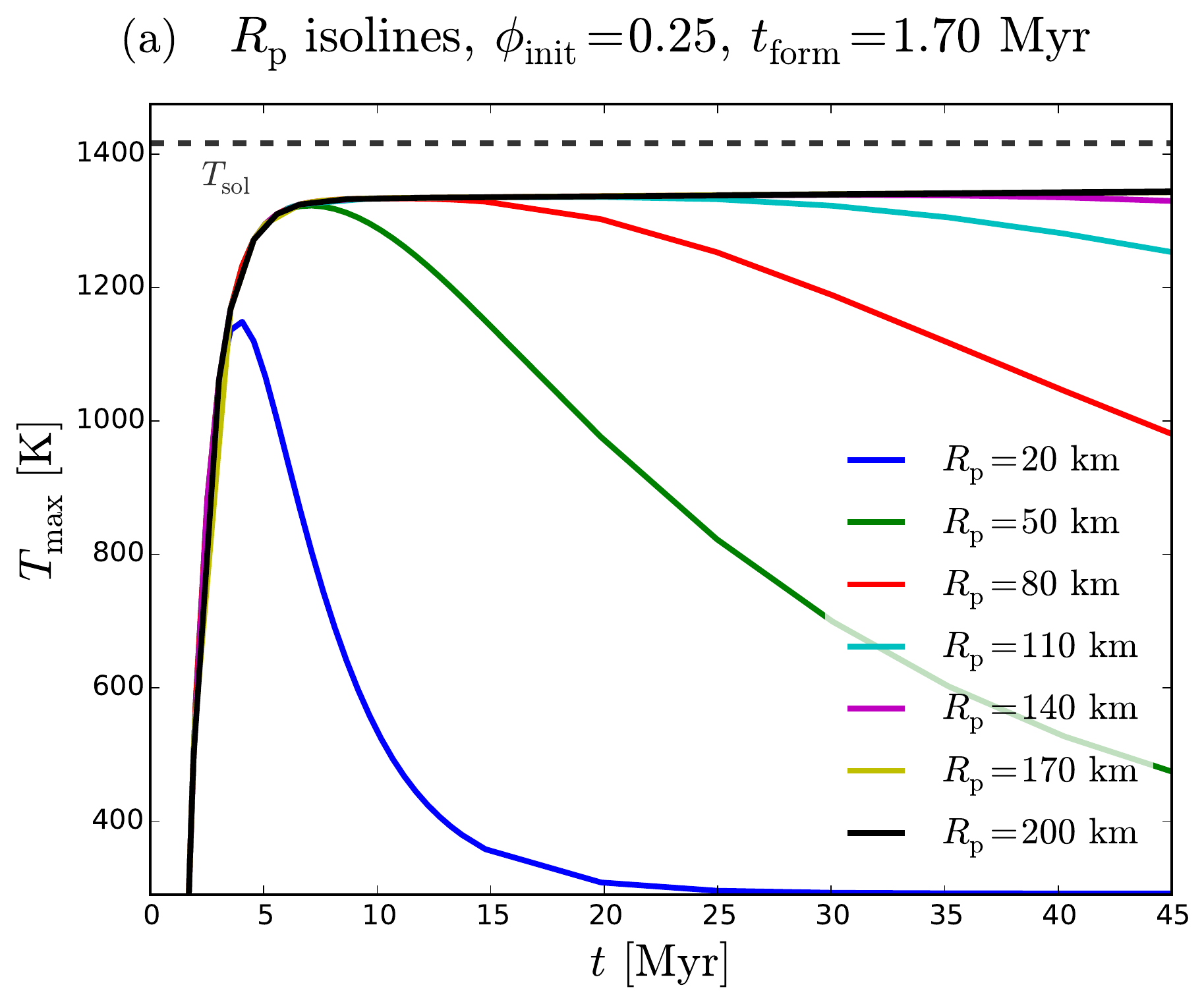}
                \includegraphics[width=0.49\textwidth]{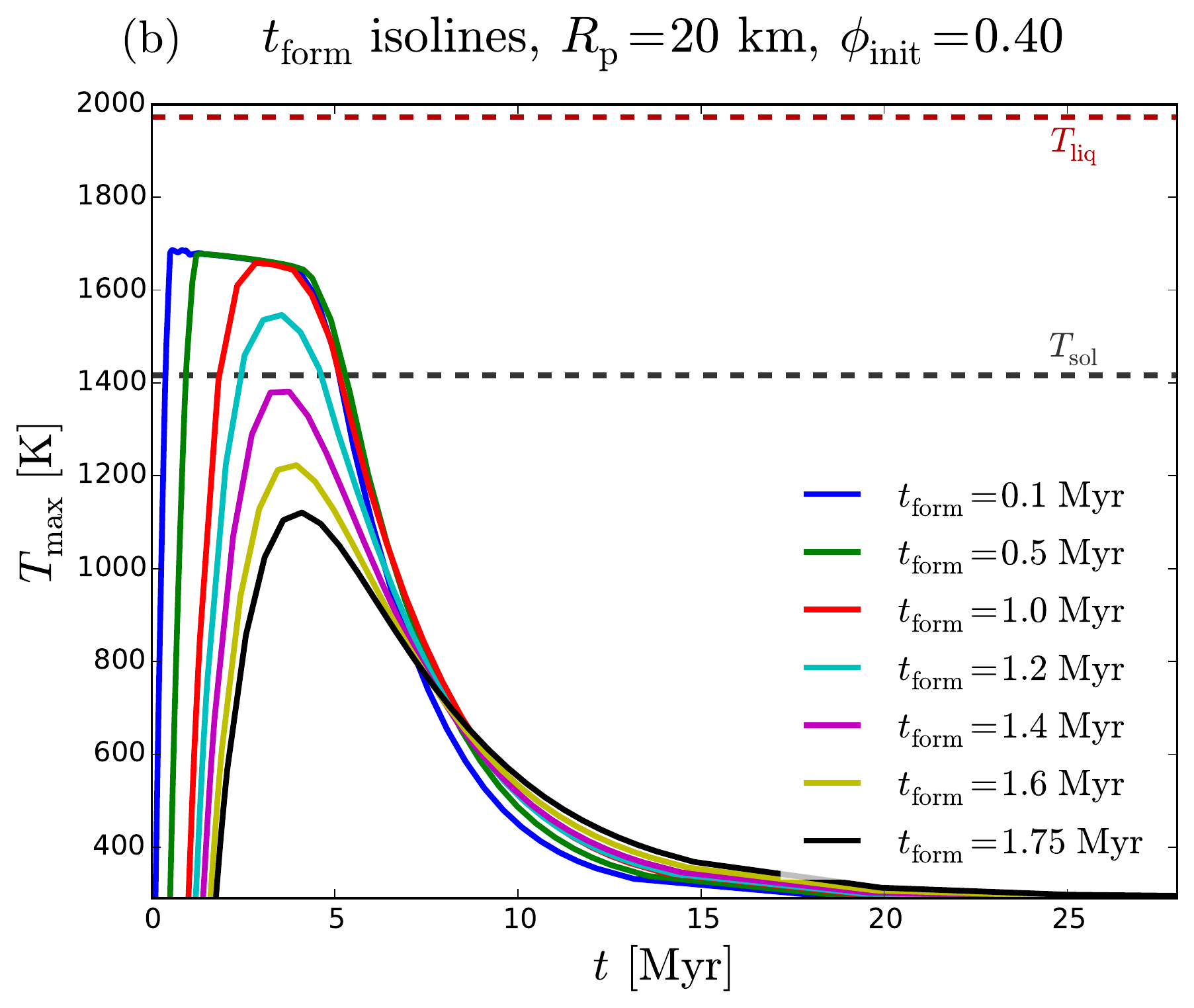}
        \caption{Peak temperature inside planetesimal versus time. $T_{\mathrm{sol}}$ and $T_{\mathrm{liq}}$ represent the solidus and liquidus temperatures, respectively. 
        \emph{(a)} Fixed $\phi_{\mathrm{init}}$, $t_{\mathrm{form}}$ and varying $R_{\mathrm{p}}$. The cooling time scale is primarily dependent on the size of body.
        \emph{(b)} Fixed $R_{\mathrm{p}}$, $\phi_{\mathrm{init}}$ and varying $t_{\mathrm{form}}$. Only models with $t_{\mathrm{form}} < 1.4$ Myr reached temperatures high enough for melting processes to occur. The $t_{\mathrm{form}} = 0.1/0.5$ Myr models were affected by the soft turbulence model from Sect. \ref{sec:melt-model}, see Sect. \ref{sec:limitations} for a discussion of the effect. 
        }
        \label{fig:Tmax_vs_time}
\end{figure*}

\begin{figure*}[tbh]
        \centering
            \includegraphics[width=0.49\textwidth]{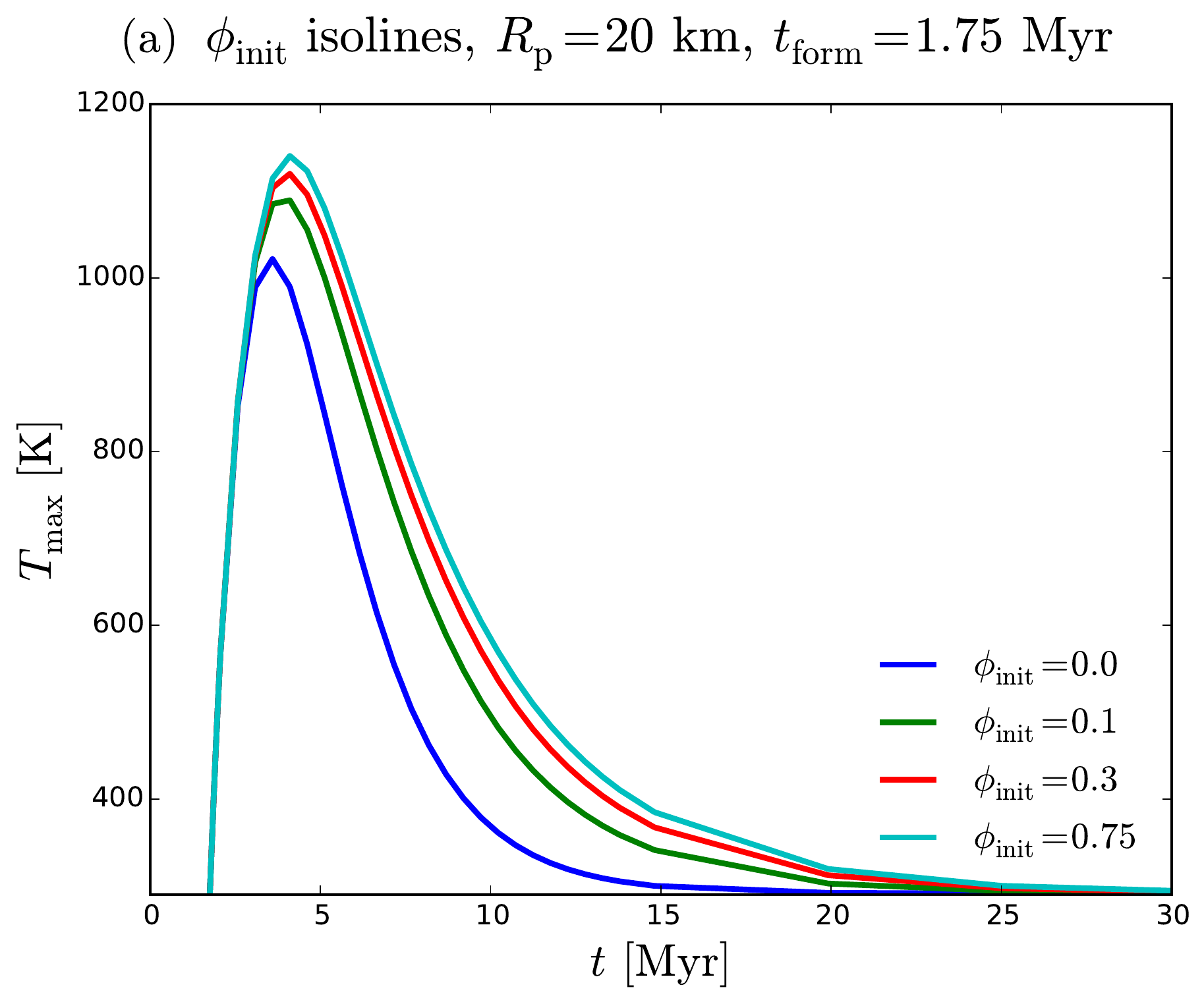}
            \includegraphics[width=0.49\textwidth]{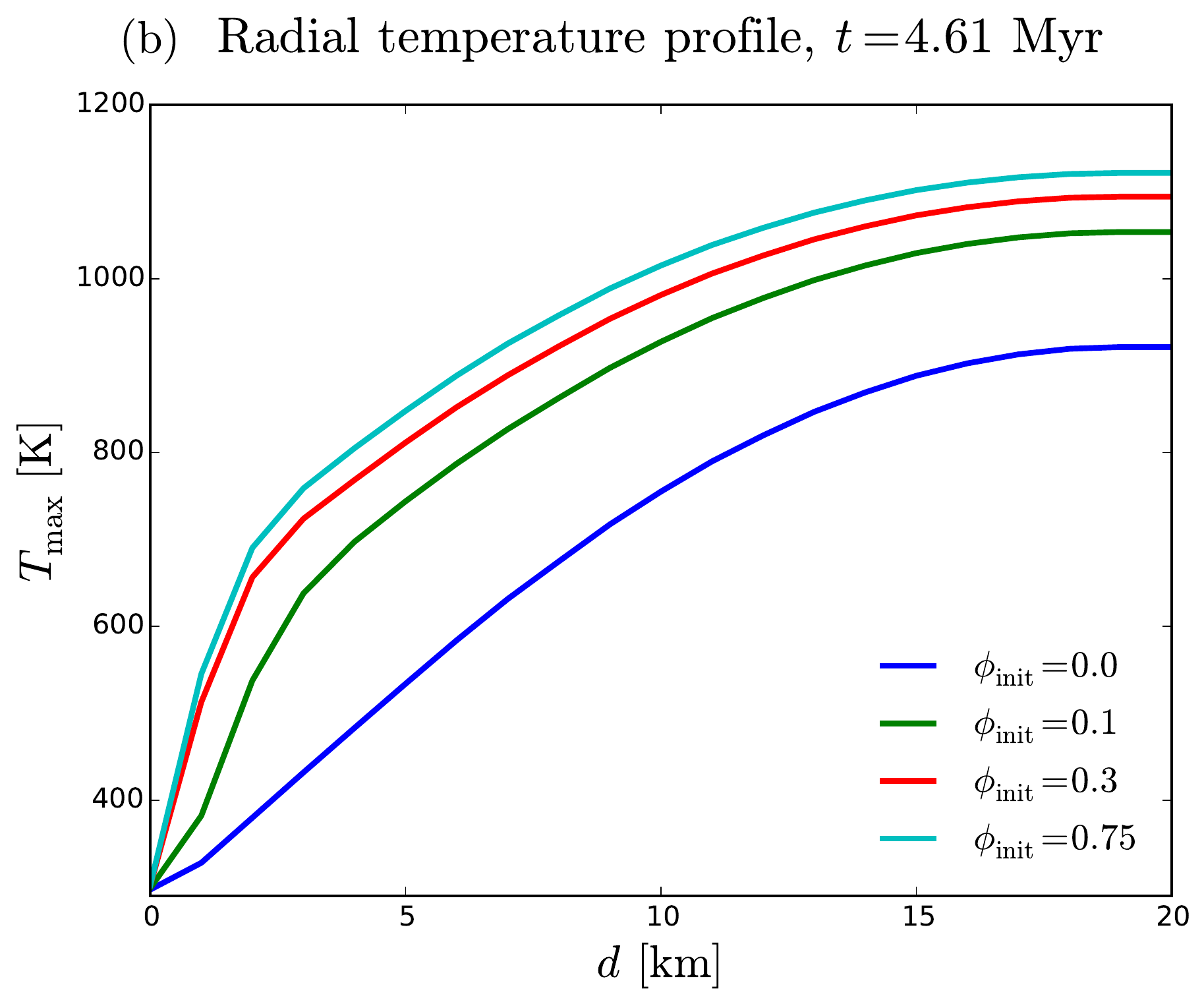}
        \caption{\emph{(a)} Peak temperature inside planetesimal versus time with fixed $R_{\mathrm{p}}$, $t_{\mathrm{form}}$ and varying $\phi_{\mathrm{init}}$. The models never reached temperatures high enough for melting processes to occur and the deviations in peak temperature were too small to inherit qualitative changes in the thermo-mechanical behavior of the simulations (compare Fig. \ref{fig:2d_grid_comp}.)
        \emph{(b)} Peak temperature profiles inside planetesimal for time $t = 4.61$ Myr for the same parameters as plot \emph{(a)}. Deviations in peak temperature were more pronounced toward the center of the planetesimal.}
        \label{fig:Tmax_porosity}
\end{figure*}

This section is devoted to an analysis of the energy reservoir in the bodies over time. To analyze the influence of each of the varied simulation parameters we construct isolines, fixing two of the three parameters (see figures \ref{fig:Tmax_vs_time} and \ref{fig:Tmax_porosity}). The models which are discussed here were among the simulations with the most extreme differences in peak temperature and are therefore best suited to show general trends in the data.

\paragraph{Influence of planetesimal radius $R_{\mathrm{p}}$}

Figure \ref{fig:Tmax_vs_time}a shows the radius isolines for all $R_{\mathrm{p}}$ values for models with $t_{\mathrm{form}} = 1.7$ Myr and $\phi_{\mathrm{init}} = 0.25$. In general, smaller bodies cooled more efficiently than their larger counterparts, which were prone to reach higher temperatures. This resulted in lower viscosities for the latter and gave them more time to develop deformed structures or convection. 

\paragraph{Influence of formation time $t_{\mathrm{form}}$}

Figure \ref{fig:Tmax_vs_time}b shows the influence of the formation time on models with $R_{\mathrm{p}} = 20$ km and $\phi_{\mathrm{init}} = 0.4$. There are two interesting characteristics to note in this plot. Firstly, the bodies with $t_{\mathrm{form}} = 0.1/0.5$ Myr showed a steep increase in temperature, compared to all other $t_{\mathrm{form}}$ isolines but reached a sudden turning point at $t \approx 7.2 \cdot 10^5$ Myr. These bodies incorporated more $^{26}$Al due to its half-life time of $t_{1/2} \approx 7.2 \cdot 10^5$ Myr. When the temperatures increased, the material transitioned to molten states and viscosities $\eta \leq \eta_{\mathrm{num}}$ occurred, the soft turbulence model set in and increased the heat flux, which permitted the body to cool at an elevated rate (see Sect. \ref{sec:limitations}).
Secondly, simulations with stronger heating sources and therefore higher peak temperatures showed steeper cooling curves than models with later formation time. In practice, the ordering of formation isolines is reverted at $t = 8$ Myr. This can be explained with the higher thermal conductivity of molten silicate states. The models with higher peak temperatures reached higher melt fractions than those with lower peak temperatures, and are therefore able to cool down more efficiently.

\paragraph{Influence of initial porosity $\phi_{\mathrm{init}}$}

Figure \ref{fig:Tmax_porosity} shows the contribution of initial porosity on peak temperature deviations in bodies with $R_{\mathrm{p}} = 20$ km and $t_{\mathrm{form}} = 1.75$ Myr. In general, higher porosity increases the voids within the granular material, effectively lowering the thermal conductivity. Therefore, models with higher initial porosity sustained their internal heat by SLRs over a longer time period. Fig. \ref{fig:Tmax_porosity}a shows an extreme case in the overall parameter range, where the maximum peak temperatures deviated by $\Delta T \approx 120$ K, not enough to achieve qualitative differences, as all peak temperatures were below the melting temperature for silicates.

To check for local variations of the temperature within specific planetesimals, we derive peak temperature profiles by assessing the maximum value from four points at the same distance from the planetesimal center. Therefore, the values in Fig.  \ref{fig:Tmax_porosity}b represent the maximum temperatures at a certain depth, which does not necessarily imply the same average value for this depth for non-axisymmetric behavior. However, irrespective of a few specific cases these are nearly undistinguishable and certainly not in the range in which these differences affect the long-term thermo-mechanical evolution. Hence, we restrict our discussion to the maximum temperature case.
The variations in peak temperature with depth were most importantly effecting small bodies, most remarkably $R_{\mathrm{p}} = 20$ km in our parameter space. Therefore, Fig. \ref{fig:Tmax_porosity}b shows the porosity isolines for the simulation with $R_{\mathrm{p}} = 20$ km and $t_{\mathrm{form}} = 1.75$ Myr at time $t = 4.61$ Myr. Going from the surface of the planetesimal to its center the temperature differences increased.

As displayed in both plots of Fig. \ref{fig:Tmax_porosity}, in such small planetesimals the peak temperatures were not enough for the onset of \emph{melting}. Thus, the temperature deviations due to porosity changes did not result in qualitative differences between the displayed models. Since the peak temperature differences between porosity isolines decrease for all other parameter combinations the porosity did not have a significant effect on the thermo-mechanical evolution of the planetesimals.


\subsection{Porous shells}
\label{sec:porosity}

\begin{figure}[bth]
    \centering
    \includegraphics[width=0.49\textwidth]{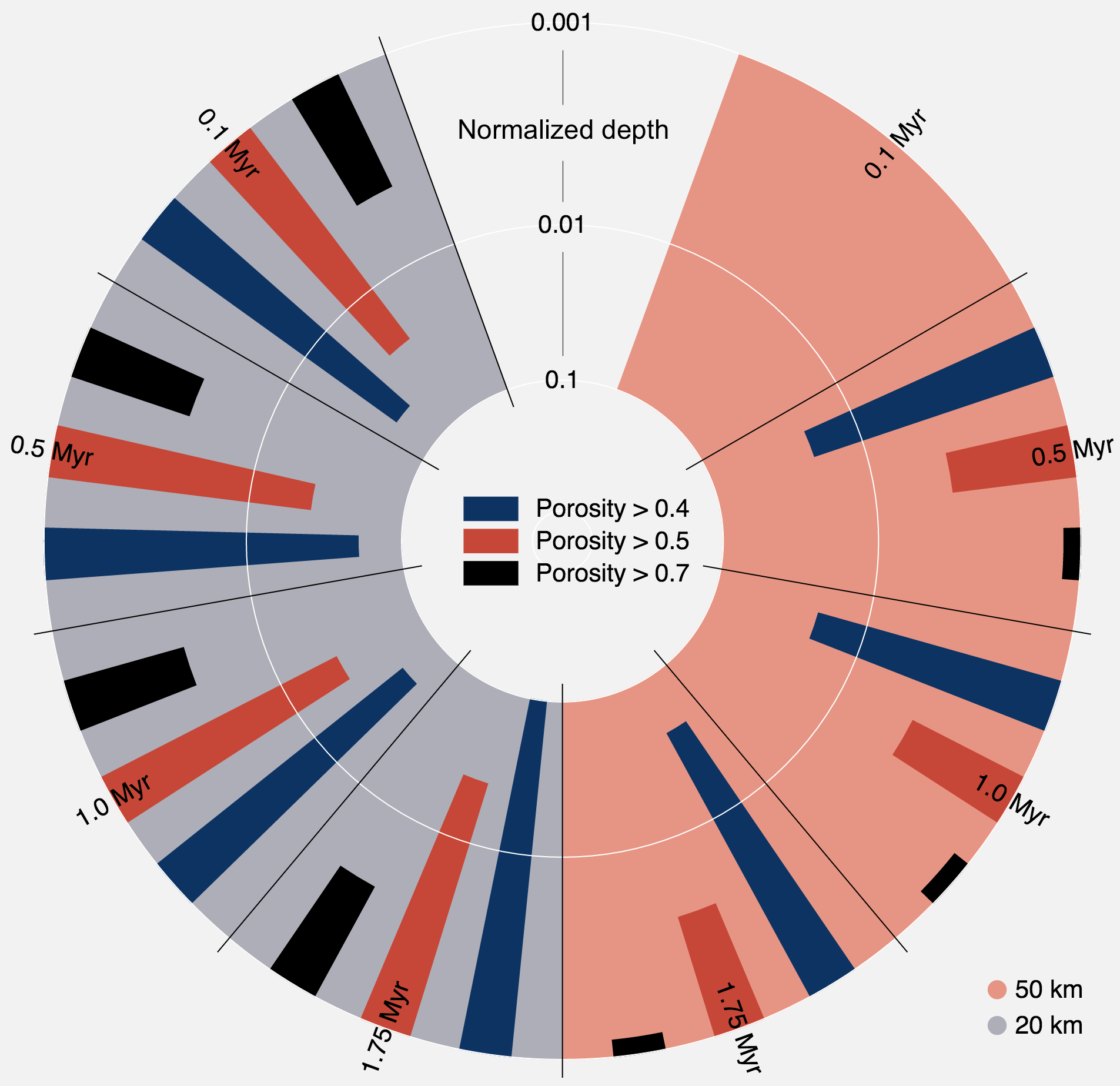}
    \caption{Remnant porous shells in evolved planetesimals with $R_{\mathrm{p}} = 20$ km (gray background) or $R_{\mathrm{p}} = 50$ km (red background), $\phi_{\mathrm{init}} = 0.75$ and varying $t_{\mathrm{form}}$. The tip of each dark blue bar represents the scaled thickness of the remnant porous shell at the end of the thermal evolution, with $\phi > 0.4$ (see Equation 12). As isostatic pressing effects decrease toward the surface, the red and black bars show depths above which the porosity was $\phi > 0.5$ or $\phi > 0.7$, respectively. The white circles represent normalized logarithmic depths within the planetesimal $d_{\mathrm{norm}} = \log(d/R_{\mathrm{p}})$ from 0.001 to 0.1. As an example, the red bar for $R_{\mathrm{p}} = 50$ km, $t_{\mathrm{form}} = 0.5$ Myr shows that for $d_{\mathrm{norm}} \lesssim 0.06$ the porosity was $\phi > 0.5$, increasing toward the surface.}
\label{fig:porous_shell}
\end{figure}

\begin{figure}[tbh]
\centering
\includegraphics[width=0.49\textwidth]{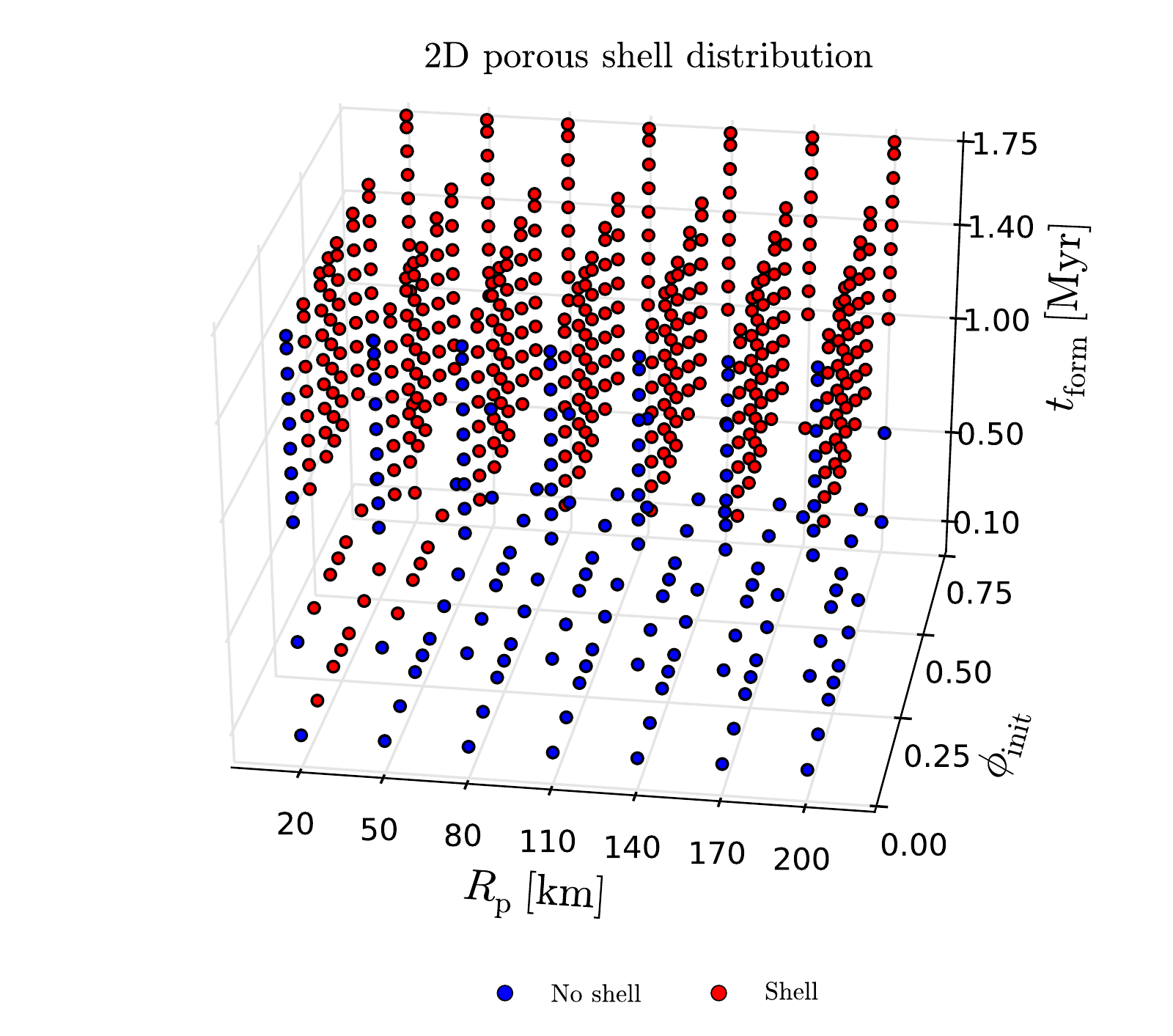}
\caption{3D parameter space showing the distribution of porous shells, with $R_{\mathrm{p}}$ in km, $\phi_{\mathrm{init}}$ non-dimensional and $t_{\mathrm{form}}$ in Myr. The retainment of a porous shell depended dominantly on formation time $t_{\mathrm{form}}$.}
\label{fig:shell_distribution}
\end{figure}

\begin{figure}[tbh]
    \centering
    \includegraphics[width=0.49\textwidth]{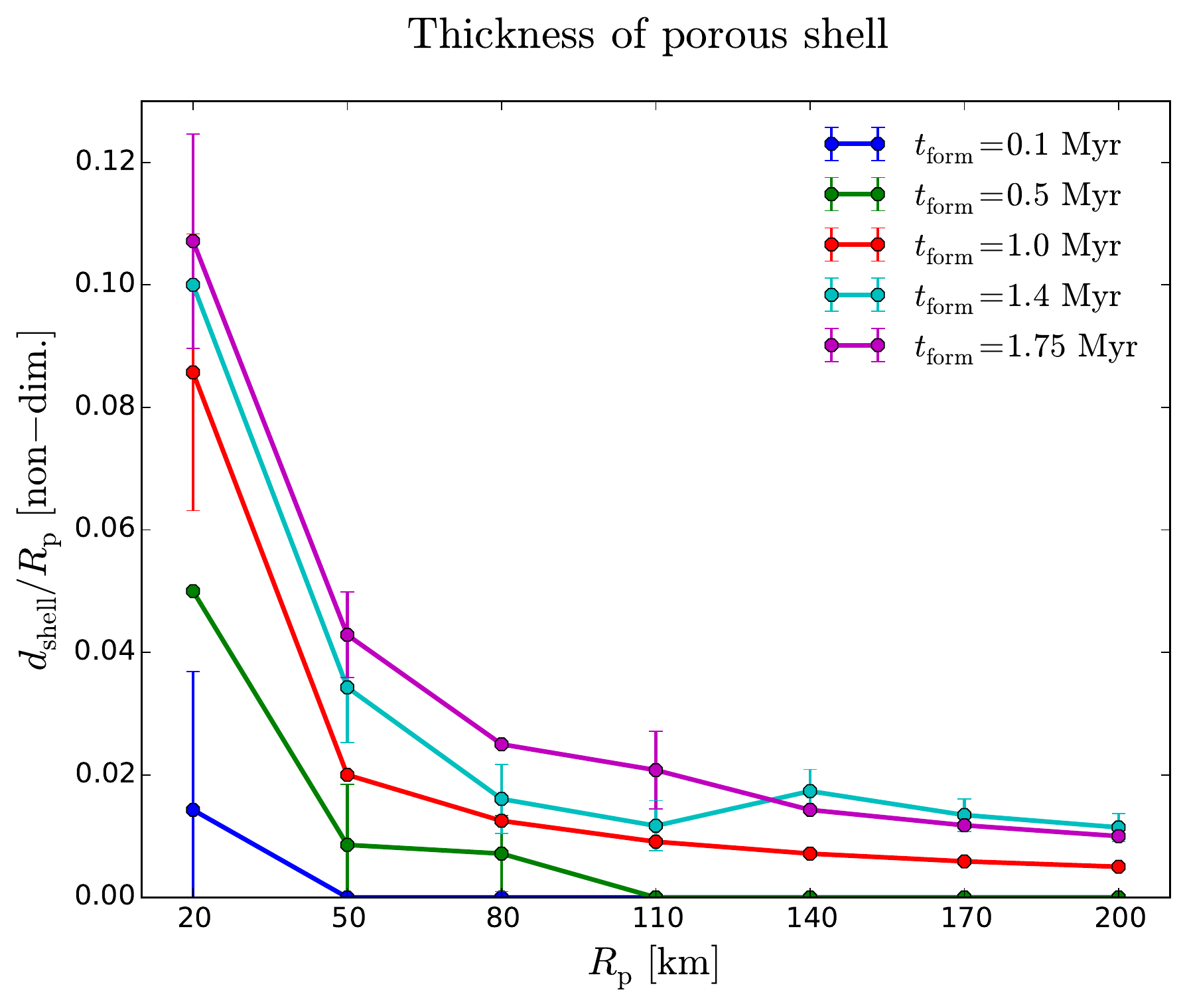}
\caption{Fraction of porous shell versus planetesimal radius after all thermo-mechanical processes have ended. The values represent the arithmetic means over the results for all initial porosities $\phi_{\mathrm{init}}$, the errorbars showing the corresponding standard deviations. The shell fraction decreased with increasing size of the body and with earlier formation time.}
\label{fig:shell_depth}
\end{figure}

Additional to the marginal effect of porosity changes on the peak temperature and the thermo-mechanical evolution, the majority of our models with initial porosity showed a porous shell feature. As illustrated for several models in Fig. \ref{fig:porous_shell}, these structures were retained during the thermo-mechanical evolution and formed because of two effects. Firstly, compaction due to self-gravity by cold pressing (Equation 12) lowered the porosity within the body close to $\phi = 0.42$  and consequently increased the density contrast between the outermost layers and the layers deeper inside the body. Secondly, during the temporal evolution of the models the temperatures deep within the planetesimals were by far higher than those close to the surface. The temperatures within the body were high enough for sintering effects, which altered the porosity value according to Equation 15. Because both effects were unimportant closer to the surface, a large subset of the model retained a porous layer throughout their whole evolution. Only the models with the most extreme heating values were hot enough to sinter or melt even their outermost layers. Fig. \ref{fig:porous_shell} shows the combined effects of planetesimal size and formation time on the extent of the porous shells and the porosity change within the shell. Sintering limited the total thickness of the shell ($d_{\mathrm{norm}}$) and compaction determined the increase in porosity toward the surface.

Figure \ref{fig:shell_distribution} illustrates the distribution of bodies with and without porous shells. Most notably, the dominant parameter determining the preservation of a porous shell was the formation time: for $t_{\mathrm{form}} \ge 1.0$ Myr all models developed such structures. Aside from the small effects of lithostatic pressure, the material distribution within the upper layers of the body did not depend on its size, since the weight on top of it was unaffected by the overall mass of the body. Therefore, these layers were only minimally affected by cold isostatic pressing. A minor effect regarding the size of the body was still observed, as models with $t_{\mathrm{form}} = 0.5$ Myr and $R_{\mathrm{p}} \le 110$ km also developed a shell, while bodies with $R_{\mathrm{p}} \ge 140$ km did not. 

The distribution of the porous shell structures within the model set remained unaffected by initial porosity $\phi_{\mathrm{init}}$ and was determined by $R_{\mathrm{p}}$ and $t_{\mathrm{form}}$. Fig. \ref{fig:shell_depth} shows the thickness of the porous shell as a fraction of the size of the body $R_{\mathrm{p}}$ for different formation times $t_{\mathrm{form}}$. The values represent an average over the results for all initial porosity values $\phi_{\mathrm{init}}$, as this parameter did not affect the final shell depths significantly. The fraction of the shell was larger for smaller bodies and for later formation times. Both vary the amount of heating sources within the body, as later formation times lowered the initial abundances of SLRs and smaller bodies cooled more efficiently and displayed lower temperatures in their interiors. Thus, sintering effects were less important.

\subsection{3D analogues}
\label{sec:3d}

\begin{figure}[tbh]
    \centering
    \includegraphics[width=0.49\textwidth]{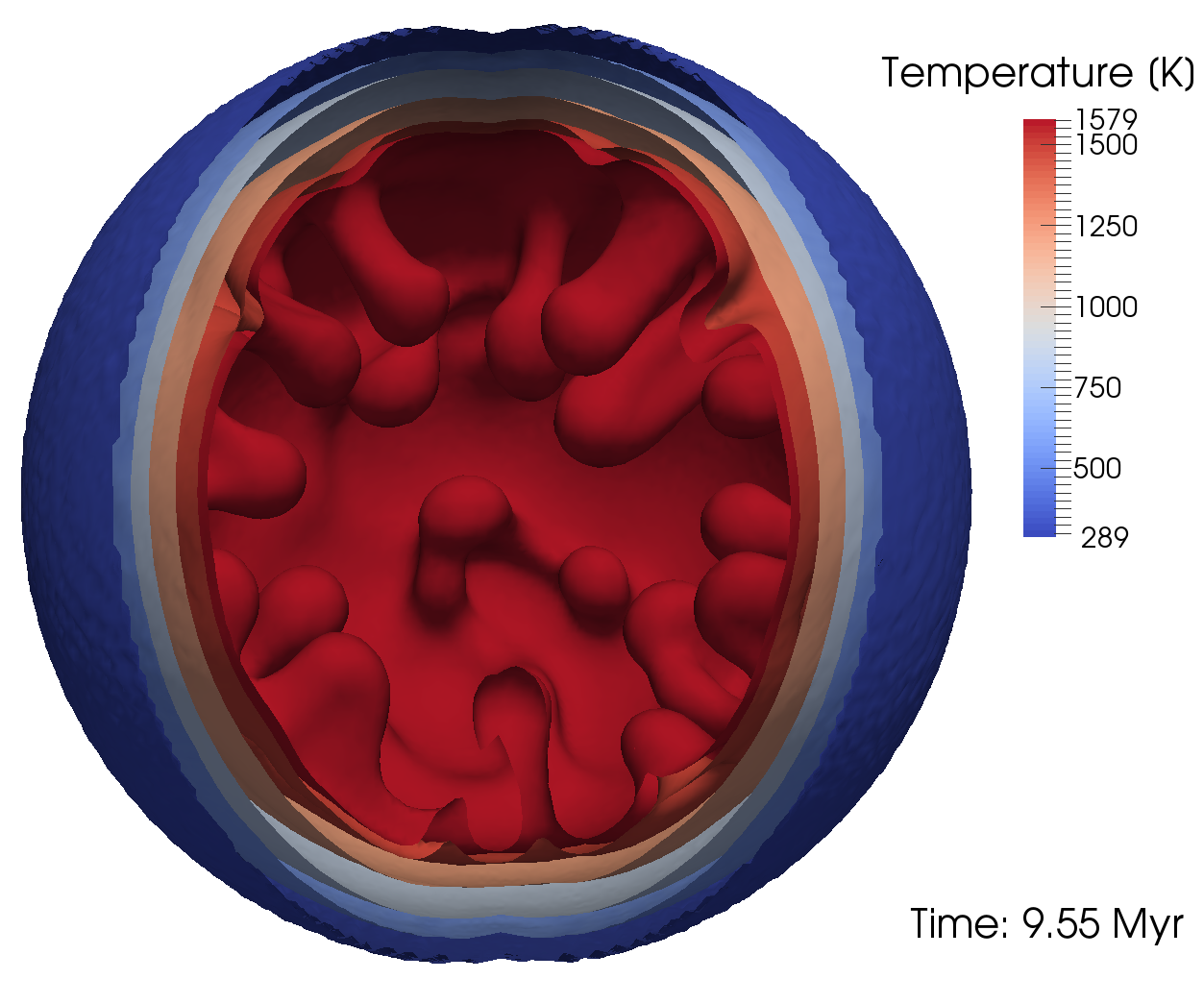}
    \caption{Density isocontours in a \emph{mixing} 3D model, with $R_{\mathrm{p}} = 110$ km, $\phi_{\mathrm{init}} = 0.25$ and $t_{\mathrm{form}} = 0.1$ Myr. The density increased from the inside (dark red, $\rho = 3100$ kg m$^{-3}$) to the outside (dark blue, $\rho = 3385.6$ kg m$^{-3}$). Therefore, the model experienced buoyancy driven mass movement.}
    \label{fig:3d_isocontours}
\end{figure}

As described in Sect. \ref{sec:initial_conditions} we additionally performed a set of 3D simulations for different parameter combinations to check for possible deviations from the 2D results. All 3D models are listed in Table \ref{tab:3d_runs}.

In principal, the selected 3D simulations confirmed the general trends we have found in two dimensions before. Smaller bodies with $R_{\mathrm{p}} \le 50$ km displayed \emph{solid} or \emph{static molten} type and developed no convection patterns, regardless of their formation time. Larger bodies were more likely to experience convectional mixing, as illustrated in Fig. \ref{fig:3d_isocontours}. Comparable to the 2D simulations the formation time was the dominant parameter for the thermo-mechanical evolution and the onset of melting processes: early formed bodies experienced stronger heating by SLRs. As expected from the 2D results we also found porous shells in the appropriate parameter ranges.

The 3D models, however, did not perfectly match the results from the 2D simulations, as can be seen for model number 624, with $R_{\mathrm{p}} = 110$ km, $\phi_{\mathrm{init}} = 0.25$ and $t_{\mathrm{form}} = 1.7$ Myr,  which evolved to a \emph{static molten} state and did not retain a porous shell. Its 2D counterpart however was \emph{solid} throughout its evolution and we found a shell at the end of its thermo-mechanical evolution. In general, as far as we can conclude from the restricted model set of 3D simulations, they seem to have experienced higher temperatures than their respective 2D analogues and thus the whole parameter space was shifted toward a higher fraction of \emph{static molten}, \emph{deformed molten} and \emph{mixing} models. As already mentioned in the introduction, this result is expected and can be attributed to the lower surface-to-volume ratio of 3D models. Hence, planetesimals in 3D experienced a lower heat flux compared to their increased volume and abundance of SLRs and therefore reached higher internal temperatures. 

All in all, our 3D models were capable of reproducing the most important structures, compositional types and porosity features of the 2D models with slightly shifted regime boundaries and therefore verified the main conclusions we have drawn before.


\section{Model limitations}
\label{sec:limitations}

The main caveat regarding the evolutionary channels from Sect. \ref{sec:therm_evo} is the lower cut-off viscosity $\eta_{\mathrm{num}}$, whereas we expect that the real viscosity at melt fractions above 0.4 drops to values orders of magnitudes smaller than the applied lower cut-off viscosity (see Sect. \ref{sec:melt-model} for examples). This especially happened for models with early formation times $t_{\mathrm{form}} = 0.1/0.5$ Myr, i.e., within the first few half-life times of $^{26}$Al. As mentioned before these low viscosities cannot be resolved numerically. 

To estimate which of our numerical models would have experienced convection, that could not be resolved, we estimate the onset time of convection based on the approach of \citet{howardconvection}. Since internal heating was important in the models, we employ the Roberts-Rayleigh number \citep{1967JFM....30...33R}, which can be used to compute the boundary layer Roberts-Rayleigh number
\begin{align}
Ra_{\delta} = \frac{\alpha g \rho_0 H_{\mathrm{r}} \delta^5}{k \kappa \eta},
\end{align}
with reference density $\rho_0$, boundary layer thickness $\delta$ and thermal diffusivity $\kappa$. For the latter we use the characteristic diffusion lengthscale $(\kappa t)^{1/2}$ and assume that the viscosity of the partial melt decreases exponentially from $10^{17}$ Pa s at $\varphi = 0.4$ to $10^{-2}$ Pa s at $\varphi = 0.6$. Solving for $t$ we obtain the relation
\begin{align}
t_{\mathrm{crit}} = \left( \frac{Ra_{\delta} k \eta}{\alpha \rho_0 g H_{\mathrm{r}}}\right)^{2/5} \kappa^{-3/5}
\label{eq:t_crit}
\end{align}
with $Ra_{\delta} \sim 30$ \citep{sotin1999three}. We use this relation to compare the time periods $\Delta t_{\eta \le \eta_{\mathrm{num}}}$, during which the viscosities are expected to drop below the numerical cut-off viscosity, with the analytical solution. Models with $t_{\mathrm{crit}} \le \Delta t_{\eta \le \eta_{\mathrm{num}}}$ are marked in Fig. \ref{fig:2d_grid_comp} (diamonds, \emph{static melt, unresolved convection}). These, together with the \emph{deformed static} class, are models for which we could not properly resolve convection. This drawback, however, did not affect the purely thermal evolution of the models, which was correctly approximated by the soft turbulence approach  \citep[as shown in][]{2001Icar..149...79T,2006MPS...41...95H,2011Icar..215..346G} and therefore all other quantities were not affected. Models for which the analytic solution is consistent with pure melting and no convection ($t_{\mathrm{crit}} > \Delta t_{\eta \le \eta_{\mathrm{num}}}$, circles in Fig. \ref{fig:2d_grid_comp}) are additionally listed in Tab. \ref{tab:2d_real_melt} and are especially found for $t_{\mathrm{form}} \approx 1.1 - 1.5$ Myr and $R_{\mathrm{p}} = 20/50$ km.

Additionally, there are some minor aspects, which could shift the trends of our results, but not crucially change the general regimes. Firstly, all planetesimals were approximated as spherical bodies. Due to accretional processes in the early formation phase, it is unlikely for planetesimals to be shaped perfectly symmetric. Irregular body structures would result in higher surface to volume ratios, hence enabling a faster cooling of the body \citep{2013MPS...48.1894D}.

Furthermore, as already discussed in \cite{2014MPS...49.1083G}, a more sophisticated approach for representing melt migration processes, cooling effects via $^{26}$Al partitioning \citep{sahijpal2007numerical} and iron-silicate-separation \citep{schubert1986thermal} would incorporate a two-phase flow model, which was not featured here. Finally we did not consider the effect of melt composition on melt density, which would influence our \emph{melting-mixing} boundary \citep{fu2014fate}.


\section{Discussion \& implications}
\label{sec:discussion}

In Sect. \ref{sec:results} we have presented the results from our set of 2D and 3D computational models of the thermo-mechanical evolution of recently formed planetesimals with varied radius, instantaneous formation time and initial porosity to gain a better understanding of the processes in the early stages of terrestrial planet formation. We now discuss the key insights of our results.

Initial porosity of the bodies was only of minor importance for the model set we have run here. Although higher initial porosity tended to lower thermal conductivity and therefore favored higher internal temperatures, the thermo-mechanical evolution was only marginally affected.

As expected, radius of the body and formation time had a strong influence on the evolution of a planetesimal. With increasing radius and decreasing formation time the bodies experienced more heating by SLRs, which resulted in higher peak temperatures and steeper heating curves. Planetesimals displaying \emph{mixing} can be expected to have experienced iron-silicate separation. The fraction of bodies prone to significant internal silicate melting was consistent with previous work on the thermal histories of planetesimals \citep{2005ASPC..341..915S}. 

With decreasing radius of the body the technical assessment of the numerical model became more important, as a thermo-mechanical regime with partially molten, but non-convectional interior was observed (static melt class in Fig. 1). In this regime with $\varphi \lesssim 0.4$ we expect the Stokes velocity $ v_{\mathrm{Stokes}} \sim g/\eta $ for iron droplets to be small, such that the time scale for differentiation is high. These melt-bearing but undifferentiated planetesimals are a potentially important link for impact splash models of chondrule formation \citep[see, e.g.,][]{2005ASPC..341..915S,2012MPS...47.2170S,2014ApJ...794...91D}. For a more stringent analysis of the importance of these models and corresponding parameter ranges we will further evaluate this connection in future work.

A subset of our models evolved to a state with highly porous outer layers, which altered the cooling history of the planetesimal. These shells occupied a larger fraction of the planetesimal radius with later formation time and smaller radius of the body. Hence, smaller and later formed objects were the most powderous bodies, which can have implications on their dynamical behavior during impact processes, as investigated by \citet{jutzi2008numerical,jutzi2009numerical}. The larger planetesimals in our dataset can either be subject to catastrophic impact events with similar-sized bodies or subject to impacts by smaller bodies. For both cases the state of the material is important for the interaction with the encountered body. All in all these effects tend to influence the dynamical history of the accretion phase of terrestrial planets and cannot be neglected for investigations of collisional growth. Additionally, the thickness of the shells could be used to relate the structure of pristine bodies in the Solar System, which did not experience catastrophic impact events after their rapid formation, to their formation time.

Many of our models reached elevated temperatures, potentially high enough to outgas existing volatile elements. When these models reached a specific boundary the resulting bodies might end up as dry bodies, unable to deliver volatile elements to the forming planets via impacts. Thus, future studies will investigate the effect of SLR heating and initial porosity on the outgassing of volatiles in small bodies and therefore might have implications for the habitability of planetary systems, when related to the delivery to accreting terrestrial planets \citep[e.g.,][]{2012Icar..221..859E,ciesla2015volatile}.

The more moderate models still showed temperatures high enough for hydration and metamorphic transformation processes, potentially creating serpentinites via an exothermic reaction. As discussed in \citet{2011Icar..213..273A} such reactions can provide energy for non-volcanic hydrothermal activity. Within certain depths of onion shell structured planetesimals, which are in accordance with our models and previous work \citep[][and references therein]{2013AREPS..41..529W}, the energy output might be in the right regime for the synthesis of primitive organic compounds, such as basic amino acids \citep{2014ApJ...783..140C}. Their synthesis is dependent on the ammonia and water content of the corresponding planetesimal and can also change with radial distance to the central star \citep{0004-637X-809-1-6}. Therefore, future studies can be directed to couple interior evolution to exterior formation conditions, i.e., the region in the protoplanetary disk and the appropriate formation time for various size classes, to gain a better understanding of the geological environment of early biological processes in our Solar System.

\section{Conclusions}
\label{sec:conclusions}

The initial state of planetesimals in the early Solar System crucially affected their thermo-mechanical evolution, which yields implications for terrestrial planet formation theories. We have conducted numerous 2D and 3D finite-difference fluid dynamics simulations of planetesimals with varying radius, formation time and initial porosity. From these we have determined the parameter space for various thermo-mechanical regimes and the influence of initial porosity. Our conclusions are the following.

\begin{itemize}
\item Typically, planetesimals with large size, early formation time and high initial porosity tended to develop convection. Small radii, late formation times and low porosities led to bodies which did not experience silicate melting.
\item A third thermo-mechanical regime with largely molten bodies without convectional mixing existed for an intermediate parameter range with a trend toward small bodies and formation times $t_{\mathrm{form}} \approx$ 1.1--1.5 Myr after CAI formation.
\item The effects of initial porosity were by far outweighed by those of planetesimal size and formation time, scarcely affecting the qualitative evolution of a planetesimal.
\item A majority of models retained a shell of highly porous material in their outer layers, which was not affected by melting and sintering processes inside the bodies. The depth of these shells increased with later formation times and decreased planetesimal size.
\end{itemize} 

With our models we were able to constrain stringent parameter ranges for the major thermo-mechanical regimes and to show that porosity is not a primary factor for the evolution of planetesimals. Future investigations will link these results to specific aspects of terrestrial planet formation, like volatile degassing and chondrule formation. Moreover, connecting these results with SLR enrichment mechanisms in stellar clusters \citep[e.g.,][]{2014MNRAS.437..946P,2016MNRAS.456.1066P}, and thus probably strongly varying abundances of SLRs, would be beneficial for a comprehensive theory of planetary assembly and habitability on interstellar or galactic scales.

\section*{Acknowledgements}
\label{sec:acknowledgements}

We thank the referee Stephen J. Mojzsis for a thorough and constructive review, which considerably helped to improve the manuscript. We thank the NCCR PlanetS and the PlanetZ platforms for an inspiring and collaborative scientific environment and Richard J. Parker and Cornelis P. Dullemond for stimulating discussions. TL was supported by ETH Research Grant ETH-17 13-1. The numerical simulations in this work were performed on the \textsc{brutus} and \textsc{euler} computing clusters of ETH Z{\"u}rich. The models were analyzed using the open source software environments \textsc{matplotlib}\footnote{\url{http://matplotlib.org}} \citep{matplotlib}, \textsc{bokeh}\footnote{\url{http://bokeh.pydata.org}} and \textsc{paraview}\footnote{\url{http://www.paraview.org}} \citep{paraview}.



\appendix

\section{Supplementary material}
\label{sec:supplementary_material}

Supplementary data associated with this article can be found, in the online version, at \url{http://dx.doi.org/10.1016/j.icarus.2016.03.004}.

\section{List of simulation runs}
\label{sec:sim_tables}

{\scriptsize


\tablehead{%
\toprule
\textsc{No.} & $R_{\mathrm{P}}$ &  $\phi_{\mathrm{init}}$ & $t_{\mathrm{form}}$& \textsc{Grid} & \textsc{Shell} & \textsc{Thermom. regime} & \textsc{Unr. Conv.}
\\ \midrule}
\tablecaption{List of all 2D simulations with radius $R_{\mathrm{P}}$ (km), formation time $t_{\mathrm{form}}$ (Myr) and initial porosity $\phi_{\mathrm{init}}$ (non-dim.). \textsc{Grid} specifies the number of nodes in the finite-difference grid, \textsc{Shell} indicates whether the corresponding model retained a porous shell after its thermo-mechanical evolution, \textsc{Thermom. regime} indicates the evolutionary channel of the model and \textsc{Unr. Conv.} states whether the model resolved the internal fluid motion.}
\label{tab:2d_runs}
\centering
\begin{supertabular}{llllllll}
001 & 20 & 0 & 0.1 & 501$^2$ & No & Static melt & Yes\\
002 & 20 & 0 & 0.5 & 501$^2$ & No & Static melt & Yes\\
003 & 20 & 0 & 1 & 501$^2$ & No & Static melt & No\\
004 & 20 & 0 & 1.1 & 501$^2$ & No & Static melt & No\\
005 & 20 & 0 & 1.2 & 501$^2$ & No & Static melt & No
 \\006 & 20 & 0 & 1.3 & 501$^2$ & No & Solid & No
 \\007 & 20 & 0 & 1.4 & 501$^2$ & No & Solid & No
 \\008 & 20 & 0 & 1.5 & 501$^2$ & No & Solid & No
 \\009 & 20 & 0 & 1.6 & 501$^2$ & No & Solid & No
 \\010 & 20 & 0 & 1.7 & 501$^2$ & No & Solid & No
 \\011 & 20 & 0 & 1.75 & 501$^2$ & No & Solid & No
 \\012 & 20 & 0.1 & 0.1 & 501$^2$ & Yes & Static melt & Yes
 \\013 & 20 & 0.1 & 0.5 & 501$^2$ & Yes & Static melt & Yes
 \\014 & 20 & 0.1 & 1 & 501$^2$ & Yes & Static melt & Yes
 \\015 & 20 & 0.1 & 1.1 & 501$^2$ & Yes & Static melt & No
 \\016 & 20 & 0.1 & 1.2 & 501$^2$ & Yes & Static melt & No
 \\017 & 20 & 0.1 & 1.3 & 501$^2$ & Yes & Static melt & No
 \\018 & 20 & 0.1 & 1.4 & 501$^2$ & Yes & Solid & No
 \\019 & 20 & 0.1 & 1.5 & 501$^2$ & Yes & Solid & No
 \\020 & 20 & 0.1 & 1.6 & 501$^2$ & Yes & Solid & No
 \\021 & 20 & 0.1 & 1.7 & 501$^2$ & Yes & Solid & No
 \\022 & 20 & 0.1 & 1.75 & 501$^2$ & Yes & Solid & No
 \\023 & 20 & 0.2 & 0.1 & 501$^2$ & Yes & Static melt & Yes
 \\024 & 20 & 0.2 & 0.5 & 501$^2$ & Yes & Static melt & Yes
 \\025 & 20 & 0.2 & 1 & 501$^2$ & Yes & Static melt & Yes
 \\026 & 20 & 0.2 & 1.1 & 501$^2$ & Yes & Static melt & No
 \\027 & 20 & 0.2 & 1.2 & 501$^2$ & Yes & Static melt & No
 \\028 & 20 & 0.2 & 1.3 & 501$^2$ & Yes & Static melt & No
 \\029 & 20 & 0.2 & 1.4 & 501$^2$ & Yes & Solid & No
 \\030 & 20 & 0.2 & 1.5 & 501$^2$ & Yes & Solid & No
 \\031 & 20 & 0.2 & 1.6 & 501$^2$ & Yes & Solid & No
 \\032 & 20 & 0.2 & 1.7 & 501$^2$ & Yes & Solid & No
 \\033 & 20 & 0.2 & 1.75 & 501$^2$ & Yes & Solid & No
 \\034 & 20 & 0.25 & 0.1 & 501$^2$ & Yes & Static melt & Yes
 \\035 & 20 & 0.25 & 0.5 & 501$^2$ & Yes & Static melt & Yes
 \\036 & 20 & 0.25 & 1 & 501$^2$ & Yes & Static melt & Yes
 \\037 & 20 & 0.25 & 1.1 & 501$^2$ & Yes & Static melt & No
 \\038 & 20 & 0.25 & 1.2 & 501$^2$ & Yes & Static melt & No
 \\039 & 20 & 0.25 & 1.3 & 501$^2$ & Yes & Static melt & No
 \\040 & 20 & 0.25 & 1.4 & 501$^2$ & Yes & Solid & No
 \\041 & 20 & 0.25 & 1.5 & 501$^2$ & Yes & Solid & No
 \\042 & 20 & 0.25 & 1.6 & 501$^2$ & Yes & Solid & No
 \\043 & 20 & 0.25 & 1.7 & 501$^2$ & Yes & Solid & No
 \\044 & 20 & 0.25 & 1.75 & 501$^2$ & Yes & Solid & No
 \\045 & 20 & 0.3 & 0.1 & 501$^2$ & Yes & Static melt & Yes
 \\046 & 20 & 0.3 & 0.5 & 501$^2$ & Yes & Static melt & Yes
 \\047 & 20 & 0.3 & 1 & 501$^2$ & Yes & Static melt & Yes
 \\048 & 20 & 0.3 & 1.1 & 501$^2$ & Yes & Static melt & No
 \\049 & 20 & 0.3 & 1.2 & 501$^2$ & Yes & Static melt & No
 \\050 & 20 & 0.3 & 1.3 & 501$^2$ & Yes & Static melt & No
 \\051 & 20 & 0.3 & 1.4 & 501$^2$ & Yes & Solid & No
 \\052 & 20 & 0.3 & 1.5 & 501$^2$ & Yes & Solid & No
 \\053 & 20 & 0.3 & 1.6 & 501$^2$ & Yes & Solid & No
 \\054 & 20 & 0.3 & 1.7 & 501$^2$ & Yes & Solid & No
 \\055 & 20 & 0.3 & 1.75 & 501$^2$ & Yes & Solid & No
 \\056 & 20 & 0.4 & 0.1 & 501$^2$ & Yes & Static melt & Yes
 \\057 & 20 & 0.4 & 0.5 & 501$^2$ & Yes & Static melt & Yes
 \\058 & 20 & 0.4 & 1 & 501$^2$ & Yes & Static melt & Yes
 \\059 & 20 & 0.4 & 1.1 & 501$^2$ & Yes & Static melt & No
 \\060 & 20 & 0.4 & 1.2 & 501$^2$ & Yes & Static melt & No
 \\061 & 20 & 0.4 & 1.3 & 501$^2$ & Yes & Static melt & No
 \\062 & 20 & 0.4 & 1.4 & 501$^2$ & Yes & Solid & No
 \\063 & 20 & 0.4 & 1.5 & 501$^2$ & Yes & Solid & No
 \\064 & 20 & 0.4 & 1.6 & 501$^2$ & Yes & Solid & No
 \\065 & 20 & 0.4 & 1.7 & 501$^2$ & Yes & Solid & No
 \\066 & 20 & 0.4 & 1.75 & 501$^2$ & Yes & Solid & No
 \\067 & 20 & 0.5 & 0.1 & 501$^2$ & Yes & Static melt & Yes
 \\068 & 20 & 0.5 & 0.5 & 501$^2$ & Yes & Static melt & Yes
 \\069 & 20 & 0.5 & 1 & 501$^2$ & Yes & Static melt & Yes
 \\070 & 20 & 0.5 & 1.1 & 501$^2$ & Yes & Static melt & Yes
 \\071 & 20 & 0.5 & 1.2 & 501$^2$ & Yes & Static melt & No
 \\072 & 20 & 0.5 & 1.3 & 501$^2$ & Yes & Static melt & No
 \\073 & 20 & 0.5 & 1.4 & 501$^2$ & Yes & Solid & No
 \\074 & 20 & 0.5 & 1.5 & 501$^2$ & Yes & Solid & No
 \\075 & 20 & 0.5 & 1.6 & 501$^2$ & Yes & Solid & No
 \\076 & 20 & 0.5 & 1.7 & 501$^2$ & Yes & Solid & No
 \\077 & 20 & 0.5 & 1.75 & 501$^2$ & Yes & Solid & No
 \\078 & 20 & 0.75 & 0.1 & 501$^2$ & Yes & Static melt & Yes
 \\079 & 20 & 0.75 & 0.5 & 501$^2$ & Yes & Static melt & Yes
 \\080 & 20 & 0.75 & 1 & 501$^2$ & Yes & Static melt & Yes
 \\081 & 20 & 0.75 & 1.1 & 501$^2$ & Yes & Static melt & Yes
 \\082 & 20 & 0.75 & 1.2 & 501$^2$ & Yes & Static melt & No
 \\083 & 20 & 0.75 & 1.3 & 501$^2$ & Yes & Static melt & No
 \\084 & 20 & 0.75 & 1.4 & 501$^2$ & Yes & Solid & No
 \\085 & 20 & 0.75 & 1.5 & 501$^2$ & Yes & Solid & No
 \\086 & 20 & 0.75 & 1.6 & 501$^2$ & Yes & Solid & No
 \\087 & 20 & 0.75 & 1.7 & 501$^2$ & Yes & Solid & No
 \\088 & 20 & 0.75 & 1.75 & 501$^2$ & Yes & Solid & No
 \\089 & 50 & 0 & 0.1 & 501$^2$ & No & Static melt & Yes
 \\090 & 50 & 0 & 0.5 & 501$^2$ & No & Static melt & Yes
 \\091 & 50 & 0 & 1 & 501$^2$ & No & Static melt & Yes
 \\092 & 50 & 0 & 1.1 & 501$^2$ & No & Static melt & Yes
 \\093 & 50 & 0 & 1.2 & 501$^2$ & No & Static melt & Yes
 \\094 & 50 & 0 & 1.3 & 501$^2$ & No & Static melt & Yes
 \\095 & 50 & 0 & 1.4 & 501$^2$ & No & Static melt & No
 \\096 & 50 & 0 & 1.5 & 501$^2$ & No & Static melt & No
 \\097 & 50 & 0 & 1.6 & 501$^2$ & No & Solid & No
 \\098 & 50 & 0 & 1.7 & 501$^2$ & No & Solid & No
 \\099 & 50 & 0 & 1.75 & 501$^2$ & No & Solid & No
 \\100 & 50 & 0.1 & 0.1 & 501$^2$ & No & Static melt & Yes
 \\101 & 50 & 0.1 & 0.5 & 501$^2$ & Yes & Static melt & Yes
 \\102 & 50 & 0.1 & 1 & 501$^2$ & Yes & Static melt & Yes
 \\103 & 50 & 0.1 & 1.1 & 501$^2$ & Yes & Static melt & Yes
 \\104 & 50 & 0.1 & 1.2 & 501$^2$ & Yes & Static melt & Yes
 \\105 & 50 & 0.1 & 1.3 & 501$^2$ & Yes & Static melt & Yes
 \\106 & 50 & 0.1 & 1.4 & 501$^2$ & Yes & Static melt & No
 \\107 & 50 & 0.1 & 1.5 & 501$^2$ & Yes & Static melt & No
 \\108 & 50 & 0.1 & 1.6 & 501$^2$ & Yes & Solid & No
 \\109 & 50 & 0.1 & 1.7 & 501$^2$ & Yes & Solid & No
 \\110 & 50 & 0.1 & 1.75 & 501$^2$ & Yes & Solid & No
 \\111 & 50 & 0.2 & 0.1 & 501$^2$ & No & Static melt & Yes
 \\112 & 50 & 0.2 & 0.5 & 501$^2$ & Yes & Static melt & Yes
 \\113 & 50 & 0.2 & 1 & 501$^2$ & Yes & Static melt & Yes
 \\114 & 50 & 0.2 & 1.1 & 501$^2$ & Yes & Static melt & Yes
 \\115 & 50 & 0.2 & 1.2 & 501$^2$ & Yes & Static melt & Yes
 \\116 & 50 & 0.2 & 1.3 & 501$^2$ & Yes & Static melt & Yes
 \\117 & 50 & 0.2 & 1.4 & 501$^2$ & Yes & Static melt & No
 \\118 & 50 & 0.2 & 1.5 & 501$^2$ & Yes & Static melt & No
 \\119 & 50 & 0.2 & 1.6 & 501$^2$ & Yes & Solid & No
 \\120 & 50 & 0.2 & 1.7 & 501$^2$ & Yes & Solid & No
 \\121 & 50 & 0.2 & 1.75 & 501$^2$ & Yes & Solid & No
 \\122 & 50 & 0.25 & 0.1 & 501$^2$ & No & Static melt & Yes
 \\123 & 50 & 0.25 & 0.5 & 501$^2$ & Yes & Static melt & Yes
 \\124 & 50 & 0.25 & 1 & 501$^2$ & Yes & Static melt & Yes
 \\125 & 50 & 0.25 & 1.1 & 501$^2$ & Yes & Static melt & Yes
 \\126 & 50 & 0.25 & 1.2 & 501$^2$ & Yes & Static melt & Yes
 \\127 & 50 & 0.25 & 1.3 & 501$^2$ & Yes & Static melt & Yes
 \\128 & 50 & 0.25 & 1.4 & 501$^2$ & Yes & Static melt & No
 \\129 & 50 & 0.25 & 1.5 & 501$^2$ & Yes & Static melt & No
 \\130 & 50 & 0.25 & 1.6 & 501$^2$ & Yes & Solid & No
 \\131 & 50 & 0.25 & 1.7 & 501$^2$ & Yes & Solid & No
 \\132 & 50 & 0.25 & 1.75 & 501$^2$ & Yes & Solid & No
 \\133 & 50 & 0.3 & 0.1 & 501$^2$ & No & Static melt & Yes
 \\134 & 50 & 0.3 & 0.5 & 501$^2$ & Yes & Static melt & Yes
 \\135 & 50 & 0.3 & 1 & 501$^2$ & Yes & Static melt & Yes
 \\136 & 50 & 0.3 & 1.1 & 501$^2$ & Yes & Static melt & Yes
 \\137 & 50 & 0.3 & 1.2 & 501$^2$ & Yes & Static melt & Yes
 \\138 & 50 & 0.3 & 1.3 & 501$^2$ & Yes & Static melt & Yes
 \\139 & 50 & 0.3 & 1.4 & 501$^2$ & Yes & Static melt & No
 \\140 & 50 & 0.3 & 1.5 & 501$^2$ & Yes & Static melt & No
 \\141 & 50 & 0.3 & 1.6 & 501$^2$ & Yes & Solid & No
 \\142 & 50 & 0.3 & 1.7 & 501$^2$ & Yes & Solid & No
 \\143 & 50 & 0.3 & 1.75 & 501$^2$ & Yes & Solid & No
 \\144 & 50 & 0.4 & 0.1 & 501$^2$ & No & Static melt & Yes
 \\145 & 50 & 0.4 & 0.5 & 501$^2$ & Yes & Def. melt & Yes
 \\146 & 50 & 0.4 & 1 & 501$^2$ & Yes & Def. melt & Yes
 \\147 & 50 & 0.4 & 1.1 & 501$^2$ & Yes & Def. melt & Yes
 \\148 & 50 & 0.4 & 1.2 & 501$^2$ & Yes & Def. melt & Yes
 \\149 & 50 & 0.4 & 1.3 & 501$^2$ & Yes & Def. melt & Yes
 \\150 & 50 & 0.4 & 1.4 & 501$^2$ & Yes & Static melt & No
 \\151 & 50 & 0.4 & 1.5 & 501$^2$ & Yes & Static melt & No
 \\152 & 50 & 0.4 & 1.6 & 501$^2$ & Yes & Solid & No
 \\153 & 50 & 0.4 & 1.7 & 501$^2$ & Yes & Solid & No
 \\154 & 50 & 0.4 & 1.75 & 501$^2$ & Yes & Solid & No
 \\155 & 50 & 0.5 & 0.1 & 501$^2$ & No & Static melt & Yes
 \\156 & 50 & 0.5 & 0.5 & 501$^2$ & No & Def. melt & Yes
 \\157 & 50 & 0.5 & 1 & 501$^2$ & Yes & Def. melt & Yes
 \\158 & 50 & 0.5 & 1.1 & 501$^2$ & Yes & Def. melt & Yes
 \\159 & 50 & 0.5 & 1.2 & 501$^2$ & Yes & Def. melt & Yes
 \\160 & 50 & 0.5 & 1.3 & 501$^2$ & Yes & Def. melt & Yes
 \\161 & 50 & 0.5 & 1.4 & 501$^2$ & Yes & Static melt & No
 \\162 & 50 & 0.5 & 1.5 & 501$^2$ & Yes & Static melt & No
 \\163 & 50 & 0.5 & 1.6 & 501$^2$ & Yes & Solid & No
 \\164 & 50 & 0.5 & 1.7 & 501$^2$ & Yes & Solid & No
 \\165 & 50 & 0.5 & 1.75 & 501$^2$ & Yes & Solid & No
 \\166 & 50 & 0.75 & 0.1 & 501$^2$ & No & Static melt & Yes
 \\167 & 50 & 0.75 & 0.5 & 501$^2$ & No & Def. melt & Yes
 \\168 & 50 & 0.75 & 1 & 501$^2$ & Yes & Def. melt & Yes
 \\169 & 50 & 0.75 & 1.1 & 501$^2$ & Yes & Def. melt & Yes
 \\170 & 50 & 0.75 & 1.2 & 501$^2$ & Yes & Def. melt & Yes
 \\171 & 50 & 0.75 & 1.3 & 501$^2$ & Yes & Def. melt & Yes
 \\172 & 50 & 0.75 & 1.4 & 501$^2$ & Yes & Static melt & No
 \\173 & 50 & 0.75 & 1.5 & 501$^2$ & Yes & Static melt & No
 \\174 & 50 & 0.75 & 1.6 & 501$^2$ & Yes & Solid & No
 \\175 & 50 & 0.75 & 1.7 & 501$^2$ & Yes & Solid & No
 \\176 & 50 & 0.75 & 1.75 & 501$^2$ & Yes & Solid & No
 \\177 & 80 & 0 & 0.1 & 501$^2$ & No & Mixing & Yes
 \\178 & 80 & 0 & 0.5 & 501$^2$ & No & Mixing & Yes
 \\179 & 80 & 0 & 1 & 501$^2$ & No & Mixing & Yes
 \\180 & 80 & 0 & 1.1 & 501$^2$ & No & Mixing & Yes
 \\181 & 80 & 0 & 1.2 & 501$^2$ & No & Mixing & Yes
 \\182 & 80 & 0 & 1.3 & 501$^2$ & No & Mixing & Yes
 \\183 & 80 & 0 & 1.4 & 501$^2$ & No & Mixing & No
 \\184 & 80 & 0 & 1.5 & 501$^2$ & No & Def. melt & No
 \\185 & 80 & 0 & 1.6 & 501$^2$ & No & Static melt & No
 \\186 & 80 & 0 & 1.7 & 501$^2$ & No & Solid & No
 \\187 & 80 & 0 & 1.75 & 501$^2$ & No & Solid & No
 \\188 & 80 & 0.1 & 0.1 & 501$^2$ & No & Mixing & Yes
 \\189 & 80 & 0.1 & 0.5 & 501$^2$ & No & Mixing & Yes
 \\190 & 80 & 0.1 & 1 & 501$^2$ & Yes & Mixing & Yes
 \\191 & 80 & 0.1 & 1.1 & 501$^2$ & Yes & Mixing & Yes
 \\192 & 80 & 0.1 & 1.2 & 501$^2$ & Yes & Mixing & Yes
 \\193 & 80 & 0.1 & 1.3 & 501$^2$ & Yes & Mixing & Yes
 \\194 & 80 & 0.1 & 1.4 & 501$^2$ & Yes & Mixing & No
 \\195 & 80 & 0.1 & 1.5 & 501$^2$ & Yes & Mixing & No
 \\196 & 80 & 0.1 & 1.6 & 501$^2$ & Yes & Static melt & No
 \\197 & 80 & 0.1 & 1.7 & 501$^2$ & Yes & Solid & No
 \\198 & 80 & 0.1 & 1.75 & 501$^2$ & Yes & Solid & No
 \\199 & 80 & 0.2 & 0.1 & 501$^2$ & No & Mixing & Yes
 \\200 & 80 & 0.2 & 0.5 & 501$^2$ & No & Mixing & Yes
 \\201 & 80 & 0.2 & 1 & 501$^2$ & Yes & Mixing & Yes
 \\202 & 80 & 0.2 & 1.1 & 501$^2$ & Yes & Mixing & Yes
 \\203 & 80 & 0.2 & 1.2 & 501$^2$ & Yes & Mixing & Yes
 \\204 & 80 & 0.2 & 1.3 & 501$^2$ & Yes & Mixing & Yes
 \\205 & 80 & 0.2 & 1.4 & 501$^2$ & Yes & Mixing & No
 \\206 & 80 & 0.2 & 1.5 & 501$^2$ & Yes & Mixing & No
 \\207 & 80 & 0.2 & 1.6 & 501$^2$ & Yes & Static melt & No
 \\208 & 80 & 0.2 & 1.7 & 501$^2$ & Yes & Solid & No
 \\209 & 80 & 0.2 & 1.75 & 501$^2$ & Yes & Solid & No
 \\210 & 80 & 0.25 & 0.1 & 501$^2$ & No & Mixing & Yes
 \\211 & 80 & 0.25 & 0.5 & 501$^2$ & No & Mixing & Yes
 \\212 & 80 & 0.25 & 1 & 501$^2$ & Yes & Mixing & Yes
 \\213 & 80 & 0.25 & 1.1 & 501$^2$ & Yes & Mixing & Yes
 \\214 & 80 & 0.25 & 1.2 & 501$^2$ & Yes & Mixing & Yes
 \\215 & 80 & 0.25 & 1.3 & 501$^2$ & Yes & Mixing & Yes
 \\216 & 80 & 0.25 & 1.4 & 501$^2$ & Yes & Mixing & No
 \\217 & 80 & 0.25 & 1.5 & 501$^2$ & Yes & Mixing & No
 \\218 & 80 & 0.25 & 1.6 & 501$^2$ & Yes & Static melt & No
 \\219 & 80 & 0.25 & 1.7 & 501$^2$ & Yes & Solid & No
 \\220 & 80 & 0.25 & 1.75 & 501$^2$ & Yes & Solid & No
 \\221 & 80 & 0.3 & 0.1 & 501$^2$ & No & Mixing & Yes
 \\222 & 80 & 0.3 & 0.5 & 501$^2$ & No & Mixing & Yes
 \\223 & 80 & 0.3 & 1 & 501$^2$ & Yes & Mixing & Yes
 \\224 & 80 & 0.3 & 1.1 & 501$^2$ & Yes & Mixing & Yes
 \\225 & 80 & 0.3 & 1.2 & 501$^2$ & Yes & Mixing & Yes
 \\226 & 80 & 0.3 & 1.3 & 501$^2$ & Yes & Mixing & Yes
 \\227 & 80 & 0.3 & 1.4 & 501$^2$ & Yes & Mixing & No
 \\228 & 80 & 0.3 & 1.5 & 501$^2$ & Yes & Mixing & No
 \\229 & 80 & 0.3 & 1.6 & 501$^2$ & Yes & Static melt & No
 \\230 & 80 & 0.3 & 1.7 & 501$^2$ & Yes & Solid & No
 \\231 & 80 & 0.3 & 1.75 & 501$^2$ & Yes & Solid & No
 \\232 & 80 & 0.4 & 0.1 & 501$^2$ & No & Mixing & Yes
 \\233 & 80 & 0.4 & 0.5 & 501$^2$ & No & Mixing & Yes
 \\234 & 80 & 0.4 & 1 & 501$^2$ & Yes & Mixing & Yes
 \\235 & 80 & 0.4 & 1.1 & 501$^2$ & Yes & Mixing & Yes
 \\236 & 80 & 0.4 & 1.2 & 501$^2$ & Yes & Mixing & Yes
 \\237 & 80 & 0.4 & 1.3 & 501$^2$ & Yes & Mixing & Yes
 \\238 & 80 & 0.4 & 1.4 & 501$^2$ & Yes & Mixing & No
 \\239 & 80 & 0.4 & 1.5 & 501$^2$ & Yes & Mixing & No
 \\240 & 80 & 0.4 & 1.6 & 501$^2$ & Yes & Static melt & No
 \\241 & 80 & 0.4 & 1.7 & 501$^2$ & Yes & Solid & No
 \\242 & 80 & 0.4 & 1.75 & 501$^2$ & Yes & Solid & No
 \\243 & 80 & 0.5 & 0.1 & 501$^2$ & No & Mixing & Yes
 \\244 & 80 & 0.5 & 0.5 & 501$^2$ & No & Mixing & Yes
 \\245 & 80 & 0.5 & 1 & 501$^2$ & Yes & Mixing & Yes
 \\246 & 80 & 0.5 & 1.1 & 501$^2$ & Yes & Mixing & Yes
 \\247 & 80 & 0.5 & 1.2 & 501$^2$ & Yes & Mixing & Yes
 \\248 & 80 & 0.5 & 1.3 & 501$^2$ & Yes & Mixing & Yes
 \\249 & 80 & 0.5 & 1.4 & 501$^2$ & Yes & Mixing & No
 \\250 & 80 & 0.5 & 1.5 & 501$^2$ & Yes & Mixing & No
 \\251 & 80 & 0.5 & 1.6 & 501$^2$ & Yes & Static melt & No
 \\252 & 80 & 0.5 & 1.7 & 501$^2$ & Yes & Solid & No
 \\253 & 80 & 0.5 & 1.75 & 501$^2$ & Yes & Solid & No
 \\254 & 80 & 0.75 & 0.1 & 501$^2$ & No & Mixing & Yes
 \\255 & 80 & 0.75 & 0.5 & 501$^2$ & No & Mixing & Yes
 \\256 & 80 & 0.75 & 1 & 501$^2$ & Yes & Mixing & Yes
 \\257 & 80 & 0.75 & 1.1 & 501$^2$ & Yes & Mixing & Yes
 \\258 & 80 & 0.75 & 1.2 & 501$^2$ & Yes & Mixing & Yes
 \\259 & 80 & 0.75 & 1.3 & 501$^2$ & Yes & Mixing & Yes
 \\260 & 80 & 0.75 & 1.4 & 501$^2$ & Yes & Mixing & No
 \\261 & 80 & 0.75 & 1.5 & 501$^2$ & Yes & Mixing & No
 \\262 & 80 & 0.75 & 1.6 & 501$^2$ & Yes & Static melt & No
 \\263 & 80 & 0.75 & 1.7 & 501$^2$ & Yes & Solid & No
 \\264 & 80 & 0.75 & 1.75 & 501$^2$ & Yes & Solid & No
 \\265 & 110 & 0 & 0.1 & 501$^2$ & No & Mixing & Yes
 \\266 & 110 & 0 & 0.5 & 501$^2$ & No & Mixing & Yes
 \\267 & 110 & 0 & 1 & 501$^2$ & No & Mixing & Yes
 \\268 & 110 & 0 & 1.1 & 501$^2$ & No & Mixing & Yes
 \\269 & 110 & 0 & 1.2 & 501$^2$ & No & Mixing & Yes
 \\270 & 110 & 0 & 1.3 & 501$^2$ & No & Mixing & Yes
 \\271 & 110 & 0 & 1.4 & 501$^2$ & No & Mixing & No
 \\272 & 110 & 0 & 1.5 & 501$^2$ & No & Mixing & No
 \\273 & 110 & 0 & 1.6 & 501$^2$ & No & Def. melt & No
 \\274 & 110 & 0 & 1.7 & 501$^2$ & No & Solid & No
 \\275 & 110 & 0 & 1.75 & 501$^2$ & No & Solid & No
 \\276 & 110 & 0.1 & 0.1 & 501$^2$ & No & Mixing & Yes
 \\277 & 110 & 0.1 & 0.5 & 501$^2$ & No & Mixing & Yes
 \\278 & 110 & 0.1 & 1 & 501$^2$ & Yes & Mixing & Yes
 \\279 & 110 & 0.1 & 1.1 & 501$^2$ & Yes & Mixing & Yes
 \\280 & 110 & 0.1 & 1.2 & 501$^2$ & Yes & Mixing & Yes
 \\281 & 110 & 0.1 & 1.3 & 501$^2$ & Yes & Mixing & Yes
 \\282 & 110 & 0.1 & 1.4 & 501$^2$ & Yes & Mixing & No
 \\283 & 110 & 0.1 & 1.5 & 501$^2$ & Yes & Mixing & No
 \\284 & 110 & 0.1 & 1.6 & 501$^2$ & Yes & Def. melt & No
 \\285 & 110 & 0.1 & 1.7 & 501$^2$ & Yes & Solid & No
 \\286 & 110 & 0.1 & 1.75 & 501$^2$ & Yes & Solid & No
 \\287 & 110 & 0.2 & 0.1 & 501$^2$ & No & Mixing & Yes
 \\288 & 110 & 0.2 & 0.5 & 501$^2$ & No & Mixing & Yes
 \\289 & 110 & 0.2 & 1 & 501$^2$ & Yes & Mixing & Yes
 \\290 & 110 & 0.2 & 1.1 & 501$^2$ & Yes & Mixing & Yes
 \\291 & 110 & 0.2 & 1.2 & 501$^2$ & Yes & Mixing & Yes
 \\292 & 110 & 0.2 & 1.3 & 501$^2$ & Yes & Mixing & Yes
 \\293 & 110 & 0.2 & 1.4 & 501$^2$ & Yes & Mixing & No
 \\294 & 110 & 0.2 & 1.5 & 501$^2$ & Yes & Mixing & No
 \\295 & 110 & 0.2 & 1.6 & 501$^2$ & Yes & Def. melt & No
 \\296 & 110 & 0.2 & 1.7 & 501$^2$ & Yes & Solid & No
 \\297 & 110 & 0.2 & 1.75 & 501$^2$ & Yes & Solid & No
 \\298 & 110 & 0.25 & 0.1 & 501$^2$ & No & Mixing & Yes
 \\299 & 110 & 0.25 & 0.5 & 501$^2$ & No & Mixing & Yes
 \\300 & 110 & 0.25 & 1 & 501$^2$ & Yes & Mixing & Yes
 \\301 & 110 & 0.25 & 1.1 & 501$^2$ & Yes & Mixing & Yes
 \\302 & 110 & 0.25 & 1.2 & 501$^2$ & Yes & Mixing & Yes
 \\303 & 110 & 0.25 & 1.3 & 501$^2$ & Yes & Mixing & Yes
 \\304 & 110 & 0.25 & 1.4 & 501$^2$ & Yes & Mixing & No
 \\305 & 110 & 0.25 & 1.5 & 501$^2$ & Yes & Mixing & No
 \\306 & 110 & 0.25 & 1.6 & 501$^2$ & Yes & Def. melt & No
 \\307 & 110 & 0.25 & 1.7 & 501$^2$ & Yes & Solid & No
 \\308 & 110 & 0.25 & 1.75 & 501$^2$ & Yes & Solid & No
 \\309 & 110 & 0.3 & 0.1 & 501$^2$ & No & Mixing & Yes
 \\310 & 110 & 0.3 & 0.5 & 501$^2$ & No & Mixing & Yes
 \\311 & 110 & 0.3 & 1 & 501$^2$ & Yes & Mixing & Yes
 \\312 & 110 & 0.3 & 1.1 & 501$^2$ & Yes & Mixing & Yes
 \\313 & 110 & 0.3 & 1.2 & 501$^2$ & Yes & Mixing & Yes
 \\314 & 110 & 0.3 & 1.3 & 501$^2$ & Yes & Mixing & Yes
 \\315 & 110 & 0.3 & 1.4 & 501$^2$ & Yes & Mixing & No
 \\316 & 110 & 0.3 & 1.5 & 501$^2$ & Yes & Mixing & No
 \\317 & 110 & 0.3 & 1.6 & 501$^2$ & Yes & Def. melt & No
 \\318 & 110 & 0.3 & 1.7 & 501$^2$ & Yes & Solid & No
 \\319 & 110 & 0.3 & 1.75 & 501$^2$ & Yes & Solid & No
 \\320 & 110 & 0.4 & 0.1 & 501$^2$ & No & Mixing & Yes
 \\321 & 110 & 0.4 & 0.5 & 501$^2$ & No & Mixing & Yes
 \\322 & 110 & 0.4 & 1 & 501$^2$ & Yes & Mixing & Yes
 \\323 & 110 & 0.4 & 1.1 & 501$^2$ & Yes & Mixing & Yes
 \\324 & 110 & 0.4 & 1.2 & 501$^2$ & Yes & Mixing & Yes
 \\325 & 110 & 0.4 & 1.3 & 501$^2$ & Yes & Mixing & Yes
 \\326 & 110 & 0.4 & 1.4 & 501$^2$ & Yes & Mixing & No
 \\327 & 110 & 0.4 & 1.5 & 501$^2$ & Yes & Mixing & No
 \\328 & 110 & 0.4 & 1.6 & 501$^2$ & Yes & Def. melt & No
 \\329 & 110 & 0.4 & 1.7 & 501$^2$ & Yes & Solid & No
 \\330 & 110 & 0.4 & 1.75 & 501$^2$ & Yes & Solid & No
 \\331 & 110 & 0.5 & 0.1 & 501$^2$ & No & Mixing & Yes
 \\332 & 110 & 0.5 & 0.5 & 501$^2$ & No & Mixing & Yes
 \\333 & 110 & 0.5 & 1 & 501$^2$ & Yes & Mixing & Yes
 \\334 & 110 & 0.5 & 1.1 & 501$^2$ & Yes & Mixing & Yes
 \\335 & 110 & 0.5 & 1.2 & 501$^2$ & Yes & Mixing & Yes
 \\336 & 110 & 0.5 & 1.3 & 501$^2$ & Yes & Mixing & Yes
 \\337 & 110 & 0.5 & 1.4 & 501$^2$ & Yes & Mixing & No
 \\338 & 110 & 0.5 & 1.5 & 501$^2$ & Yes & Mixing & No
 \\339 & 110 & 0.5 & 1.6 & 501$^2$ & Yes & Def. melt & No
 \\340 & 110 & 0.5 & 1.7 & 501$^2$ & Yes & Solid & No
 \\341 & 110 & 0.5 & 1.75 & 501$^2$ & Yes & Solid & No
 \\342 & 110 & 0.75 & 0.1 & 501$^2$ & No & Mixing & Yes
 \\343 & 110 & 0.75 & 0.5 & 501$^2$ & No & Mixing & Yes
 \\344 & 110 & 0.75 & 1 & 501$^2$ & Yes & Mixing & Yes
 \\345 & 110 & 0.75 & 1.1 & 501$^2$ & Yes & Mixing & Yes
 \\346 & 110 & 0.75 & 1.2 & 501$^2$ & Yes & Mixing & Yes
 \\347 & 110 & 0.75 & 1.3 & 501$^2$ & Yes & Mixing & Yes
 \\348 & 110 & 0.75 & 1.4 & 501$^2$ & Yes & Mixing & No
 \\349 & 110 & 0.75 & 1.5 & 501$^2$ & Yes & Mixing & No
 \\350 & 110 & 0.75 & 1.6 & 501$^2$ & Yes & Def. melt & No
 \\351 & 110 & 0.75 & 1.7 & 501$^2$ & Yes & Solid & No
 \\352 & 110 & 0.75 & 1.75 & 501$^2$ & Yes & Solid & No
 \\353 & 140 & 0 & 0.1 & 501$^2$ & No & Mixing & No
 \\354 & 140 & 0 & 0.5 & 501$^2$ & No & Mixing & No
 \\355 & 140 & 0 & 1 & 501$^2$ & No & Mixing & No
 \\356 & 140 & 0 & 1.1 & 501$^2$ & No & Mixing & No
 \\357 & 140 & 0 & 1.2 & 501$^2$ & No & Mixing & No
 \\358 & 140 & 0 & 1.3 & 501$^2$ & No & Mixing & No
 \\359 & 140 & 0 & 1.4 & 501$^2$ & No & Mixing & No
 \\360 & 140 & 0 & 1.5 & 501$^2$ & No & Mixing & No
 \\361 & 140 & 0 & 1.6 & 501$^2$ & No & Mixing & No
 \\362 & 140 & 0 & 1.7 & 501$^2$ & No & Solid & No
 \\363 & 140 & 0 & 1.75 & 501$^2$ & No & Solid & No
 \\364 & 140 & 0.1 & 0.1 & 501$^2$ & No & Mixing & No
 \\365 & 140 & 0.1 & 0.5 & 501$^2$ & No & Mixing & No
 \\366 & 140 & 0.1 & 1 & 501$^2$ & Yes & Mixing & No
 \\367 & 140 & 0.1 & 1.1 & 501$^2$ & Yes & Mixing & No
 \\368 & 140 & 0.1 & 1.2 & 501$^2$ & Yes & Mixing & No
 \\369 & 140 & 0.1 & 1.3 & 501$^2$ & Yes & Mixing & No
 \\370 & 140 & 0.1 & 1.4 & 501$^2$ & Yes & Mixing & No
 \\371 & 140 & 0.1 & 1.5 & 501$^2$ & Yes & Mixing & No
 \\372 & 140 & 0.1 & 1.6 & 501$^2$ & Yes & Mixing & No
 \\373 & 140 & 0.1 & 1.7 & 501$^2$ & Yes & Solid & No
 \\374 & 140 & 0.1 & 1.75 & 501$^2$ & Yes & Solid & No
 \\375 & 140 & 0.2 & 0.1 & 501$^2$ & No & Mixing & No
 \\376 & 140 & 0.2 & 0.5 & 501$^2$ & No & Mixing & No
 \\377 & 140 & 0.2 & 1 & 501$^2$ & Yes & Mixing & No
 \\378 & 140 & 0.2 & 1.1 & 501$^2$ & Yes & Mixing & No
 \\379 & 140 & 0.2 & 1.2 & 501$^2$ & Yes & Mixing & No
 \\380 & 140 & 0.2 & 1.3 & 501$^2$ & Yes & Mixing & No
 \\381 & 140 & 0.2 & 1.4 & 501$^2$ & Yes & Mixing & No
 \\382 & 140 & 0.2 & 1.5 & 501$^2$ & Yes & Mixing & No
 \\383 & 140 & 0.2 & 1.6 & 501$^2$ & Yes & Mixing & No
 \\384 & 140 & 0.2 & 1.7 & 501$^2$ & Yes & Solid & No
 \\385 & 140 & 0.2 & 1.75 & 501$^2$ & Yes & Solid & No
 \\386 & 140 & 0.25 & 0.1 & 501$^2$ & No & Mixing & No
 \\387 & 140 & 0.25 & 0.5 & 501$^2$ & No & Mixing & No
 \\388 & 140 & 0.25 & 1 & 501$^2$ & Yes & Mixing & No
 \\389 & 140 & 0.25 & 1.1 & 501$^2$ & Yes & Mixing & No
 \\390 & 140 & 0.25 & 1.2 & 501$^2$ & Yes & Mixing & No
 \\391 & 140 & 0.25 & 1.3 & 501$^2$ & Yes & Mixing & No
 \\392 & 140 & 0.25 & 1.4 & 501$^2$ & Yes & Mixing & No
 \\393 & 140 & 0.25 & 1.5 & 501$^2$ & Yes & Mixing & No
 \\394 & 140 & 0.25 & 1.6 & 501$^2$ & Yes & Mixing & No
 \\395 & 140 & 0.25 & 1.7 & 501$^2$ & Yes & Solid & No
 \\396 & 140 & 0.25 & 1.75 & 501$^2$ & Yes & Solid & No
 \\397 & 140 & 0.3 & 0.1 & 501$^2$ & No & Mixing & No
 \\398 & 140 & 0.3 & 0.5 & 501$^2$ & No & Mixing & No
 \\399 & 140 & 0.3 & 1 & 501$^2$ & Yes & Mixing & No
 \\400 & 140 & 0.3 & 1.1 & 501$^2$ & Yes & Mixing & No
 \\401 & 140 & 0.3 & 1.2 & 501$^2$ & Yes & Mixing & No
 \\402 & 140 & 0.3 & 1.3 & 501$^2$ & Yes & Mixing & No
 \\403 & 140 & 0.3 & 1.4 & 501$^2$ & Yes & Mixing & No
 \\404 & 140 & 0.3 & 1.5 & 501$^2$ & Yes & Mixing & No
 \\405 & 140 & 0.3 & 1.6 & 501$^2$ & Yes & Mixing & No
 \\406 & 140 & 0.3 & 1.7 & 501$^2$ & Yes & Solid & No
 \\407 & 140 & 0.3 & 1.75 & 501$^2$ & Yes & Solid & No
 \\408 & 140 & 0.4 & 0.1 & 501$^2$ & No & Mixing & No
 \\409 & 140 & 0.4 & 0.5 & 501$^2$ & No & Mixing & No
 \\410 & 140 & 0.4 & 1 & 501$^2$ & Yes & Mixing & No
 \\411 & 140 & 0.4 & 1.1 & 501$^2$ & Yes & Mixing & No
 \\412 & 140 & 0.4 & 1.2 & 501$^2$ & Yes & Mixing & No
 \\413 & 140 & 0.4 & 1.3 & 501$^2$ & Yes & Mixing & No
 \\414 & 140 & 0.4 & 1.4 & 501$^2$ & Yes & Mixing & No
 \\415 & 140 & 0.4 & 1.5 & 501$^2$ & Yes & Mixing & No
 \\416 & 140 & 0.4 & 1.6 & 501$^2$ & Yes & Mixing & No
 \\417 & 140 & 0.4 & 1.7 & 501$^2$ & Yes & Solid & No
 \\418 & 140 & 0.4 & 1.75 & 501$^2$ & Yes & Solid & No
 \\419 & 140 & 0.5 & 0.1 & 501$^2$ & No & Mixing & No
 \\420 & 140 & 0.5 & 0.5 & 501$^2$ & No & Mixing & No
 \\421 & 140 & 0.5 & 1 & 501$^2$ & Yes & Mixing & No
 \\422 & 140 & 0.5 & 1.1 & 501$^2$ & Yes & Mixing & No
 \\423 & 140 & 0.5 & 1.2 & 501$^2$ & Yes & Mixing & No
 \\424 & 140 & 0.5 & 1.3 & 501$^2$ & Yes & Mixing & No
 \\425 & 140 & 0.5 & 1.4 & 501$^2$ & Yes & Mixing & No
 \\426 & 140 & 0.5 & 1.5 & 501$^2$ & Yes & Mixing & No
 \\427 & 140 & 0.5 & 1.6 & 501$^2$ & Yes & Mixing & No
 \\428 & 140 & 0.5 & 1.7 & 501$^2$ & Yes & Solid & No
 \\429 & 140 & 0.5 & 1.75 & 501$^2$ & Yes & Solid & No
 \\430 & 140 & 0.75 & 0.1 & 501$^2$ & No & Mixing & No
 \\431 & 140 & 0.75 & 0.5 & 501$^2$ & No & Mixing & No
 \\432 & 140 & 0.75 & 1 & 501$^2$ & Yes & Mixing & No
 \\433 & 140 & 0.75 & 1.1 & 501$^2$ & Yes & Mixing & No
 \\434 & 140 & 0.75 & 1.2 & 501$^2$ & Yes & Mixing & No
 \\435 & 140 & 0.75 & 1.3 & 501$^2$ & Yes & Mixing & No
 \\436 & 140 & 0.75 & 1.4 & 501$^2$ & Yes & Mixing & No
 \\437 & 140 & 0.75 & 1.5 & 501$^2$ & Yes & Mixing & No
 \\438 & 140 & 0.75 & 1.6 & 501$^2$ & Yes & Mixing & No
 \\439 & 140 & 0.75 & 1.7 & 501$^2$ & Yes & Solid & No
 \\440 & 140 & 0.75 & 1.75 & 501$^2$ & Yes & Solid & No
 \\441 & 170 & 0 & 0.1 & 501$^2$ & No & Mixing & No
 \\442 & 170 & 0 & 0.5 & 501$^2$ & No & Mixing & No
 \\443 & 170 & 0 & 1 & 501$^2$ & No & Mixing & No
 \\444 & 170 & 0 & 1.1 & 501$^2$ & No & Mixing & No
 \\445 & 170 & 0 & 1.2 & 501$^2$ & No & Mixing & No
 \\446 & 170 & 0 & 1.3 & 501$^2$ & No & Mixing & No
 \\447 & 170 & 0 & 1.4 & 501$^2$ & No & Mixing & No
 \\448 & 170 & 0 & 1.5 & 501$^2$ & No & Mixing & No
 \\449 & 170 & 0 & 1.6 & 501$^2$ & No & Mixing & No
 \\450 & 170 & 0 & 1.7 & 501$^2$ & No & Solid & No
 \\451 & 170 & 0 & 1.75 & 501$^2$ & No & Solid & No
 \\452 & 170 & 0.1 & 0.1 & 501$^2$ & No & Mixing & No
 \\453 & 170 & 0.1 & 0.5 & 501$^2$ & No & Mixing & No
 \\454 & 170 & 0.1 & 1 & 501$^2$ & Yes & Mixing & No
 \\455 & 170 & 0.1 & 1.1 & 501$^2$ & Yes & Mixing & No
 \\456 & 170 & 0.1 & 1.2 & 501$^2$ & Yes & Mixing & No
 \\457 & 170 & 0.1 & 1.3 & 501$^2$ & Yes & Mixing & No
 \\458 & 170 & 0.1 & 1.4 & 501$^2$ & Yes & Mixing & No
 \\459 & 170 & 0.1 & 1.5 & 501$^2$ & Yes & Mixing & No
 \\460 & 170 & 0.1 & 1.6 & 501$^2$ & Yes & Mixing & No
 \\461 & 170 & 0.1 & 1.7 & 501$^2$ & Yes & Solid & No
 \\462 & 170 & 0.1 & 1.75 & 501$^2$ & Yes & Solid & No
 \\463 & 170 & 0.2 & 0.1 & 501$^2$ & No & Mixing & No
 \\464 & 170 & 0.2 & 0.5 & 501$^2$ & No & Mixing & No
 \\465 & 170 & 0.2 & 1 & 501$^2$ & Yes & Mixing & No
 \\466 & 170 & 0.2 & 1.1 & 501$^2$ & Yes & Mixing & No
 \\467 & 170 & 0.2 & 1.2 & 501$^2$ & Yes & Mixing & No
 \\468 & 170 & 0.2 & 1.3 & 501$^2$ & Yes & Mixing & No
 \\469 & 170 & 0.2 & 1.4 & 501$^2$ & Yes & Mixing & No
 \\470 & 170 & 0.2 & 1.5 & 501$^2$ & Yes & Mixing & No
 \\471 & 170 & 0.2 & 1.6 & 501$^2$ & Yes & Mixing & No
 \\472 & 170 & 0.2 & 1.7 & 501$^2$ & Yes & Solid & No
 \\473 & 170 & 0.2 & 1.75 & 501$^2$ & Yes & Solid & No
 \\474 & 170 & 0.25 & 0.1 & 501$^2$ & No & Mixing & No
 \\475 & 170 & 0.25 & 0.5 & 501$^2$ & No & Mixing & No
 \\476 & 170 & 0.25 & 1 & 501$^2$ & Yes & Mixing & No
 \\477 & 170 & 0.25 & 1.1 & 501$^2$ & Yes & Mixing & No
 \\478 & 170 & 0.25 & 1.2 & 501$^2$ & Yes & Mixing & No
 \\479 & 170 & 0.25 & 1.3 & 501$^2$ & Yes & Mixing & No
 \\480 & 170 & 0.25 & 1.4 & 501$^2$ & Yes & Mixing & No
 \\481 & 170 & 0.25 & 1.5 & 501$^2$ & Yes & Mixing & No
 \\482 & 170 & 0.25 & 1.6 & 501$^2$ & Yes & Mixing & No
 \\483 & 170 & 0.25 & 1.7 & 501$^2$ & Yes & Def. melt & No
 \\484 & 170 & 0.25 & 1.75 & 501$^2$ & Yes & Solid & No
 \\485 & 170 & 0.3 & 0.1 & 501$^2$ & No & Mixing & No
 \\486 & 170 & 0.3 & 0.5 & 501$^2$ & No & Mixing & No
 \\487 & 170 & 0.3 & 1 & 501$^2$ & Yes & Mixing & No
 \\488 & 170 & 0.3 & 1.1 & 501$^2$ & Yes & Mixing & No
 \\489 & 170 & 0.3 & 1.2 & 501$^2$ & Yes & Mixing & No
 \\490 & 170 & 0.3 & 1.3 & 501$^2$ & Yes & Mixing & No
 \\491 & 170 & 0.3 & 1.4 & 501$^2$ & Yes & Mixing & No
 \\492 & 170 & 0.3 & 1.5 & 501$^2$ & Yes & Mixing & No
 \\493 & 170 & 0.3 & 1.6 & 501$^2$ & Yes & Mixing & No
 \\494 & 170 & 0.3 & 1.7 & 501$^2$ & Yes & Def. melt & No
 \\495 & 170 & 0.3 & 1.75 & 501$^2$ & Yes & Solid & No
 \\496 & 170 & 0.4 & 0.1 & 501$^2$ & No & Mixing & No
 \\497 & 170 & 0.4 & 0.5 & 501$^2$ & No & Mixing & No
 \\498 & 170 & 0.4 & 1 & 501$^2$ & Yes & Mixing & No
 \\499 & 170 & 0.4 & 1.1 & 501$^2$ & Yes & Mixing & No
 \\500 & 170 & 0.4 & 1.2 & 501$^2$ & Yes & Mixing & No
 \\501 & 170 & 0.4 & 1.3 & 501$^2$ & Yes & Mixing & No
 \\502 & 170 & 0.4 & 1.4 & 501$^2$ & Yes & Mixing & No
 \\503 & 170 & 0.4 & 1.5 & 501$^2$ & Yes & Mixing & No
 \\504 & 170 & 0.4 & 1.6 & 501$^2$ & Yes & Mixing & No
 \\505 & 170 & 0.4 & 1.7 & 501$^2$ & Yes & Def. melt & No
 \\506 & 170 & 0.4 & 1.75 & 501$^2$ & Yes & Solid & No
 \\507 & 170 & 0.5 & 0.1 & 501$^2$ & No & Mixing & No
 \\508 & 170 & 0.5 & 0.5 & 501$^2$ & No & Mixing & No
 \\509 & 170 & 0.5 & 1 & 501$^2$ & Yes & Mixing & No
 \\510 & 170 & 0.5 & 1.1 & 501$^2$ & Yes & Mixing & No
 \\511 & 170 & 0.5 & 1.2 & 501$^2$ & Yes & Mixing & No
 \\512 & 170 & 0.5 & 1.3 & 501$^2$ & Yes & Mixing & No
 \\513 & 170 & 0.5 & 1.4 & 501$^2$ & Yes & Mixing & No
 \\514 & 170 & 0.5 & 1.5 & 501$^2$ & Yes & Mixing & No
 \\515 & 170 & 0.5 & 1.6 & 501$^2$ & Yes & Mixing & No
 \\516 & 170 & 0.5 & 1.7 & 501$^2$ & Yes & Def. melt & No
 \\517 & 170 & 0.5 & 1.75 & 501$^2$ & Yes & Solid & No
 \\518 & 170 & 0.75 & 0.1 & 501$^2$ & No & Mixing & No
 \\519 & 170 & 0.75 & 0.5 & 501$^2$ & Yes & Mixing & No
 \\520 & 170 & 0.75 & 1 & 501$^2$ & Yes & Mixing & No
 \\521 & 170 & 0.75 & 1.1 & 501$^2$ & Yes & Mixing & No
 \\522 & 170 & 0.75 & 1.2 & 501$^2$ & Yes & Mixing & No
 \\523 & 170 & 0.75 & 1.3 & 501$^2$ & Yes & Mixing & No
 \\524 & 170 & 0.75 & 1.4 & 501$^2$ & Yes & Mixing & No
 \\525 & 170 & 0.75 & 1.5 & 501$^2$ & Yes & Mixing & No
 \\526 & 170 & 0.75 & 1.6 & 501$^2$ & Yes & Mixing & No
 \\527 & 170 & 0.75 & 1.7 & 501$^2$ & Yes & Def. melt & No
 \\528 & 170 & 0.75 & 1.75 & 501$^2$ & Yes & Solid & No
 \\529 & 200 & 0 & 0.1 & 501$^2$ & No & Mixing & No
 \\530 & 200 & 0 & 0.5 & 501$^2$ & No & Mixing & No
 \\531 & 200 & 0 & 1 & 501$^2$ & No & Mixing & No
 \\532 & 200 & 0 & 1.1 & 501$^2$ & No & Mixing & No
 \\533 & 200 & 0 & 1.2 & 501$^2$ & No & Mixing & No
 \\534 & 200 & 0 & 1.3 & 501$^2$ & No & Mixing & No
 \\535 & 200 & 0 & 1.4 & 501$^2$ & No & Mixing & No
 \\536 & 200 & 0 & 1.5 & 501$^2$ & No & Mixing & No
 \\537 & 200 & 0 & 1.6 & 501$^2$ & No & Mixing & No
 \\538 & 200 & 0 & 1.7 & 501$^2$ & No & Mixing & No
 \\539 & 200 & 0 & 1.75 & 501$^2$ & No & Solid & No
 \\540 & 200 & 0.1 & 0.1 & 501$^2$ & No & Mixing & No
 \\541 & 200 & 0.1 & 0.5 & 501$^2$ & No & Mixing & No
 \\542 & 200 & 0.1 & 1 & 501$^2$ & Yes & Mixing & No
 \\543 & 200 & 0.1 & 1.1 & 501$^2$ & Yes & Mixing & No
 \\544 & 200 & 0.1 & 1.2 & 501$^2$ & Yes & Mixing & No
 \\545 & 200 & 0.1 & 1.3 & 501$^2$ & Yes & Mixing & No
 \\546 & 200 & 0.1 & 1.4 & 501$^2$ & Yes & Mixing & No
 \\547 & 200 & 0.1 & 1.5 & 501$^2$ & Yes & Mixing & No
 \\548 & 200 & 0.1 & 1.6 & 501$^2$ & Yes & Mixing & No
 \\549 & 200 & 0.1 & 1.7 & 501$^2$ & Yes & Mixing & No
 \\550 & 200 & 0.1 & 1.75 & 501$^2$ & Yes & Solid & No
 \\551 & 200 & 0.2 & 0.1 & 501$^2$ & No & Mixing & No
 \\552 & 200 & 0.2 & 0.5 & 501$^2$ & No & Mixing & No
 \\553 & 200 & 0.2 & 1 & 501$^2$ & Yes & Mixing & No
 \\554 & 200 & 0.2 & 1.1 & 501$^2$ & Yes & Mixing & No
 \\555 & 200 & 0.2 & 1.2 & 501$^2$ & Yes & Mixing & No
 \\556 & 200 & 0.2 & 1.3 & 501$^2$ & Yes & Mixing & No
 \\557 & 200 & 0.2 & 1.4 & 501$^2$ & Yes & Mixing & No
 \\558 & 200 & 0.2 & 1.5 & 501$^2$ & Yes & Mixing & No
 \\559 & 200 & 0.2 & 1.6 & 501$^2$ & Yes & Mixing & No
 \\560 & 200 & 0.2 & 1.7 & 501$^2$ & Yes & Mixing & No
 \\561 & 200 & 0.2 & 1.75 & 501$^2$ & Yes & Solid & No
 \\562 & 200 & 0.25 & 0.1 & 501$^2$ & No & Mixing & No
 \\563 & 200 & 0.25 & 0.5 & 501$^2$ & No & Mixing & No
 \\564 & 200 & 0.25 & 1 & 501$^2$ & Yes & Mixing & No
 \\565 & 200 & 0.25 & 1.1 & 501$^2$ & Yes & Mixing & No
 \\566 & 200 & 0.25 & 1.2 & 501$^2$ & Yes & Mixing & No
 \\567 & 200 & 0.25 & 1.3 & 501$^2$ & Yes & Mixing & No
 \\568 & 200 & 0.25 & 1.4 & 501$^2$ & Yes & Mixing & No
 \\569 & 200 & 0.25 & 1.5 & 501$^2$ & Yes & Mixing & No
 \\570 & 200 & 0.25 & 1.6 & 501$^2$ & Yes & Mixing & No
 \\571 & 200 & 0.25 & 1.7 & 501$^2$ & Yes & Mixing & No
 \\572 & 200 & 0.25 & 1.75 & 501$^2$ & Yes & Solid & No
 \\573 & 200 & 0.3 & 0.1 & 501$^2$ & No & Mixing & No
 \\574 & 200 & 0.3 & 0.5 & 501$^2$ & No & Mixing & No
 \\575 & 200 & 0.3 & 1 & 501$^2$ & Yes & Mixing & No
 \\576 & 200 & 0.3 & 1.1 & 501$^2$ & Yes & Mixing & No
 \\577 & 200 & 0.3 & 1.2 & 501$^2$ & Yes & Mixing & No
 \\578 & 200 & 0.3 & 1.3 & 501$^2$ & Yes & Mixing & No
 \\579 & 200 & 0.3 & 1.4 & 501$^2$ & Yes & Mixing & No
 \\580 & 200 & 0.3 & 1.5 & 501$^2$ & Yes & Mixing & No
 \\581 & 200 & 0.3 & 1.6 & 501$^2$ & Yes & Mixing & No
 \\582 & 200 & 0.3 & 1.7 & 501$^2$ & Yes & Mixing & No
 \\583 & 200 & 0.3 & 1.75 & 501$^2$ & Yes & Solid & No
 \\584 & 200 & 0.4 & 0.1 & 501$^2$ & No & Mixing & No
 \\585 & 200 & 0.4 & 0.5 & 501$^2$ & No & Mixing & No
 \\586 & 200 & 0.4 & 1 & 501$^2$ & Yes & Mixing & No
 \\587 & 200 & 0.4 & 1.1 & 501$^2$ & Yes & Mixing & No
 \\588 & 200 & 0.4 & 1.2 & 501$^2$ & Yes & Mixing & No
 \\589 & 200 & 0.4 & 1.3 & 501$^2$ & Yes & Mixing & No
 \\590 & 200 & 0.4 & 1.4 & 501$^2$ & Yes & Mixing & No
 \\591 & 200 & 0.4 & 1.5 & 501$^2$ & Yes & Mixing & No
 \\592 & 200 & 0.4 & 1.6 & 501$^2$ & Yes & Mixing & No
 \\593 & 200 & 0.4 & 1.7 & 501$^2$ & Yes & Mixing & No
 \\594 & 200 & 0.4 & 1.75 & 501$^2$ & Yes & Solid & No
 \\595 & 200 & 0.5 & 0.1 & 501$^2$ & No & Mixing & No
 \\596 & 200 & 0.5 & 0.5 & 501$^2$ & No & Mixing & No
 \\597 & 200 & 0.5 & 1 & 501$^2$ & Yes & Mixing & No
 \\598 & 200 & 0.5 & 1.1 & 501$^2$ & Yes & Mixing & No
 \\599 & 200 & 0.5 & 1.2 & 501$^2$ & Yes & Mixing & No
 \\600 & 200 & 0.5 & 1.3 & 501$^2$ & Yes & Mixing & No
 \\601 & 200 & 0.5 & 1.4 & 501$^2$ & Yes & Mixing & No
 \\602 & 200 & 0.5 & 1.5 & 501$^2$ & Yes & Mixing & No
 \\603 & 200 & 0.5 & 1.6 & 501$^2$ & Yes & Mixing & No
 \\604 & 200 & 0.5 & 1.7 & 501$^2$ & Yes & Mixing & No
 \\605 & 200 & 0.5 & 1.75 & 501$^2$ & Yes & Solid & No
 \\606 & 200 & 0.75 & 0.1 & 501$^2$ & No & Mixing & No
 \\607 & 200 & 0.75 & 0.5 & 501$^2$ & No & Mixing & No
 \\608 & 200 & 0.75 & 1 & 501$^2$ & Yes & Mixing & No
 \\609 & 200 & 0.75 & 1.1 & 501$^2$ & Yes & Mixing & No
 \\610 & 200 & 0.75 & 1.2 & 501$^2$ & Yes & Mixing & No
 \\611 & 200 & 0.75 & 1.3 & 501$^2$ & Yes & Mixing & No
 \\612 & 200 & 0.75 & 1.4 & 501$^2$ & Yes & Mixing & No
 \\613 & 200 & 0.75 & 1.5 & 501$^2$ & Yes & Mixing & No
 \\614 & 200 & 0.75 & 1.6 & 501$^2$ & Yes & Mixing & No
 \\615 & 200 & 0.75 & 1.7 & 501$^2$ & Yes & Mixing & No
 \\616 & 200 & 0.75 & 1.75 & 501$^2$ & Yes & Solid & No\\
\end{supertabular}%


\centering

\tablefirsthead{\toprule \textsc{No.} & $R_{\mathrm{P}}$ &  $\phi_{\mathrm{init}}$ & $t_{\mathrm{form}}$ & \textsc{Grid} & \textsc{Shell} & \textsc{Thermom. regime}  \\ \midrule}
\tablehead{%
\toprule
\textsc{No.} & $R_{\mathrm{P}}$&  $\phi_{\mathrm{init}}$ & $t_{\mathrm{form}}$& \textsc{Grid} & \textsc{Shell} & \textsc{Thermom. regime} 
\\ \midrule}
\tablecaption{List of all 3D simulations with radius $R_{\mathrm{P}}$ (km), formation time $t_{\mathrm{form}}$ (Myr) and initial porosity $\phi_{\mathrm{init}}$ (non-dim.). \textsc{Grid} specifies the number of nodes in the finite-difference grid, \textsc{Shell} indicates whether the corresponding model retained a porous shell after its thermo-mechanical evolution ended and \textsc{Thermom. regime} indicates the evolutionary channel of the model.}
\label{tab:3d_runs}
\begin{supertabular}{lllllll}
    617 & 20 & 0.4 & 0.1 & 261$^3$ & Yes & Static melt      \\
    618 & 50 & 0.3 & 1.75 & 261$^3$ & Yes & Solid           \\
    619 & 50 & 0.3 & 1.75 & 261$^3$ & Yes & Solid           \\
    620 & 50 & 0.25 & 0.5 & 261$^3$ & No  & Def. melt       \\
    621 & 50 & 0.25 & 1.5 & 261$^3$ & Yes & Static melt     \\
    622 & 50 & 0.5 & 1.5 & 261$^3$ & Yes & Static melt      \\
    623 & 110 & 0.25 & 0.1 & 261$^3$ & No & Mixing          \\
    624 & 110 & 0.25 & 1.7 & 261$^3$ & No & Static melt     \\
    625 & 140 & 0.2 & 0.5 & 261$^3$ & No & Mixing           \\
    626 & 140 & 0.4 & 1.0 & 261$^3$ & No & Mixing           \\
    627 & 170 & 0.4 & 1.3 & 261$^3$ & No & Mixing           \\
\end{supertabular}%


\tablefirsthead{\toprule \textsc{No.} & $R_{\mathrm{P}}$ &  $\phi_{\mathrm{init}}$ & $t_{\mathrm{form}}$ & \textsc{Grid} & \textsc{Shell} \\ \midrule}
\tablehead{%
\toprule
\textsc{No.} & $R_{\mathrm{P}}$&  $\phi_{\mathrm{init}}$ & $t_{\mathrm{form}}$& \textsc{Grid} & \textsc{Shell}
\\ \midrule}
\tablecaption{List of all 2D simulations of thermo-mechanical type \emph{static melt}, for which the numerical model is consistent with the analytical solution. Parameters are radius $R_{\mathrm{P}}$ (km), formation time $t_{\mathrm{form}}$ (Myr), initial porosity $\phi_{\mathrm{init}}$ (non-dim.), \textsc{Grid} specifies the number of nodes in the finite-difference grid, \textsc{Shell} indicates whether the corresponding model retained a porous shell after its thermo-mechanical evolution.}
\label{tab:2d_real_melt}
\begin{supertabular}{llllll}
    003 & 20 & 0 & 1 & 501$^2$ & No \\
004 & 20 & 0 & 1.1 & 501$^2$ & No \\
005 & 20 & 0 & 1.2 & 501$^2$ & No \\
015 & 20 & 0.1 & 1.1 & 501$^2$ & Yes \\
016 & 20 & 0.1 & 1.2 & 501$^2$ & Yes \\
017 & 20 & 0.1 & 1.3 & 501$^2$ & Yes \\
026 & 20 & 0.2 & 1.1 & 501$^2$ & Yes \\
027 & 20 & 0.2 & 1.2 & 501$^2$ & Yes \\
028 & 20 & 0.2 & 1.3 & 501$^2$ & Yes \\
037 & 20 & 0.25 & 1.1 & 501$^2$ & Yes \\
038 & 20 & 0.25 & 1.2 & 501$^2$ & Yes \\
039 & 20 & 0.25 & 1.3 & 501$^2$ & Yes \\
048 & 20 & 0.3 & 1.1 & 501$^2$ & Yes \\
049 & 20 & 0.3 & 1.2 & 501$^2$ & Yes \\
050 & 20 & 0.3 & 1.3 & 501$^2$ & Yes \\
059 & 20 & 0.4 & 1.1 & 501$^2$ & Yes \\
060 & 20 & 0.4 & 1.2 & 501$^2$ & Yes \\
061 & 20 & 0.4 & 1.3 & 501$^2$ & Yes \\
071 & 20 & 0.5 & 1.2 & 501$^2$ & Yes \\
072 & 20 & 0.5 & 1.3 & 501$^2$ & Yes \\
082 & 20 & 0.75 & 1.2 & 501$^2$ & Yes \\
083 & 20 & 0.75 & 1.3 & 501$^2$ & Yes \\
095 & 50 & 0 & 1.4 & 501$^2$ & No \\
096 & 50 & 0 & 1.5 & 501$^2$ & No \\
106 & 50 & 0.1 & 1.4 & 501$^2$ & Yes \\
107 & 50 & 0.1 & 1.5 & 501$^2$ & Yes \\
117 & 50 & 0.2 & 1.4 & 501$^2$ & Yes \\
118 & 50 & 0.2 & 1.5 & 501$^2$ & Yes \\
128 & 50 & 0.25 & 1.4 & 501$^2$ & Yes \\
129 & 50 & 0.25 & 1.5 & 501$^2$ & Yes \\
139 & 50 & 0.3 & 1.4 & 501$^2$ & Yes \\
140 & 50 & 0.3 & 1.5 & 501$^2$ & Yes \\
150 & 50 & 0.4 & 1.4 & 501$^2$ & Yes \\
151 & 50 & 0.4 & 1.5 & 501$^2$ & Yes \\
161 & 50 & 0.5 & 1.4 & 501$^2$ & Yes \\
162 & 50 & 0.5 & 1.5 & 501$^2$ & Yes \\
172 & 50 & 0.75 & 1.4 & 501$^2$ & Yes \\
173 & 50 & 0.75 & 1.5 & 501$^2$ & Yes \\
185 & 80 & 0 & 1.6 & 501$^2$ & No \\
196 & 80 & 0.1 & 1.6 & 501$^2$ & Yes \\
207 & 80 & 0.2 & 1.6 & 501$^2$ & Yes \\
218 & 80 & 0.25 & 1.6 & 501$^2$ & Yes \\
229 & 80 & 0.3 & 1.6 & 501$^2$ & Yes \\
240 & 80 & 0.4 & 1.6 & 501$^2$ & Yes \\
251 & 80 & 0.5 & 1.6 & 501$^2$ & Yes \\
262 & 80 & 0.75 & 1.6 & 501$^2$ & Yes \\

\end{supertabular}%

} 

\begingroup
\section*{References}
\bibliographystyle{elsarticle-harv} 
\bibliography{ref_por,ref_clusters}

\begin{thebibliography}{74}
\expandafter\ifx\csname natexlab\endcsname\relax\def\natexlab#1{#1}\fi
\expandafter\ifx\csname url\endcsname\relax
  \def\url#1{\texttt{#1}}\fi
\expandafter\ifx\csname urlprefix\endcsname\relax\def\urlprefix{URL }\fi

\bibitem[{{Abramov} and {Mojzsis}(2011)}]{2011Icar..213..273A}
{Abramov}, O., {Mojzsis}, S.~J., 2011. {Abodes for life in carbonaceous
  asteroids?} \icarus 213, 273--279.

\bibitem[{Ahrens et~al.(2005)Ahrens, Geveci, and Law}]{paraview}
Ahrens, J., Geveci, B., Law, C., 2005. Paraview: An end-user tool for
  large-data visualization. Elsevier, ISBN-13: 978-0123875822, 2005.

\bibitem[{{Barr} and {Canup}(2008)}]{2008Icar..198..163B}
{Barr}, A.~C., {Canup}, R.~M., 2008. {Constraints on gas giant satellite
  formation from the interior states of partially differentiated satellites}.
  \icarus 198, 163--177.

\bibitem[{{Barshay} and {Lewis}(1976)}]{1976ARAA..14...81B}
{Barshay}, S.~S., {Lewis}, J.~S., 1976. {Chemistry of primitive solar
  material}. \araa 14, 81--94.

\bibitem[{Bottinga and Weill(1972)}]{bottinga1972viscosity}
Bottinga, Y., Weill, D.~F., 1972. The viscosity of magmatic silicate liquids: a
  model calculation. Am. J. Sci. 272, 438--475.

\bibitem[{{Chambers}(2010)}]{2010Icar..208..505C}
{Chambers}, J.~E., 2010. {Planetesimal formation by turbulent concentration}.
  \icarus 208, 505--517.

\bibitem[{Ciesla et~al.(2015)Ciesla, Mulders, Pascucci, and
  Apai}]{ciesla2015volatile}
Ciesla, F.~J., Mulders, G.~D., Pascucci, I., Apai, D., 2015. Volatile delivery
  to planets from water-rich planetesimals around low mass stars. \apj 804, 9.

\bibitem[{{Cobb} and {Pudritz}(2014)}]{2014ApJ...783..140C}
{Cobb}, A.~K., {Pudritz}, R.~E., 2014. {Nature's Starships. I. Observed
  Abundances and Relative Frequencies of Amino Acids in Meteorites}. \apj 783,
  140--152.

\bibitem[{Cobb et~al.(2015)Cobb, Pudritz, and Pearce}]{0004-637X-809-1-6}
Cobb, A.~K., Pudritz, R.~E., Pearce, B. K.~D., 2015. {Nature's Starships. II:
  Simulating the Synthesis of Amino Acids in Meteorite Parent Bodies}. \apj
  809, 6.

\bibitem[{{Costa} et~al.(2009){Costa}, {Caricchi}, and
  {Bagdassarov}}]{2009GGG....10.3010C}
{Costa}, A., {Caricchi}, L., {Bagdassarov}, N., 2009. {A model for the rheology
  of particle-bearing suspensions and partially molten rocks}. Geochem.
  Geophys. Geosys 10, 3010.

\bibitem[{{Crameri} et~al.(2012){Crameri}, {Schmeling}, {Golabek}, {Duretz},
  {Orendt}, {Buiter}, {May}, {Kaus}, {Gerya}, and
  {Tackley}}]{2012GeoJI.189...38C}
{Crameri}, F., {Schmeling}, H., {Golabek}, G.~J., {Duretz}, T., {Orendt}, R.,
  {Buiter}, S.~J.~H., {May}, D.~A., {Kaus}, B.~J.~P., {Gerya}, T.~V.,
  {Tackley}, P.~J., 2012. {A comparison of numerical surface topography
  calculations in geodynamic modelling: an evaluation of the 'sticky air'
  method}. Geophys. J. Int. 189, 38--54.

\bibitem[{{Cuzzi} et~al.(2008){Cuzzi}, {Hogan}, and
  {Shariff}}]{2008ApJ...687.1432C}
{Cuzzi}, J.~N., {Hogan}, R.~C., {Shariff}, K., 2008. {Toward Planetesimals:
  Dense Chondrule Clumps in the Protoplanetary Nebula}. \apj 687, 1432--1447.

\bibitem[{{Davison} et~al.(2013){Davison}, {O'Brien}, {Ciesla}, and
  {Collins}}]{2013MPS...48.1894D}
{Davison}, T.~M., {O'Brien}, D.~P., {Ciesla}, F.~J., {Collins}, G.~S., 2013.
  {The early impact histories of meteorite parent bodies}. Meteorit. Planet.
  Sci. 48, 1894--1918.

\bibitem[{{Dullemond} et~al.(2014){Dullemond}, {Stammler}, and
  {Johansen}}]{2014ApJ...794...91D}
{Dullemond}, C.~P., {Stammler}, S.~M., {Johansen}, A., 2014. {Forming
  Chondrules in Impact Splashes. I. Radiative Cooling Model}. \apj 794,
  91--103.

\bibitem[{{Elkins-Tanton} et~al.(2011){Elkins-Tanton}, {Weiss}, and
  {Zuber}}]{2011E&PSL.305....1E}
{Elkins-Tanton}, L.~T., {Weiss}, B.~P., {Zuber}, M.~T., 2011. {Chondrites as
  samples of differentiated planetesimals}. Earth Planet. Sci. Lett. 305,
  1--10.

\bibitem[{{Elser} et~al.(2012){Elser}, {Meyer}, and
  {Moore}}]{2012Icar..221..859E}
{Elser}, S., {Meyer}, M.~R., {Moore}, B., 2012. {On the origin of elemental
  abundances in the terrestrial planets}. \icarus 221, 859--874.

\bibitem[{Fu and Elkins-Tanton(2014)}]{fu2014fate}
Fu, R.~R., Elkins-Tanton, L.~T., 2014. The fate of magmas in planetesimals and
  the retention of primitive chondritic crusts. Earth Planet. Sci. Lett. 390,
  128--137.

\bibitem[{{Gail} et~al.(2015){Gail}, {Henke}, and
  {Trieloff}}]{2015AA...576A..60G}
{Gail}, H.-P., {Henke}, S., {Trieloff}, M., 2015. {Thermal evolution and
  sintering of chondritic planetesimals. II. Improved treatment of the
  compaction process}. \aap 576, A60.

\bibitem[{{Gerya} and {Yuen}(2003)}]{2003PEPI..140..293G}
{Gerya}, T.~V., {Yuen}, D.~A., 2003. {Characteristics-based marker-in-cell
  method with conservative finite-differences schemes for modeling geological
  flows with strongly variable transport properties}. Phys. Earth Planet. Int.
  140, 293--318.

\bibitem[{{Gerya} and {Yuen}(2007)}]{2007PEPI..163...83G}
{Gerya}, T.~V., {Yuen}, D.~A., 2007. {Robust characteristics method for
  modelling multiphase visco-elasto-plastic thermo-mechanical problems}. Phys.
  Earth Planet. Int. 163, 83--105.

\bibitem[{{Ghosh} and {McSween}(1998)}]{1998Icar..134..187G}
{Ghosh}, A., {McSween}, H.~Y., 1998. {A Thermal Model for the Differentiation
  of Asteroid 4 Vesta, Based on Radiogenic Heating}. \icarus 134, 187--206.

\bibitem[{{Golabek} et~al.(2014){Golabek}, {Bourdon}, and
  {Gerya}}]{2014MPS...49.1083G}
{Golabek}, G.~J., {Bourdon}, B., {Gerya}, T.~V., 2014. {Numerical models of the
  thermomechanical evolution of planetesimals: Application to the
  acapulcoite-lodranite parent body}. Meteorit. Planet. Sci. 49, 1083--1099.

\bibitem[{{Golabek} et~al.(2011){Golabek}, {Keller}, {Gerya}, {Zhu}, {Tackley},
  and {Connolly}}]{2011Icar..215..346G}
{Golabek}, G.~J., {Keller}, T., {Gerya}, T.~V., {Zhu}, G., {Tackley}, P.~J.,
  {Connolly}, J.~A.~D., 2011. {Origin of the martian dichotomy and Tharsis from
  a giant impact causing massive magmatism}. \icarus 215, 346--357.

\bibitem[{{Goldreich} et~al.(2004){Goldreich}, {Lithwick}, and
  {Sari}}]{2004ARAA..42..549G}
{Goldreich}, P., {Lithwick}, Y., {Sari}, R., 2004. {Planet Formation by
  Coagulation: A Focus on Uranus and Neptune}. \araa 42, 549--601.

\bibitem[{{Greenberg} et~al.(1978){Greenberg}, {Hartmann}, {Chapman}, and
  {Wacker}}]{1978Icar...35....1G}
{Greenberg}, R., {Hartmann}, W.~K., {Chapman}, C.~R., {Wacker}, J.~F., 1978.
  {Planetesimals to planets - Numerical simulation of collisional evolution}.
  \icarus 35, 1--26.

\bibitem[{{Henke} et~al.(2012){Henke}, {Gail}, {Trieloff}, {Schwarz}, and
  {Kleine}}]{2012AA...537A..45H}
{Henke}, S., {Gail}, H.-P., {Trieloff}, M., {Schwarz}, W.~H., {Kleine}, T.,
  2012. {Thermal evolution and sintering of chondritic planetesimals}. \aap
  537, A45.

\bibitem[{{Herzberg} et~al.(2000){Herzberg}, {Raterron}, and
  {Zhang}}]{2000GGG.....1.1051H}
{Herzberg}, C., {Raterron}, P., {Zhang}, J., 2000. {New experimental
  observations on the anhydrous solidus for peridotite KLB-1}. Geochem.
  Geophys. Geosyst. 1, 1051--14.

\bibitem[{{Hevey} and {Sanders}(2006)}]{2006MPS...41...95H}
{Hevey}, P.~J., {Sanders}, I.~S., 2006. {A model for planetesimal meltdown by
  $^{26}$Al and its implications for meteorite parent bodies}. Meteorit.
  Planet. Sci. 41, 95--106.

\bibitem[{{Howard}(1964)}]{howardconvection}
{Howard}, L.~N., 1964. {Convection at high Rayleigh number}. Proceedings of the
  11th International Congress in Applied Mechanics. Springer, Berlin, pp.
  1109--1115.

\bibitem[{Hunter(2007)}]{matplotlib}
Hunter, J.~D., 2007. Matplotlib: A 2d graphics environment. Computing In
  Science \& Engineering 9, 90--95.

\bibitem[{{Jacobsen} et~al.(2008){Jacobsen}, {Yin}, {Moynier}, {Amelin},
  {Krot}, {Nagashima}, {Hutcheon}, and {Palme}}]{jacobsen200826}
{Jacobsen}, B., {Yin}, Q.-Z., {Moynier}, F., {Amelin}, Y., {Krot}, A.~N.,
  {Nagashima}, K., {Hutcheon}, I.~D., {Palme}, H., 2008. {$^{26}$Al- $^{26}$Mg
  and $^{207}$Pb- $^{206}$Pb systematics of Allende CAIs: Canonical solar
  initial $^{26}$Al/ $^{27}$Al ratio reinstated}. Earth Planet. Sci. Lett. 272,
  353--364.

\bibitem[{{Johansen} et~al.(2015){Johansen}, {Mac Low}, {Lacerda}, and
  {Bizzarro}}]{2015SciA....115109J}
{Johansen}, A., {Mac Low}, M.-M., {Lacerda}, P., {Bizzarro}, M., 2015. {Growth
  of asteroids, planetary embryos, and Kuiper belt objects by chondrule
  accretion}. Sci. Adv. 1, 1500109.

\bibitem[{{Johansen} et~al.(2007){Johansen}, {Oishi}, {Mac Low}, {Klahr},
  {Henning}, and {Youdin}}]{2007Natur.448.1022J}
{Johansen}, A., {Oishi}, J.~S., {Mac Low}, M.-M., {Klahr}, H., {Henning}, T.,
  {Youdin}, A., 2007. {Rapid planetesimal formation in turbulent circumstellar
  disks}. \nat 448, 1022--1025.

\bibitem[{Jutzi et~al.(2008)Jutzi, Benz, and Michel}]{jutzi2008numerical}
Jutzi, M., Benz, W., Michel, P., 2008. Numerical simulations of impacts
  involving porous bodies: I. implementing sub-resolution porosity in a 3d sph
  hydrocode. Icarus 198, 242--255.

\bibitem[{Jutzi et~al.(2009)Jutzi, Michel, Hiraoka, Nakamura, and
  Benz}]{jutzi2009numerical}
Jutzi, M., Michel, P., Hiraoka, K., Nakamura, A.~M., Benz, W., 2009. Numerical
  simulations of impacts involving porous bodies: {II}. comparison with
  laboratory experiments. Icarus 201, 802--813.

\bibitem[{{Kleine} et~al.(2009){Kleine}, {Touboul}, {Bourdon}, {Nimmo},
  {Mezger}, {Palme}, {Jacobsen}, {Yin}, and {Halliday}}]{2009GeCoA..73.5150K}
{Kleine}, T., {Touboul}, M., {Bourdon}, B., {Nimmo}, F., {Mezger}, K., {Palme},
  H., {Jacobsen}, S.~B., {Yin}, Q.-Z., {Halliday}, A.~N., 2009. {Hf-W
  chronology of the accretion and early evolution of asteroids and terrestrial
  planets}. \gca 73, 5150--5188.

\bibitem[{{Kraichnan}(1962)}]{1962PhFl....5.1374K}
{Kraichnan}, R.~H., 1962. {Turbulent Thermal Convection at Arbitrary Prandtl
  Number}. Phys. Fluids 5, 1374--1389.

\bibitem[{{Liebske} et~al.(2005){Liebske}, {Schmickler}, {Terasaki}, {Poe},
  {Suzuki}, {Funakoshi}, {Ando}, and {Rubie}}]{2005EPSL.240..589L}
{Liebske}, C., {Schmickler}, B., {Terasaki}, H., {Poe}, B.~T., {Suzuki}, A.,
  {Funakoshi}, K.-I., {Ando}, R., {Rubie}, D.~C., 2005. {Viscosity of
  peridotite liquid up to 13 GPa: Implications for magma ocean viscosities}.
  Earth Planet. Sci. Lett. 240, 589--604.

\bibitem[{{Mackwell}(1991)}]{1991GeoRL..18.2027M}
{Mackwell}, S.~J., 1991. {High-temperature rheology of enstatite: Implications
  for creep in the mantle}. \grl 18, 2027--2030.

\bibitem[{{Mishra} et~al.(2016){Mishra}, {Marhas}, and
  {Sameer}}]{2016E&PSL.436...71M}
{Mishra}, R.~K., {Marhas}, K.~K., {Sameer}, Feb. 2016. {Abundance of $^{60}$Fe
  inferred from nanoSIMS study of QUE 97008 (L3.05) chondrules}. Earth Planet.
  Sci. Lett. 436, 71--81.

\bibitem[{{Morbidelli} et~al.(2009){Morbidelli}, {Bottke}, {Nesvorn{\'y}}, and
  {Levison}}]{2009Icar..204..558M}
{Morbidelli}, A., {Bottke}, W.~F., {Nesvorn{\'y}}, D., {Levison}, H.~F., 2009.
  {Asteroids were born big}. \icarus 204, 558--573.

\bibitem[{{Morbidelli} et~al.(2015){Morbidelli}, {Lambrechts}, {Jacobson}, and
  {Bitsch}}]{2015Icar..258..418M}
{Morbidelli}, A., {Lambrechts}, M., {Jacobson}, S., {Bitsch}, B., 2015. {The
  great dichotomy of the Solar System: Small terrestrial embryos and massive
  giant planet cores}. \icarus 258, 418--429.

\bibitem[{{Neumann} et~al.(2014){Neumann}, {Breuer}, and
  {Spohn}}]{2014AA...567A.120N}
{Neumann}, W., {Breuer}, D., {Spohn}, T., 2014. {Modelling of compaction in
  planetesimals}. \aap 567, A120.

\bibitem[{{Parker} et~al.(2014){Parker}, {Church}, {Davies}, and
  {Meyer}}]{2014MNRAS.437..946P}
{Parker}, R.~J., {Church}, R.~P., {Davies}, M.~B., {Meyer}, M.~R., 2014.
  {Supernova enrichment and dynamical histories of solar-type stars in
  clusters}. \mnras 437, 946--958.

\bibitem[{{Parker} and {Dale}(2016)}]{2016MNRAS.456.1066P}
{Parker}, R.~J., {Dale}, J.~E., 2016. {Did the Solar system form in a
  sequential triggered star formation event?} \mnras 456, 1066--1072.

\bibitem[{{Pinkerton} and {Stevenson}(1992)}]{1992JVGR...53...47P}
{Pinkerton}, H., {Stevenson}, R.~J., 1992. {Methods of determining the
  rheological properties of magmas at sub-liquidus temperatures}. J. Volcanol.
  Geotherm. Res. 53, 47--66.

\bibitem[{{Qin} et~al.(2008){Qin}, {Dauphas}, {Wadhwa}, {Masarik}, and
  {Janney}}]{2008EPSL.273...94Q}
{Qin}, L., {Dauphas}, N., {Wadhwa}, M., {Masarik}, J., {Janney}, P.~E., 2008.
  {Rapid accretion and differentiation of iron meteorite parent bodies inferred
  from $^{182}$Hf-$^{182}$W chronometry and thermal modeling}. Earth Planet.
  Sci. Lett. 273, 94--104.

\bibitem[{Ranalli(1995)}]{ranalli1995rheology}
Ranalli, G., 1995. Rheology of the Earth. Chapman and Hall, New York.

\bibitem[{{Roberts}(1967)}]{1967JFM....30...33R}
{Roberts}, P.~H., 1967. {Convection in horizontal layers with internal heat
  generation. Theory}. Fluid Mech. 30, 33--49.

\bibitem[{{Rubie} et~al.(2003){Rubie}, {Melosh}, {Reid}, {Liebske}, and
  {Righter}}]{2003EPSL.205..239R}
{Rubie}, D.~C., {Melosh}, H.~J., {Reid}, J.~E., {Liebske}, C., {Righter}, K.,
  2003. {Mechanisms of metal-silicate equilibration in the terrestrial magma
  ocean}. Earth Planet. Sci. Lett. 205, 239--255.

\bibitem[{Sahijpal et~al.(2007)Sahijpal, Soni, and
  Gupta}]{sahijpal2007numerical}
Sahijpal, S., Soni, P., Gupta, G., 2007. Numerical simulations of the
  differentiation of accreting planetesimals with $^{26}${A}l and $^{60}${F}e
  as the heat sources. Meteorit. Planet. Sci. 42, 1529--1548.

\bibitem[{{Sanders} and {Scott}(2012)}]{2012MPS...47.2170S}
{Sanders}, I.~S., {Scott}, E.~R.~D., 2012. {The origin of chondrules and
  chondrites: Debris from low-velocity impacts between molten planetesimals?}
  Meteorit. Planet. Sci. 47, 2170--2192.

\bibitem[{{Sanders} and {Taylor}(2005)}]{2005ASPC..341..915S}
{Sanders}, I.~S., {Taylor}, G.~J., 2005. {Implications of $^{26}$Al in Nebular
  Dust: Formation of Chondrules by Disruption of Molten Planetesimals}. In:
  {Krot}, A.~N., {Scott}, E.~R.~D., {Reipurth}, B. (Eds.), Chondrites and the
  Protoplanetary Disk. Vol. 341 of Astronomical Society of the Pacific
  Conference Series. pp. 915--932.

\bibitem[{{Schmeling} et~al.(2008){Schmeling}, {Babeyko}, {Enns}, {Faccenna},
  {Funiciello}, {Gerya}, {Golabek}, {Grigull}, {Kaus}, {Morra}, {Schmalholz},
  and {van Hunen}}]{2008PEPI..171..198S}
{Schmeling}, H., {Babeyko}, A.~Y., {Enns}, A., {Faccenna}, C., {Funiciello},
  F., {Gerya}, T., {Golabek}, G.~J., {Grigull}, S., {Kaus}, B.~J.~P., {Morra},
  G., {Schmalholz}, S.~M., {van Hunen}, J., 2008. {A benchmark comparison of
  spontaneous subduction models-Towards a free surface}. Phys. Earth Planet.
  Int. 171, 198--223.

\bibitem[{Schubert et~al.(1986)Schubert, Spohn, and
  Reynolds}]{schubert1986thermal}
Schubert, G., Spohn, T., Reynolds, R.~T., 1986. Thermal histories, compositions
  and internal structures of the moons of the solar system. In: Satellites.
  Arizona University Press, pp. 224--292.

\bibitem[{{Siggia}(1994)}]{1994AnRFM..26..137S}
{Siggia}, E.~D., 1994. {High Rayleigh number convection}. Annu. Rev. Fluid
  Mech. 26, 137--168.

\bibitem[{Solomatov(2015)}]{solomatov2015magma}
Solomatov, V.~S., 2015. Magma oceans and primordial mantle differentiation.
  Treatise on Geophysics 2nd ed., pp. 81--104.

\bibitem[{Sotin and Labrosse(1999)}]{sotin1999three}
Sotin, C., Labrosse, S., 1999. Three-dimensional thermal convection in an
  iso-viscous, infinite prandtl number fluid heated from within and from below:
  applications to the transfer of heat through planetary mantles. Phys. Earth
  Planet. Int. 112, 171--190.

\bibitem[{{Stolper} et~al.(1981){Stolper}, {Hager}, {Walker}, and
  {Hays}}]{1981JGR....86.6261S}
{Stolper}, E., {Hager}, B.~H., {Walker}, D., {Hays}, J.~F., 1981. {Melt
  segregation from partially molten source regions - The importance of melt
  density and source region size}. \jgr 86, 6261--6271.

\bibitem[{{Suzuki} et~al.(1998){Suzuki}, {Ohtani}, and
  {Kato}}]{1998PEPI..107...53S}
{Suzuki}, A., {Ohtani}, E., {Kato}, T., 1998. {Density and thermal expansion of
  a peridotite melt at high pressure}. Phys. Earth Planet. Int. 107, 53--61.

\bibitem[{{Tackley} et~al.(2001){Tackley}, {Schubert}, {Glatzmaier}, {Schenk},
  {Ratcliff}, and {Matas}}]{2001Icar..149...79T}
{Tackley}, P.~J., {Schubert}, G., {Glatzmaier}, G.~A., {Schenk}, P.,
  {Ratcliff}, J.~T., {Matas}, J.-P., 2001. {Three-Dimensional Simulations of
  Mantle Convection in Io}. \icarus 149, 79--93.

\bibitem[{{Tang} and {Dauphas}(2012)}]{2012E&PSL.359..248Ta}
{Tang}, H., {Dauphas}, N., 2012. {Abundance, distribution, and origin of
  $^{60}$Fe in the solar protoplanetary disk}. Earth Planet. Sci. Lett. 359,
  248--263.

\bibitem[{{Tarduno} et~al.(2012){Tarduno}, {Cottrell}, {Nimmo}, {Hopkins},
  {Voronov}, {Erickson}, {Blackman}, {Scott}, and
  {McKinley}}]{2012Sci...338..939T}
{Tarduno}, J.~A., {Cottrell}, R.~D., {Nimmo}, F., {Hopkins}, J., {Voronov}, J.,
  {Erickson}, A., {Blackman}, E., {Scott}, E.~R.~D., {McKinley}, R., 2012.
  {Evidence for a Dynamo in the Main Group Pallasite Parent Body}. Science 338,
  939--942.

\bibitem[{{Thrane} et~al.(2006){Thrane}, {Bizzarro}, and
  {Baker}}]{2006ApJ...646L.159T}
{Thrane}, K., {Bizzarro}, M., {Baker}, J.~A., 2006. {Extremely Brief Formation
  Interval for Refractory Inclusions and Uniform Distribution of $^{26}$Al in
  the Early Solar System}. \apjl 646, L159--L162.

\bibitem[{{Tkalcec} and {Brenker}(2014)}]{2014MPS...49.1202Ta}
{Tkalcec}, B.~J., {Brenker}, F.~E., 2014. {Plastic deformation of olivine-rich
  diogenites and implications for mantle processes on the diogenite parent
  body}. Meteorit. Planet. Sci. 49, 1202--1213.

\bibitem[{{Tkalcec} et~al.(2013){Tkalcec}, {Golabek}, and
  {Brenker}}]{2013NatGe...6...93T}
{Tkalcec}, B.~J., {Golabek}, G.~J., {Brenker}, F.~E., 2013. {Solid-state
  plastic deformation in the dynamic interior of a differentiated asteroid}.
  Nature Geosci. 6, 93--97.

\bibitem[{{Tr{\o}nnes} and {Frost}(2002)}]{2002EPSL.197..117T}
{Tr{\o}nnes}, R.~G., {Frost}, D.~J., 2002. {Peridotite melting and mineral-melt
  partitioning of major and minor elements at 22-24.5 GPa}. Earth Planet. Sci.
  Lett. 197, 117--131.

\bibitem[{Turcotte and Schubert(2014)}]{turcotte2014geodynamics}
Turcotte, D.~L., Schubert, G., 2014. Geodynamics, 3rd ed. Cambridge University
  Press, p. 636.

\bibitem[{{Wade} and {Wood}(2005)}]{2005EPSL.236...78W}
{Wade}, J., {Wood}, B.~J., 2005. {Core formation and the oxidation state of the
  Earth}. Earth Planet. Sci. Lett. 236, 78--95.

\bibitem[{{Weidenschilling}(1977)}]{1977MNRAS.180...57W}
{Weidenschilling}, S.~J., 1977. {Aerodynamics of solid bodies in the solar
  nebula}. \mnras 180, 57--70.

\bibitem[{{Weidenschilling} and {Cuzzi}(2006)}]{2006mess.book..473W}
{Weidenschilling}, S.~J., {Cuzzi}, J.~N., 2006. {Accretion Dynamics and
  Timescales: Relation to Chondrites. Meteorites and the Early Solar System II,
  1}. pp. 473--485.

\bibitem[{{Weiss} and {Elkins-Tanton}(2013)}]{2013AREPS..41..529W}
{Weiss}, B.~P., {Elkins-Tanton}, L.~T., 2013. {Differentiated Planetesimals and
  the Parent Bodies of Chondrites}. Annu. Rev. Earth Planet. Sci. 41, 529--560.

\bibitem[{{Yomogida} and {Matsui}(1984)}]{1984EPSL..68...34Y}
{Yomogida}, K., {Matsui}, T., 1984. {Multiple parent bodies of ordinary
  chondrites}. Earth Planet. Sci. Lett. 68, 34--42.

\bibitem[{{Zsom} et~al.(2010){Zsom}, {Ormel}, {G{\"u}ttler}, {Blum}, and
  {Dullemond}}]{2010AA...513A..57Z}
{Zsom}, A., {Ormel}, C.~W., {G{\"u}ttler}, C., {Blum}, J., {Dullemond}, C.~P.,
  2010. {The outcome of protoplanetary dust growth: pebbles, boulders, or
  planetesimals? II. Introducing the bouncing barrier}. \aap 513, A57.

\end{thebibliography}

\endgroup

\end{document}